\documentclass[twocolumn,showpacs,showkeys,preprintnumbers,superscriptaddress,amsmath,floatfix,amssymb,aps,secnumarabic,nofootinbib,floatfix]{revtex4-1}
\usepackage[colorlinks=true]{hyperref}
\usepackage[utf8]{inputenc}
\usepackage{graphicx}
\usepackage{placeins}
\usepackage{braket}
\usepackage{amssymb}
\usepackage{amsmath}
\usepackage{bbold}
\usepackage{epsfig}
\usepackage{float}
\usepackage{nicefrac}
\usepackage{xcolor}
\usepackage{color}
\usepackage{bm}
\usepackage{subfigure}
\usepackage{chngcntr}

\def\({\left(}
\def\){\right)}
\def\[{\left[}
\def\]{\right]}

\newcommand{\vev}[1]{ \langle \, #1 \, \rangle }

\newcommand{\beq} {\begin{eqnarray}}
\newcommand{\eeq} {\end{eqnarray}}
\newcommand{\bml} {\begin{multiline}}
\newcommand{\eml} {\end{multiline}}
\newcommand{\nn}{ \nonumber} 

\newcommand{\comment}[1]{}

\newcommand{\Tr}{ {\rm Tr} \, }

\newcommand{\bpsi}{{\bar \psi}}

\begin{document}
\sloppy

\title{Bridging the gap between numerics and experiment in free standing graphene}

\author{Maksim~Ulybyshev}
\email{maksim.ulybyshev@uni-wuerzburg.de}
\affiliation{Institute for Theoretical Physics, Julius-Maximilians-Universit\"at W\"urzburg,
   97074 W\"urzburg, Germany}

\author{Savvas~Zafeiropoulos}
\email{Savvas.Zafeiropoulos@cpt.univ-mrs.fr}
\affiliation{ Aix Marseille Univ, Universit\'e de Toulon, CNRS, CPT, Marseille, France}

\author{Christopher~Winterowd}
\email{winterowd@itp.uni-frankfurt.de}
\affiliation{Johann Wolfgang Goethe-Universit\"at Frankfurt am Main,Frankfurt am Main, Germany}

\author{Fakher~Assaad}
\email{Fakher.Assaad@physik.uni-wuerzburg.de}
\affiliation{Institute for Theoretical Physics, Julius-Maximilians-Universit\"at W\"urzburg,
   97074 W\"urzburg, Germany}
\affiliation{W\"urzburg-Dresden Cluster of Excellence ct.qmat, Julius-Maximilians-Universit\"at W\"urzburg,
   97074 W\"urzburg, Germany}

\begin{abstract}

We report results of large-scale quantum Monte Carlo (QMC) simulations of graphene. Using cutting-edge algorithmic improvements,  we are  able to consider spatial volumes, corresponding to 20808  electrons, that allow  us to access energy scales of direct relevance to  experiments.  Using constrained random phase approximation (cRPA)  estimates of short-ranged interactions combined with a  Coulomb tail, we are able to successfully confront  numerical and experimental estimates of the Fermi velocity renormalization, thus fixing the parameters of microscopic Hamiltonian to exactly reproduce the experimental data. Subsequent comparison of the QMC results with perturbation theory not only show the non-Fermi liquid character of graphene, but also prove the importance of lattice-scale physics and higher-order perturbative corrections beyond RPA for the quantitative description of the experimental data for the Fermi velocity renormalization in suspended graphene. The same Monte Carlo technique is subsequently employed to investigate electrical transport by computing the renormalization of the optical conductivity in freestanding graphene. The large lattice sizes and low temperatures accessible in our QMC simulations enable a clear resolution of the Dirac plateau in the spectral function, even in the presence of interactions. We demonstrate that, despite the strong renormalization of the single-particle dispersion, the optical conductivity remains essentially constant, in agreement with current experimental observations. This finding points toward the possibility of extending the theorem on the absence of optical conductivity renormalization for the Hubbard model on the hexagonal lattice to the case of long-range interactions.
 
\end{abstract}

\maketitle

\section*{\label{sec:Intro}Introduction}

Since its experimental discovery in 2004 \cite{Novoselov666}, graphene has  
attracted the attention of both the condensed-matter and the high-energy physics community. This is partly due to the fact that its low-energy electronic excitations can be described by a variant of quantum electrodynamics (QED)\cite{InteractionRMP2012}. As the electronic properties of graphene can be probed experimentally, it provides an ideal playground to test our ability to quantitatively check our theoretical concepts and computational techniques developed for strongly correlated Quantum Field Theories (QFTs). For instance, one can test the famous argument put forth by Dyson regarding the breakdown of the QED perturbative expansion \cite{PhysRev.85.631}. Another long-standing problem was the possible realization of spontaneous chiral symmetry breaking in this effective QED, especially in freestanding graphene, where screening by the substrate is completely removed and the electron–electron interaction is maximal \cite{PhysRevB.81.075429, PhysRevB.82.121403, PhysRevB.75.235423, PhysRevB.77.075115, PhysRevLett.102.026802, KHVESHCHENKO2004323}.

We start from the general action of the low-energy effective field theory (EFT), which includes both the long-range Coulomb interaction and short-range couplings:  
\begin{eqnarray}
S= && \int \mathrm{d}t \mathrm{d}^2x ( i \bar \psi_a \gamma^0 \partial_0 \psi_a   + i v_F \bar \psi_a \gamma^i \partial_i \psi_a - e \bar \psi_a \gamma^0 \psi_a A_0 ) \nonumber \\  && + \frac{1}{2}\int \mathrm{d}t \mathrm{d}^3x(\partial_i A_0)^2 + S_{U} + S_{V}, \label{eq:action} \\  && 
 S_U = G_U \int \mathrm{d}^2x \left\{  (\bpsi_\alpha \psi_\alpha )^2 +  (\bpsi_\alpha \gamma_0 \psi_\alpha )^2 \right. \nonumber \\  &&  \left. - \frac{1}{2}(\bpsi_\alpha \gamma_i \gamma_3 \psi_\alpha )^2 - \frac{1}{2}(\bpsi_\alpha \gamma_0 \gamma_i \gamma_3 \psi_\alpha )^2 \right\}, \label{eq:actionSU}  \\  && S_V = G_V \int \mathrm{d}^2x \left\{  (\bpsi_\alpha \gamma_0 \psi_\alpha )^2  -  (\bpsi_\alpha \psi_\alpha )^2 \right\},\label{eq:actionSV}
\end{eqnarray}
where $\psi_a$ is a two-flavor, four-component Dirac spinor.  The effective fine-structure constant is rescaled by the ratio of the speed of light to the Fermi velocity $v_F$, which is roughly $300$.  Actions $S_U$ and $S_V$ describe the effect of Hubbard interaction and nearest-neighbor interaction interaction, with constants $G_U$ and $G_V$ being proportional to respective microscopic interactions. Derivation of these terms is included in the Supplementary material \cite{Supp_mat} and largely follows \cite{PhysRevB.79.085116}.  These additional interactions account for the physics of Gross-Neveu (GN) transition, where the Dirac semimetal transforms into Mott insulator with antiferromagnetic (AFM) ordering \cite{PhysRevLett.97.146401}. However, this physics can not be directly observed on physical graphene despite the ratio of the on-site interaction to hopping ($U=9.3$ eV vs $t=2.7$ eV) being very close to critical value of 3.8 reported for GN transition on hexagonal lattice \cite{PhysRevX.3.031010}. The reason is that the nearest-neighbor interaction and the whole Coulomb tail pushes the GN transition to much larger interactions, which can not be achieved even in free-standing graphene \cite{PhysRevLett.111.056801}. Thus the accessible interaction effects are dominated by the Coulomb tail, which, with the enhanced fine structure constant, effectively models strongly-correlated QED. In conventional QED, the perturbative series in the fine-structure constant is asymptotic and gives increasingly accurate results up to a very large order, roughly the value of the inverse fine-structure constant ($\approx 137$).  It is thus thought that the perturbative series for effective QED in graphene will display its asymptotic behavior at much lower order than ordinary QED, demonstrating the inadequacy of perturbation theory. Ideally, the deviation between the perturbative results calculated within the low-energy continuum theory and experimental data should give a clear indication of the onset of this divergence \cite{PhysRevB.89.235431}. The search for such indications is the primary subject of this paper.  We are thus concentrating mainly on the effects of long-range Coulomb tail, trying to numerically quantify the viability of the perturbative approach in the description of the electronic properties of free-standing graphene, where the interactions should play maximum role.

However, before diving into comparison of perturbative series with the the results from non-perturbative QMC simulations, we also numerically check the role of the local interactions $S_U$ and $S_V$ in EFT. This issue constitutes the second subject studied in this paper: how accurately freestanding graphene can be described by the strongly-correlated QED neglecting the local interaction and the lattice-scale physics in general (e.g. deviation of the electronic dispersion from the Dirac cone at large energies). The question is nontrivial, since loop corrections become increasingly important with growing coupling strength. And lattice-scale physics can start to influence these loop integrals, because in reality they should be performed over a finite Brillouin zone rather than over infinite momentum space.

In order to quantify effects of both short-range and long-range Coulomb interactions and lattice-scale physics, we actually perform QMC simulations using the microscopic tight-binding Hamiltonian with long-range Coulomb interaction:   
\begin{eqnarray}
\hat{{H}} = -\kappa \sum_{\sigma, \langle x,y\rangle} (  \hat a^\dag_{\sigma, x} \hat a_{\sigma, y} +  \mbox{h.c} ) + \frac{1}{2}\sum_{x,y} V_{x,y}\hat q_x \hat q_y, 
\label{eq:H}
\end{eqnarray}
where $\hat a^\dag_{\sigma, x}$, $\sigma=\uparrow, \downarrow$ are creation operators for the electrons on $\pi$-orbitals, $\hat q_x$ is the electron charge operator, and $\kappa=2.7$ eV is the nearest-neighbor hopping parameter \cite{RevModPhys.82.2673}. The next-nearest neighbor hopping matrix elements are neglected since they are an order of magnitude smaller. The matrix describing two-body interactions, $V_{x,y}$, provides a general description of the electron-electron coupling. A key unresolved issue concerns the parametrization of the interaction part of the microscopic Hamiltonian. While the hopping amplitudes are reliably established through numerous Density Functional Theory (DFT) calculations, the electron–electron interaction potentials have a much larger uncertainty. In particular, they likely require renormalization to account for screening effects arising from the $\sigma$-orbitals \cite{Wehling_2011}.  Thus, in order to fix the parameters of the electron–electron interaction, we compare with the experimental data the renormalization of physical quantities computed directly from the microscopic Hamiltonian \ref{eq:H}, focusing on the renormalization of the Fermi velocity $v_F$. The focus on the renormalization of $v_F$ is motivated by the fact that its infrared limit has been  investigated in several experiments \cite{EliasNaturePhysics2011, Yu3282} on suspended graphene. In addition, there exists a substantial body of theoretical literature addressing $v_F$ renormalization, both within perturbative approaches \cite{PhysRevB.59.R2474, SonPhysRevB2007, PhysRevLett.118.026403, PhysRevB.89.235431, PhysRevLett.98.216801, DasSarma:PhysRevB2007} and using non-perturbative approximations \cite{Stauber_2017fuj, PhysRevB.93.235425}.

Technically, explicit verification of the microscopic Hamiltonian through comparison with experiment was made possible by the use of a nonstandard Quantum Monte Carlo technique, namely the Hybrid Monte Carlo algorithm. In this approach, unlike in BSS-QMC, the fermionic determinant is treated stochastically \cite{PhysRevLett.102.026802, PhysRevLett.111.056801, 2013arXiv1304.1711D, PhysRevB.84.075123}. This algorithm has been primarily used for gauge theories in lattice Quantum Chromodynamics. However, even in tight-binding models for lattice fermions, it sometimes exhibits more favorable scaling with system size than the standard BSS-QMC technique, which allowed us to simulate systems with spatial sizes of up to $102\times102$ lattice unit cells.
introduction, now 
After successful verification of the microscopic Hamiltonian through comparison with experimental data, we turn to a comparison of QMC results with perturbation theory. As mentioned above, in order to distinguish between lattice effects and genuine deficiencies of perturbative expansions, we compare the QMC data with perturbative results obtained both in EFT and in lattice perturbation theory (LPT) constructed directly for the microscopic Hamiltonian \ref{eq:H}. By using QMC data instead of experimental results, we can provide a much cleaner comparison, since QMC data are not only more precise but also do not involve a number of possible systematic corrections that are difficult to account for theoretically (a more detailed discussion is given in Section \ref{sec:perturbative}).  A comparison of the QMC data for the renormalization of the Fermi velocity with perturbative theories reveals that both lattice-scale effects and higher-order perturbative corrections play an important role. First, the low-energy EFT results differ from those obtained in LPT starting already at the random phase approximation (RPA) level. Second, even the RPA-level diagrams within lattice perturbation theory are not sufficient to describe the QMC data, with some qualitative differences remaining between the two.

\begin{figure}[!t]
    \centering
	\includegraphics[angle=0,width=0.3\textwidth]{ 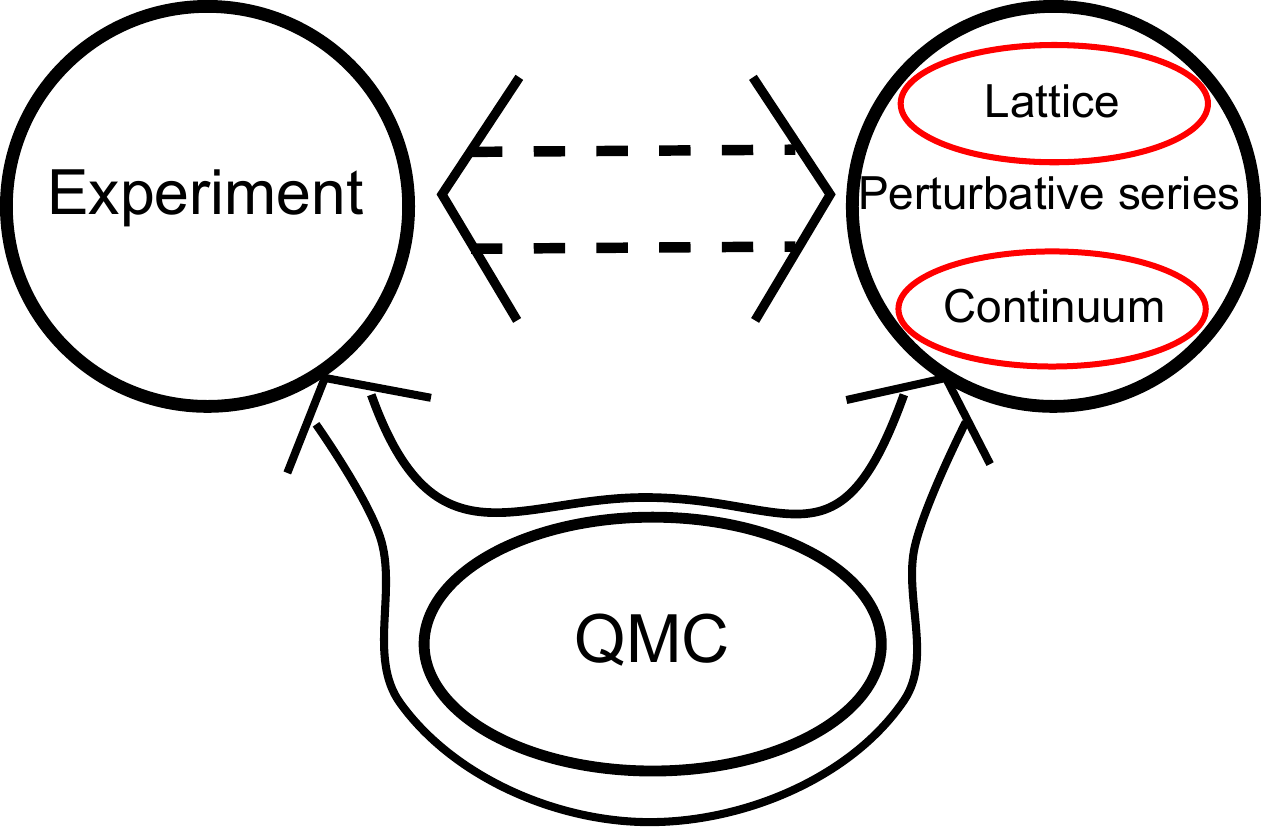}
	\caption{A schematic depiction of the relations between the different theoretical approaches which we are using to describe the electronic properties of free-standing graphene.}
\label{fig:scheme}
\end{figure}	

Summarizing the outline of our paper (see the scheme \ref{fig:scheme}), a direct comparison of experimental data with the results of effective low-energy field theory can hardly disentangle corrections arising from lattice-scale physics, systematic effects in experimental data that are not accounted for in theory, and a genuine breakdown of the perturbative series. Instead, we first verify the microscopic Hamiltonian by comparing QMC results with experiment and then compare the QMC results separately with LPT and effective QED calculations. This strategy provides a much cleaner comparison and allows us to disentangle lattice-physics effects from the onset of divergence in the perturbative series. It also enables us to clearly demonstrate the deficiencies of perturbation theory, at least at the level of RPA diagrams.

Section \ref{sec:experiment} is devoted to the verification of the microscopic Hamiltonian, Section \ref{sec:perturbative} presents the comparison of QMC data with LPT and effective QED calculations, and Section \ref{sec:LPTvsEFT} discusses the lattice effects that lead to differences between effective QED and LPT results at the RPA level. Finally, in Section \ref{sec:conductivity} and in the Summary, we discuss further perspectives of the employed techniques. As an example of their further applications, we study electronic transport in suspended graphene. In particular, we demonstrate the complete absence of renormalization of the optical conductivity even in suspended graphene.

\section {\label{sec:experiment}Comparison of QMC results with experiment}

We perform QMC calculations for the interacting tight-binding Hamiltonian \ref{eq:H} with the matrix of two-body interactions $V_{x,y}$ tuned in a way to model the electronic properties of suspended graphene. The salient features of $V_{x,y}$ are the on-site interaction $V_{x,x}=U_0$ and the long-range Coulomb tail, $V_{x,y}=\gamma U_0/(2|\vec R_{x,y}|)$, where $R_{x,y}$ is the distance between lattice sites in units of the spacing of the hexagonal lattice. In order to separately study both the effects of long-range and short-range interactions and to reproduce the experimental data for suspended graphene, we use three different interaction matrices $V_{x,y}$: 
\begin{itemize}
    \item  variant I uses short-range couplings computed with the cRPA method in \cite{Wehling_2011} and the strength of Coulomb tail is defined by the next-to-next-to-nearest coupling \cite{PhysRevLett.111.056801}, thus $\gamma=1.548$; 
    \item  variant II uses the same short-range couplings, but the long-range Coulomb tail is formed by a more complicated set of functions which fit the dielectric permittivity function $\varepsilon(k)$ \cite{Wehling_2011}, $\gamma=2.181$ in this case; 
    \item  variant III is generally the same as variant I except for a shift of the nearest-neighbor (from 5.5 eV to 6.7 eV) and next-to-nearest-neighbor (from 4.1 eV to 4.5 eV) couplings. 
\end{itemize}
The details of potential variant II are explained in \cite{PhysRevB.89.195429}, and it should yield the closest match to experimental data. The difference between the results obtained with potential variants I and III is used to elucidate the role of the short-range couplings.

The QMC calculations were performed on $102\times102$ and $48\times48$ lattices at three different temperatures $T=0.03125$ eV, $T=0.0625$ eV and $T=0.125$ eV in order to identify both finite-size and finite-temperature effects. Initially, we compute single-particle spectral functions (examples of them can be found in the Supplementary material), and peaks of those spectral functions give us the renormalized dispersion relation $E(k)$.  Due to the fact that we are simulating a finite volume, the resolution in momentum is limited by the lattice size. Thus, the numerical differentiation was used for the computation of $v_F$ (see the supplementary materials for the details). Another important consideration is that the experimental measurements for $v_F$ in suspended graphene \cite{EliasNaturePhysics2011, Yu3282} were done by tuning the electron density. Thus, the experimental curve represents the dependence of the Fermi velocity on the filling factor. Unfortunately, it is well known that QMC simulations performed on a bipartite lattice with electron–electron interactions suffer from the sign problem away from half filling. The average sign decreases exponentially with decreasing temperature and increasing lattice volume, which renders simulations on lattices as large as those employed in this work completely unfeasible outside half filling.  Moreover, the severity of the sign problem depends on the specific choice of the Hubbard–Stratonovich transformation \cite{PhysRevD.101.014508}. In the case of a spin-density–coupled auxiliary field, the sign problem is less severe, allowing simulations on smaller lattices at relatively high temperatures. However, the presence of long-range interactions forces us to couple the auxiliary fields to the charge density, which further exacerbates the sign problem. As a result, the simulations could only be performed at half filling. Nevertheless, in order to enable a direct comparison with experimental data, we convert the distance from the Dirac point obtained in the QMC simulations into an effective filling factor. This conversion neglects the interplay between filling and interactions; however, these effects can be separated reasonably well in the comparisons presented below.

\begin{figure}[t!]
   \centering
   \subfigure[] {\label{fig:Vf_comparisonA}\includegraphics[width=0.5\textwidth,clip]{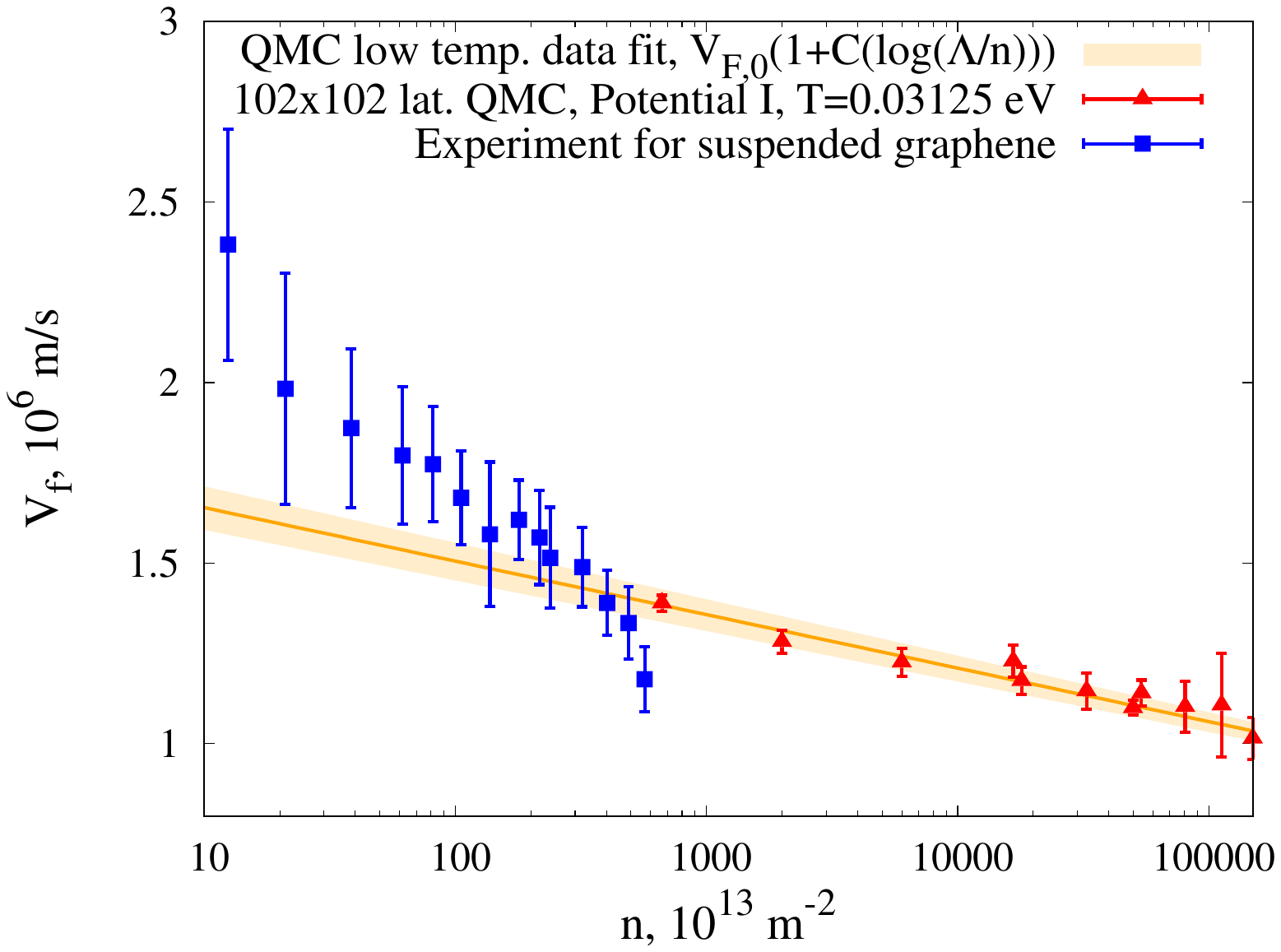}}
   \subfigure[]  {\label{fig:Vf_comparisonB}\includegraphics[width=0.5\textwidth,clip]{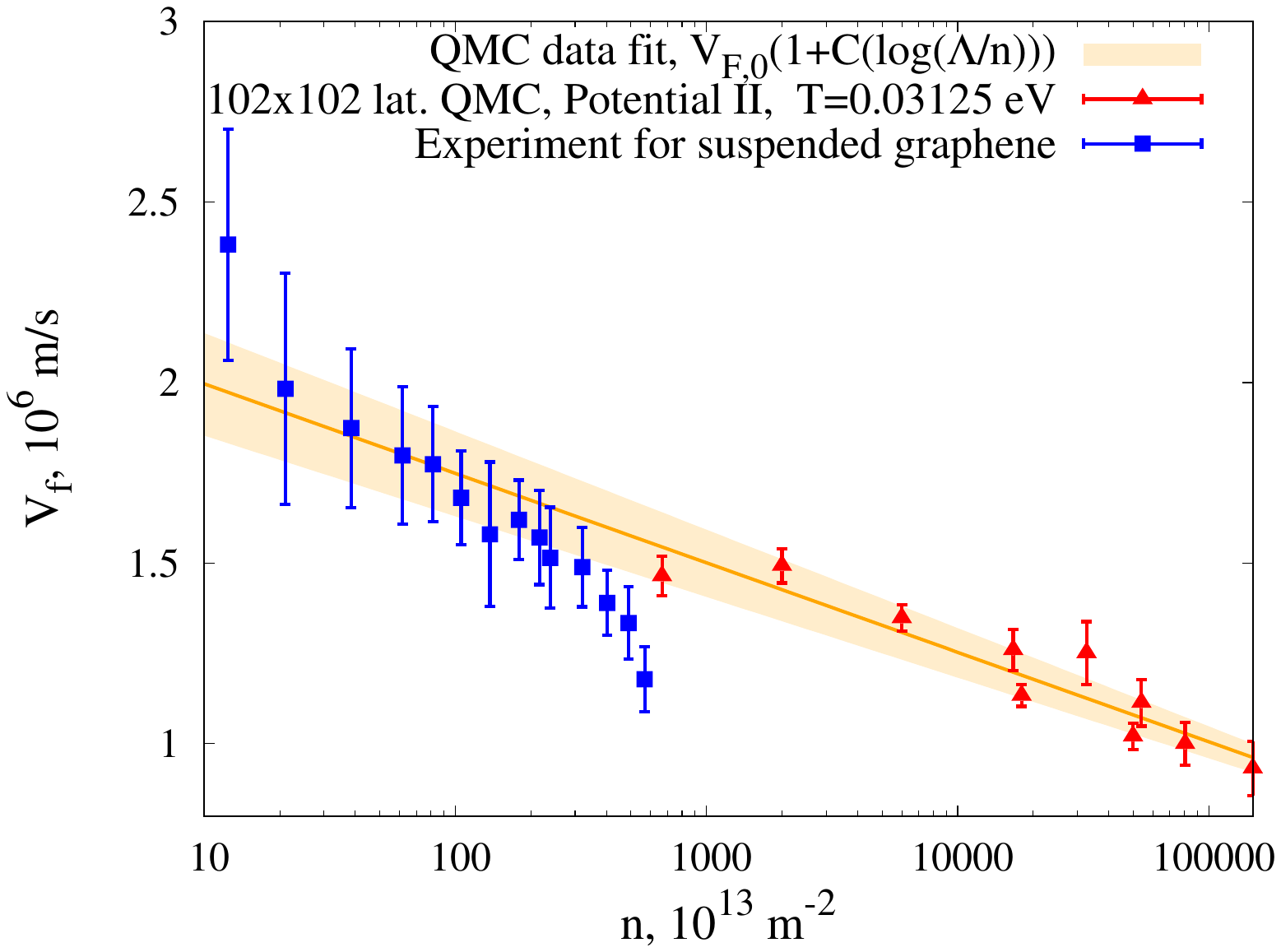}}
      \caption{Comparison of experimental results for suspended graphene with QMC data obtained with potential variants I \textbf{(a)} and II \textbf{(b)}. The latter corresponds to suspended graphene, while the former has a reduced Coulomb tail (see Section \ref{sec:experiment} for the details). The Fermi velocity is computed via finite differences from the dispersion relation on the lattice along K-K and K-M lines in the Brillouin zone (BZ). QMC results are computed at the lowest temperature $T=0.03125$ eV, corresponding to inverse temperatures $\beta \kappa = 86.4$ in  units of inverse hopping. The shaded area corresponds to the error of the logarithmic fit.}
	\label{fig:Vf_comparison}
\end{figure}

A comparison with experiment is shown in Fig. \ref{fig:Vf_comparison}, where the QMC data is displayed alongside the experimental data from \cite{EliasNaturePhysics2011}. The renormalized Fermi velocity is plotted as a function of density $n$, for two variants of the two-body interaction potential and the largest lattice and the lowest temperature used in QMC simulations. QMC data was fitted with the logarithmic function
\begin{equation}
v_F(k)=v_{F,0}(1+C \ln \Lambda/k),
\label{eq:v_f}
\end{equation}
where $v_{F,0}$ is non-renormalized value of the Fermi velocity. In order to outline the logarithmic nature of the renormalization more clearly, we used logarithmic scale in density $n$, thus the logarithmic function from \ref{eq:v_f} appears as a straight line in the figure. This arrangement also highlights the most relevant points at low densities.

Unlike previous QMC studies \cite{Tang570}, the lattice size appears to be large enough to clearly observe the non-linear dispersion relation which can be reliably fitted with the logarithmic function. Moreover, our sample sizes are just enough to get in touch with the scales probed in experiments.

The logarithmic fitting of the QMC data with its error bars displayed in Fig. \ref{fig:Vf_comparisonA} shows that the Coulomb tail corresponding to $\gamma=1.548$ is too small to reproduce the experimental data. This is understandable, as the Coulomb tail is reduced here in comparison with that in suspended graphene.  In contrast, the logarithmic fit of QMC data for the larger Coulomb tail  (Fig. \ref{fig:Vf_comparisonB}) demonstrates a remarkably good agreement with the experimental data points in the limit of small densities. One can also notice that the experimental points systematically shift below from the straight line representing the logarithm curve for densities larger then roughly $n=2\times10^{15}\,m^{-2}$. Our conjecture is that we observe exactly the effect mentioned above: since in experiment the actual measurements are done at a finite gate voltage, which is equivalent to the introduction of a chemical potential, the finite density of charge carriers can impose additional screening on the Coulomb interaction. This reason for the discrepancies between experimental results and theoretical computations at half filling was already mentioned in the literature \cite{PhysRevB.89.235431,Stauber_2017fuj}. This reasoning is also consistent with the observed effect that, in the limit of low densities, the additional screening disappears, leading to improved agreement between experiment and QMC results on Fig. \ref{fig:Vf_comparisonB}.

Another observation is that the extrapolated QMC data and the experimental results agree not only in the coefficient in front of the logarithm (i.e., the slope of the curve in Fig. \ref{fig:Vf_comparison}), but also in the vertical offset, which is essentially the cutoff $\Lambda$ in the fitting function \ref{eq:v_f}. This point is important because, as we will show below, the ultraviolet cutoff is sensitive to the short-range part of the electron–electron interaction. Thus, our comparison verifies not only the long-range Coulomb tail (which determines the slope of the curve), but also the short-range part of the interaction matrix in the microscopic Hamiltonian $\ref{eq:H}$. 

In summary, we conclude that our QMC results  provide a strong numerical evidence that the Fermi velocity is indeed logarithmically renormalized even in fully non-perturbative calculations beyond simple one-loop diagrams. Furthermore, the tight-binding Hamiltonian with the interaction matrix defined by the potentials of variant II, whose short-range part is taken from the cRPA calculations of \cite{Wehling_2011}, provides a reliable microscopic model for describing the electronic properties of suspended graphene.

\section {\label{sec:perturbative}Comparison of QMC results with perturbative expansions}

\begin{figure}[!t]
    \centering
	\includegraphics[angle=0,width=0.3\textwidth]{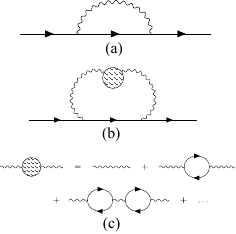}
	\caption{Diagrams used in perturbative calculations: (a) one-loop self energy; (b) RPA approximation with (c) dressed boson propagator.}
\label{fig:RPA}
\end{figure}

Having verified the microscopic Hamiltonian, we now turn to an analysis of the performance of perturbative expansions by comparing the QMC results with various versions of perturbation theory. A direct comparison with QMC data allows us to systematically investigate the influence of different parameters, which are difficult to control in real experiments. For example, we can independently vary the interaction potentials at short and long distances and also carefully investigate temperature effects.

 Due to the fact that we are simulating a finite volume, the resolution in momentum is limited by the lattice size. Thus, the numerical differentiation needed for the computation of $v_F$ brings about additional systematic errors (see supplementary material for examples). For this reason, in order to make precise numerical comparison of various approaches, we use the dispersion relation $E(k)$ directly. Indeed, the logarithmic renormalization of the Fermi velocity \ref{eq:v_f} also leads to the logarithmic renormalization of the energy itself: 
\begin{equation}
\frac{E(k)}{E_{0}(k)}=[1+C(1+\ln \Lambda/k)],
\label{eq:E_renorm}
\end{equation}
where $E_{0}(k)$ is the free dispersion relation. Thus we have a well-defined fitting function for the QMC and LPT data sets.

\begin{figure}[t!]
   \centering
   \subfigure[] {\label{fig:Dirac_comparisonA}\includegraphics[width=0.21\textwidth,clip]{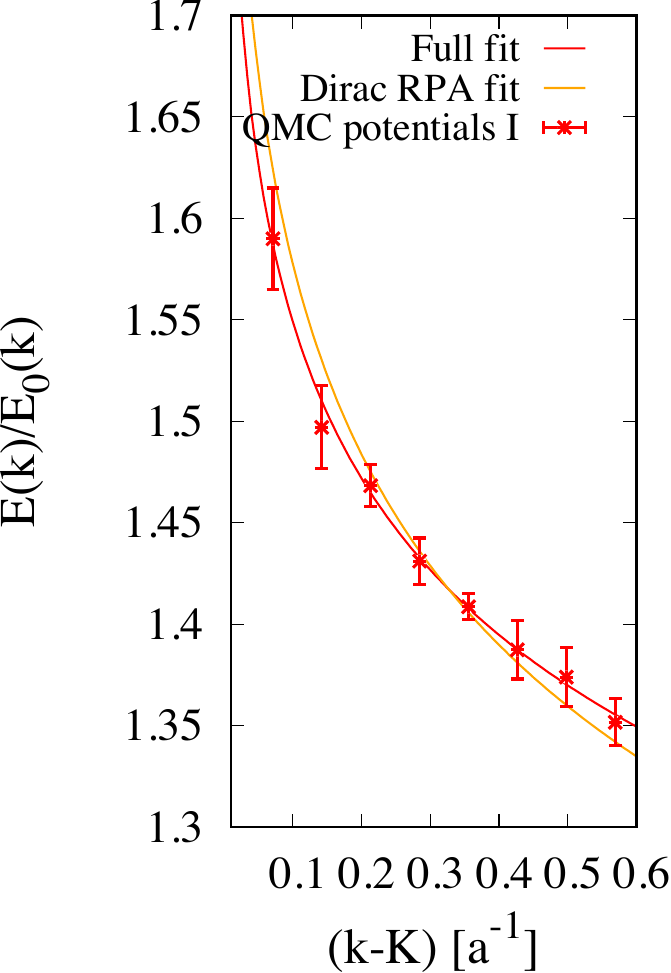}}
   \subfigure[]  {\label{fig:Dirac_comparisonB}\includegraphics[width=0.2\textwidth,clip]{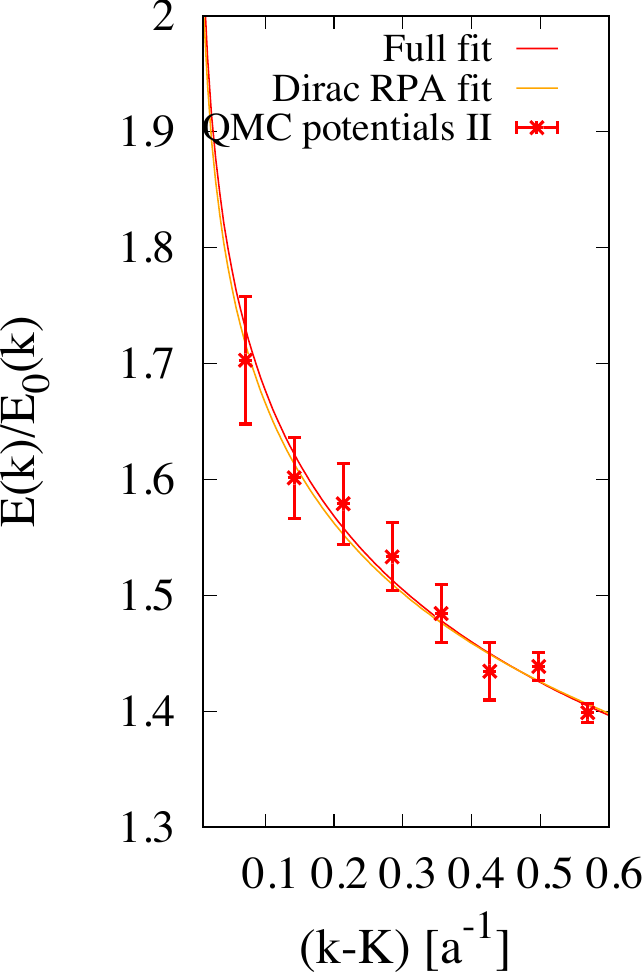}}
   \subfigure[] {\label{fig:Dirac_comparisonA}\includegraphics[width=0.2\textwidth,clip]{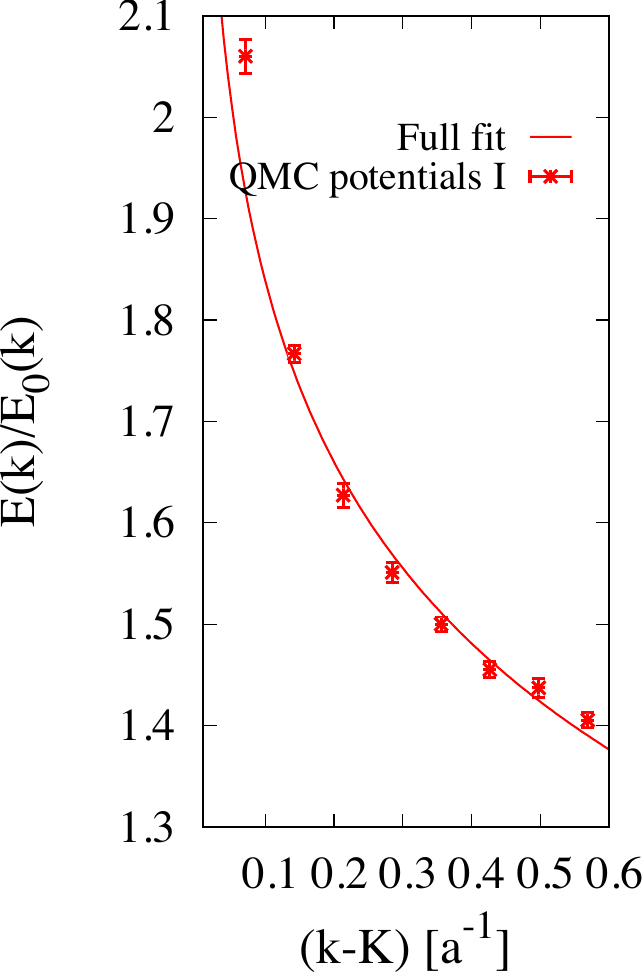}}
   \subfigure[]  {\label{fig:Dirac_comparisonB}\includegraphics[width=0.2\textwidth,clip]{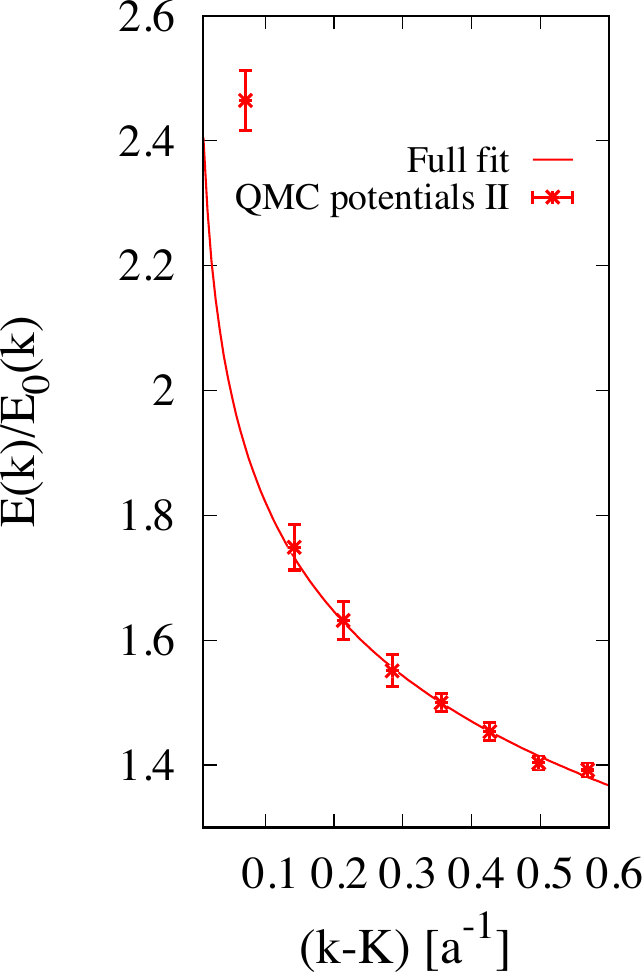}}
      \caption{Comparison of QMC data with EFT perturbation theory for potential variants I (\textbf{(a)} and \textbf{(c)}) and II (\textbf{(b)} and \textbf{(d)}). The latter corresponds to suspended graphene, while the former has a reduced Coulomb tail (see Section \ref{sec:experiment} for the details). QMC data are plotted along the K-K line in BZ. Lattice size is $102\times102$ and temperatures are  $T=0.03125$ eV (\textbf{(a)} and \textbf{(b)}) and $T=0.0625$ eV (\textbf{(c)} and \textbf{(d)}). Two types of fitting functions are used: 1) full fit according to the eq. \ref{eq:E_renorm}, where we tune both coefficient $C$ and cutoff $\Lambda$; 2) constrained fit, where the coefficient $C$ is fixed at its value from zero temperature RPA-level EFT expression. Corresponding fit coefficients for low temperature data are shown in table \ref{tab:data_tableDirac}. First point is omitted in the logarithmic fit of higher temperature data (see Section \ref{sec:perturbative} for the discussion on finite-temperature effects).}
	\label{fig:Dirac_comparison}
\end{figure}

\begin{table}[t!]
    \centering
    \begin{tabular}{|c|c|c|}
        \hline
                                 & Potential var. I & Potentials var. II  \\ \hline
       $C$ from QMC fit & $0.111 \pm 0.003$  &   $0.156 \pm 0.012$  \\ \hline
       $C$ from EFT (RPA)              & $0.13559$ &  $0.14944$   \\ \hline
    \end{tabular}
    \caption{$C$ coefficients in logarithmic fit (eq. \ref{eq:E_renorm}) for low temperature  QMC data (Fig. \ref{fig:Dirac_comparisonA} and \ref{fig:Dirac_comparisonB}) and corresponding values from RPA-level EFT calculations.}
    \label{tab:data_tableDirac}
\end{table}

As was already mentioned above, we use two different versions of the perturbation theory. One is the standard weak-coupling expansion of the EFT defined by the action $\ref{eq:action}$. Renormalized dispersion is obtained from fermion self-energy computed in one-loop (Fig. \ref{fig:RPA}\textcolor{red}{(a)} or Random Phase approximation (RPA), which differs from the former by the insertion of dressed bosonic propagator in the self-energy loop (Fig. \ref{fig:RPA}\textcolor{red}{(b,c)}). In both approximations the energy $E(k)$ is renormalized logarithmically, with expressions for coefficient $C$ obtained in several previous papers \cite{SonPhysRevB2007, GONZALEZ1994595}. For convenience, we reproduced corresponding expressions in Supplementary material. 

This theory however, completely neglects the lattice structure, since the Dirac dispersion relation is valid only in the vicinity of K points in BZ. For this reason, we have also employed the perturbation theory based on the initial microscopic Hamiltonian \ref{eq:H}. It can be formulated using exactly the same diagrammatic technique as the EFT perturbations, if one starts from he action which emerges in the Euclidean path integral for the partition function after the Hubbard-Stratonovich transformation: 
\begin{eqnarray}
\label{eq:PartitionFunctionLinearized}
&&Z = \int \mathcal{D} \phi \mathcal{D} \bar\psi \mathcal{D} \psi e^{-S[\bar\psi, \psi, \phi]}, \\ \label{eq:ActionSumofTerms}
&&S = S_0 + S_{\rm {int.}} + S_{\rm{B}}, \\ \nonumber && S = \int^{\beta}_0 \mathrm{d}\tau \bigg[ -\sum_{x,a} \bar\psi_{x,a} \partial_{\tau} \psi_{x,a} + \sum_{x,a;y,a} \bar\psi_{x,a} (H_0)_{x,a;y,b} \psi_{y,b} \\ \label{eq:ActionRealSpace} && \qquad \qquad \qquad + i \sum_{x,a,b}\phi_x \bar\psi_{x,a} \sigma^z_{a,b}\psi_{x,b} - \frac{1}{2} \sum_{x,y}\phi_x V^{-1}_{x,y}\phi_y \bigg],
\end{eqnarray} 
where $\psi_a$ are Grassmann fields for electrons ($a=1$) and holes ($a=2$) and $\phi$ is a charge-coupled scalar field. $H_0$ is the matrix of tight-binding Hamiltonian.  Similar approach has been already used in the comparison of QMC data with perturbative calculations in \cite{razmadze2024hubbardinteractionfinitet} on small clusters (up to 12 sites) for Hubbard model on hexagonal lattice. 
 
We first compare the QMC data with EFT; the results are shown in Fig. \ref{fig:Dirac_comparison}. The logarithmic renormalization of the energy is clearly visible. At first glance the use of zero temperature RPA-level expressions (though not the one-loop expressions, which yield values of the coefficient $C$ approximately three times larger) appears to be justified for the fitting of low-temperature QMC data. For potentials variant II, the agreement is very good, while for potentials variant I the value of the coefficient $C$ from EFT lies outside of the error bars of $C$ coefficient obtained from fitting the QMC data, although they remain relatively close (see table  \ref{tab:data_tableDirac}). However, even this situation is somewhat concerning, since potentials variant I corresponds to a weaker Coulomb tail and thus a smaller value of $\alpha$. Thus, theoretically, it should exhibit better agreement with perturbation theory than potentials variant II. Moreover, the apparent agreement between EFT and QMC data in Fig. \ref{fig:Dirac_comparison} is largely driven by the presence of a free parameter - the cutoff $\Lambda$ - in the EFT expression. Indeed, subsequent analysis shows that the seemingly good agreement between EFT and QMC data is in fact an artifact resulting from neglected lattice-scale effects.

\begin{figure}[]
   \centering
   \subfigure[] {\label{fig:energy_comparisonA}\includegraphics[width=0.45\textwidth,clip]{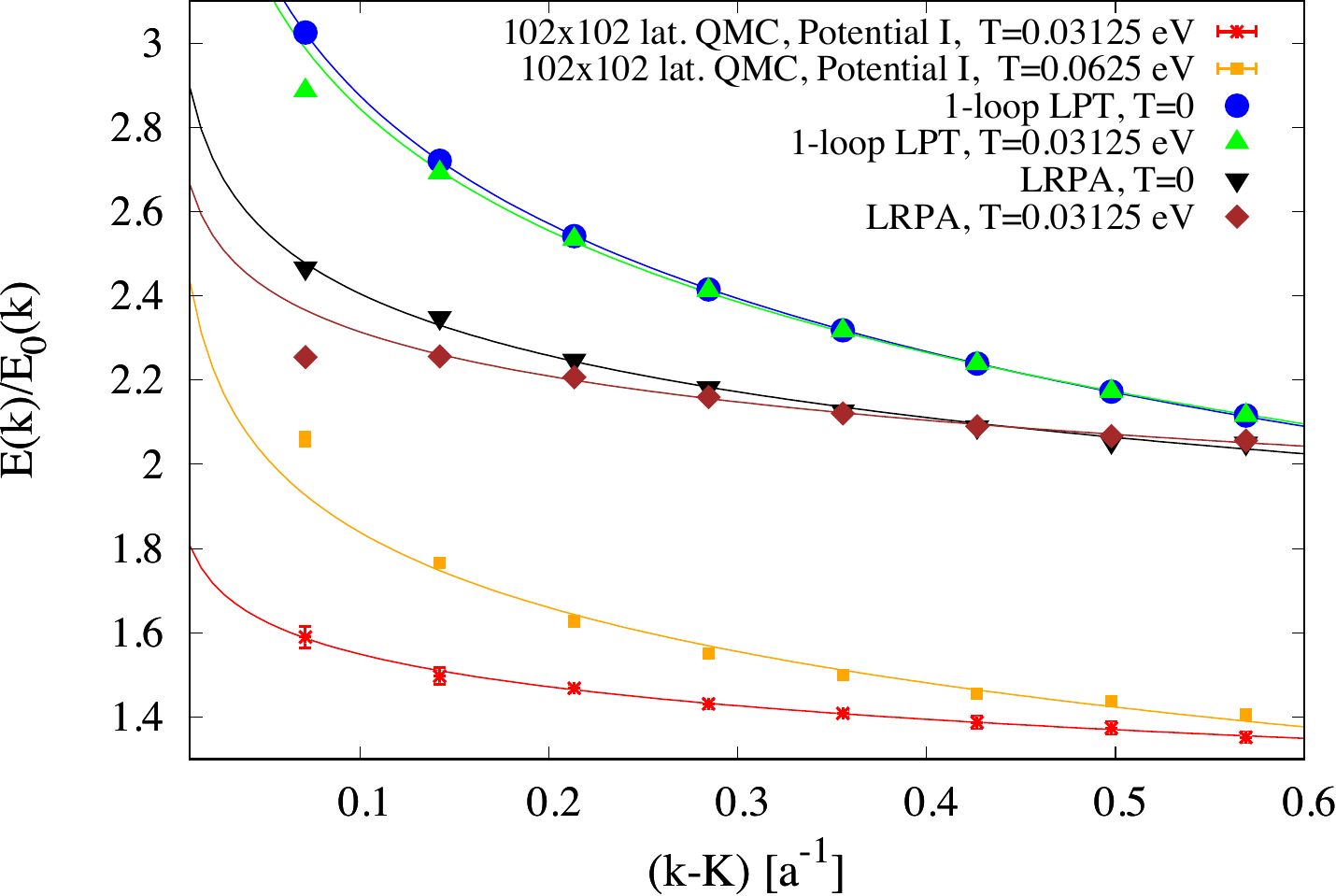}}
   \subfigure[]  {\label{fig:energy_comparisonB}\includegraphics[width=0.45\textwidth,clip]{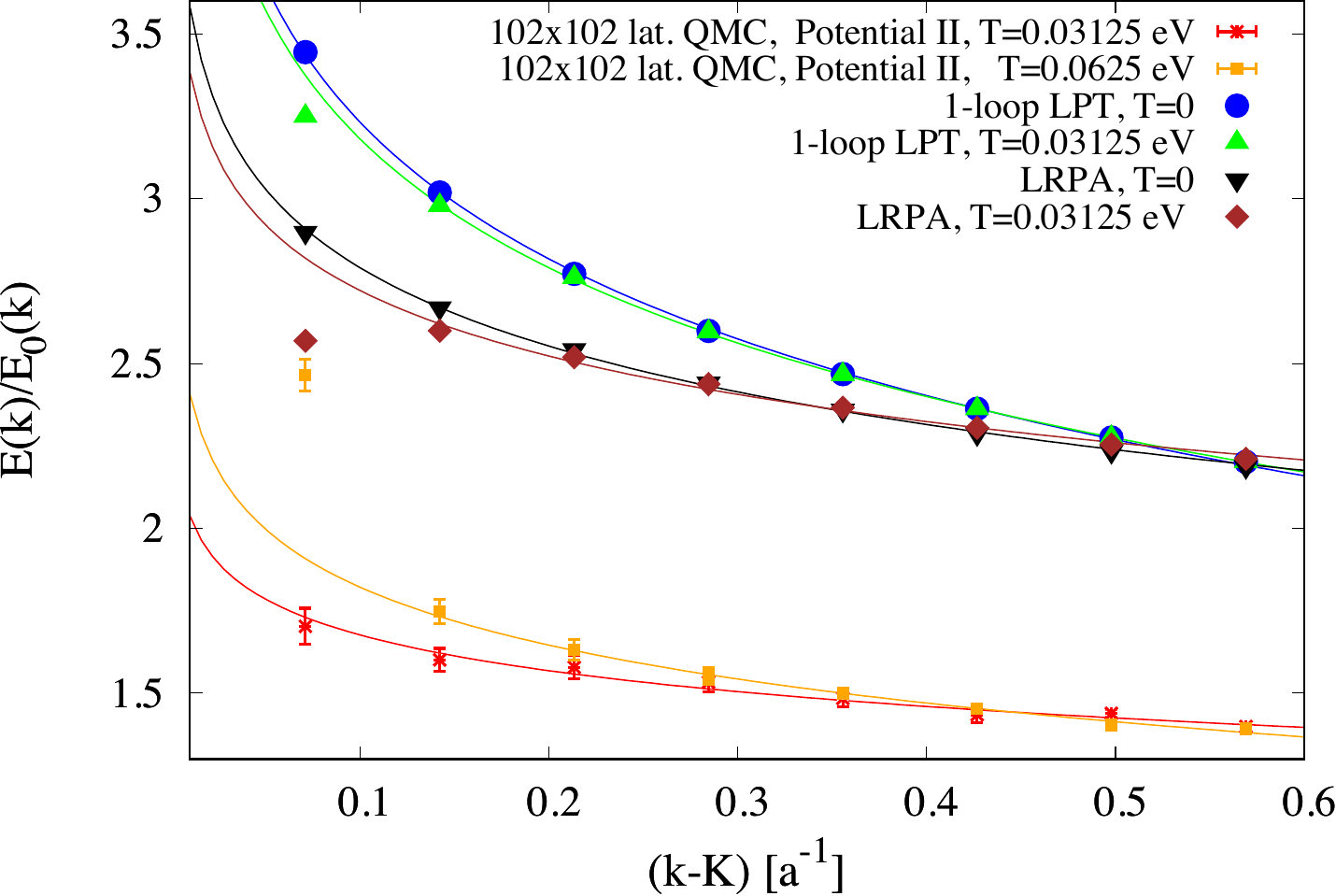}}
     \caption{Comparison of QMC data with perturbative results on the lattice (zero temperature and finite-temperature one-loop LPT, also zero temperature and finite-temperature lattice random phase approximation (LRPA)). Figure \textbf{(a)} shows data for Potential I and \textbf{(b)} shows data for Potential II. Both LPT and QMC results are obtained on a $102\times102$ lattice and we consider the K-K profile in the BZ. The fitting parameters $C$ are shown in the table \ref{tab:data_tableLPT}. The momentum closest to the Dirac point is omitted in the fit of the finite-temperature LPT data and in the fit of higher-temperature QMC data. }
	\label{fig:energy_comparison}
\end{figure}

A comparison with lattice perturbation theory (LPT) results is shown in Fig. \ref{fig:energy_comparison} for the same two variants of the electron–electron interaction. Since LPT explicitly incorporates the lattice structure, it does not contain free parameters such as the cutoff $\Lambda$. Therefore, the QMC and LPT data should coincide if LPT remains valid at the chosen interaction strength (indeed, QMC and LPT results converge to the same values when the interaction strength is reduced — see section \textcolor{red}{2.1} of the Supplementary Materials for the corresponding checks). However, our data indicate that at interaction strengths close to those relevant for suspended graphene, LPT is no longer quantitatively adequate for describing the renormalization of $v_F$. Both the one-loop and RPA-level results deviate significantly from the QMC data. This discrepancy is also reflected in the coefficients in front of the logarithm listed in Table \ref{tab:data_tableLPT}.

Another effect is that the coefficients in front of the logarithm differ significantly between LPT and EFT already at the RPA level. This indicates that lattice-scale physics becomes important as soon as the interaction strength is sufficient for polarization loops to play a significant role in the diagrams. This effect is investigated in more detail in the next section.

Comparison between the QMC and LPT data also reveals another effect, this time a qualitative one, signaling the breakdown of the perturbative series (at least at the RPA level). When we compare QMC data obtained at two different temperatures, the renormalization of $v_F$ clearly increases with temperature. At higher temperature, the points at the lowest momenta deviate from the logarithmic fit: they are shifted upwards relative to the logarithmic curve for both large and small Coulomb tails. This feature of the data cannot be attributed to finite-size effects. It is most pronounced at low energies, where lowering the temperature would be expected to enhance such effects. However, we observe the opposite behavior: the low-temperature data for the $102\times102$ lattice are perfectly aligned with the logarithmic fit for both sets of interaction potentials.

A similar effect can also be observed on the $48\times48$ lattice when comparing the two higher temperatures, $T=0.0625$ eV and $T=0.125$ eV (Fig. \textcolor{red}{S4} in the Supplementary Materials). When both the inverse temperature and the lattice size are reduced by a factor of two, we again observe an enhancement of the first data point relative to the logarithmic curve, compared to the same lattice at lower temperature. Notably, the subsequent data points do not exhibit significant finite-temperature effects, in full agreement with the $102\times102$ lattice results. We therefore conclude that these finite-temperature effects, which shift the dispersion relation (and hence also $v_F$) upward relative to the logarithmic curve, should in principle be observable in experiments.

\begin{table}[t!]
    \centering
    \begin{tabular}{|c|c|c|}
        \hline
                                 & Potential var. I & Potentials var. II  \\ \hline
        QMC T=0.03125 eV  & $0.111 \pm 0.003$  &   $0.156 \pm 0.012$  \\ \hline
        QMC T=0.0625 eV    & $0.258 \pm 0.014$ &  $0.254 \pm  0.013$   \\ \hline
        1-loop LPT T=0     & $0.438$     &   $0.599$  \\ \hline
        1-loop LPT T=0.0625eV & $0.417$ &  $0.564$   \\ \hline
        LPT RPA T=0        &  $0.212$    &   $0.343$  \\ \hline
        LPT RPA T=0.0625eV & $0.151$    &  $0.287$   \\ \hline
    \end{tabular}
    \caption{$C$ coefficients in logarithmic fit \ref{eq:E_renorm} for various datasets in Fig.  \ref{fig:energy_comparison}. }
    \label{tab:data_tableLPT}
\end{table}

\begin{figure}[!t]
    \centering
	\includegraphics[angle=0,width=0.45\textwidth]{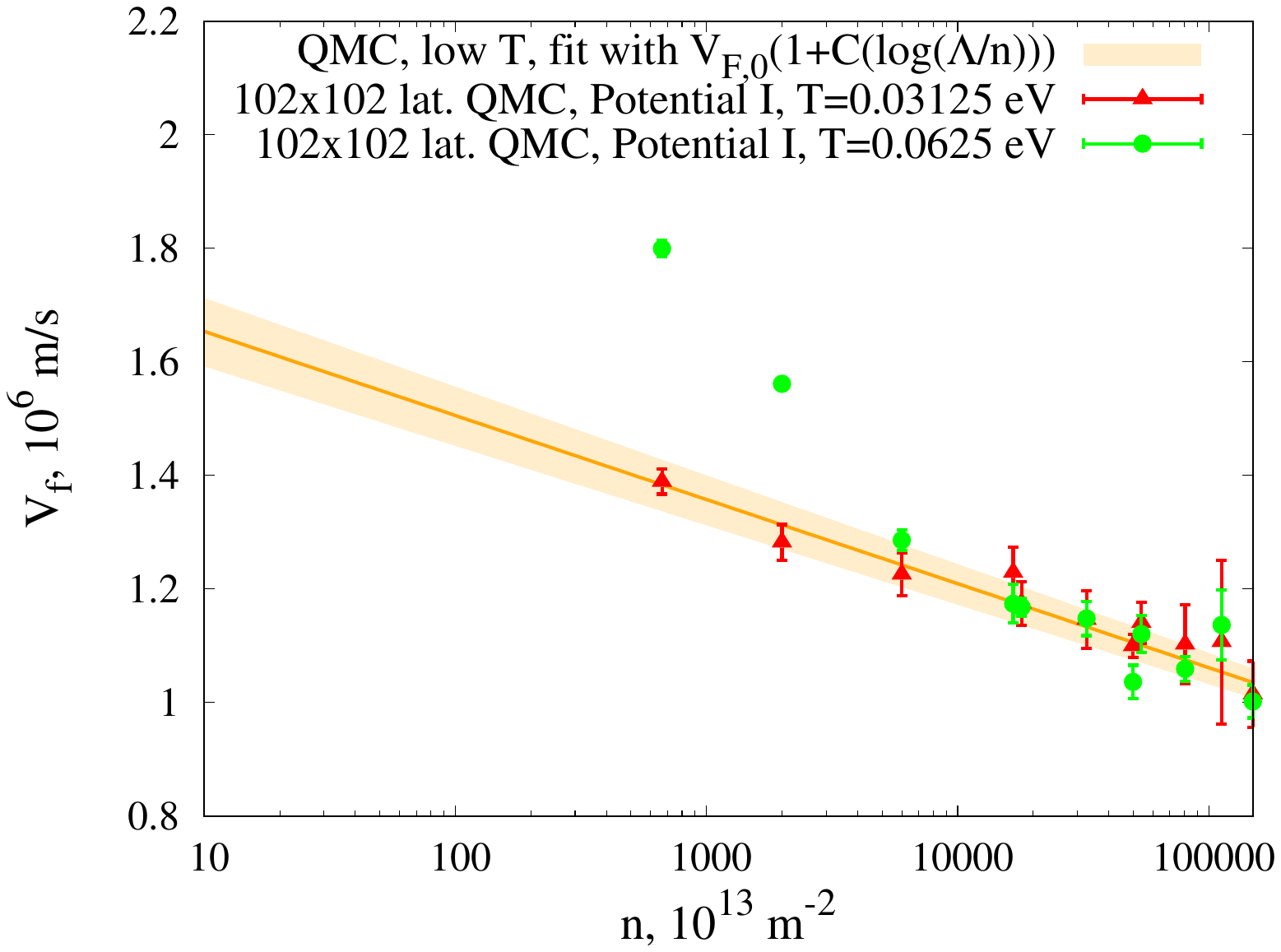}
	\caption{Finite temperature effects in the Fermi velocity renormalization.}
\label{fig:Vf_T}
\end{figure}

The finite-temperature effects in the LPT data, however, exhibit the opposite behavior. At both the RPA and one-loop levels, finite temperature also leads to a breakdown of the logarithmic renormalization of $v_F$ at the lowest momenta, but in this case the renormalization decreases with increasing temperature. Thus, finite-temperature effects provide a qualitative signature of the importance of higher-order corrections beyond the RPA level.  The corresponding effect on $v_F$ is shown in Fig. \ref{fig:Vf_T}. A comparison of QMC simulations performed at two different temperatures demonstrates that the logarithmic renormalization of the Fermi velocity breaks down once the measurements are carried out at fillings corresponding to energies $E_0(k)\leq 2\ldots3T$, with the Fermi velocity increasing more rapidly than predicted by the logarithmic curve.

One can therefore conclude that the apparent validity of the RPA-level EFT perturbations is just a numerical artifact of neglected lattice effects. On the other hand, comparison between LPT and QMC data unambiguously demonstrates the importance of higher-order corrections for a quantitative description of the renormalization of the Fermi velocity in suspended graphene. This effect could, in principle, be confirmed experimentally, provided that measurements are performed over suitable ranges of temperature and carrier density. On the theoretical side, achieving a more accurate perturbative description likely requires the inclusion of vertex corrections in LPT. These have been shown to play an important role in Schwinger–Dyson (SD) studies \cite{PhysRevB.94.125102}. Within the SD framework, an accurate description of both dynamical mass generation and Fermi velocity renormalization necessitates the use of a gauge-invariant vertex. Its longitudinal part is constrained by the Ward identity, while its transverse part must be free of kinematic singularities \cite{PhysRevD.22.2542}. Combined with the inclusion of dynamical screening via the Lindhard function, this approach has been shown to yield more reliable results.

\section {\label{sec:LPTvsEFT}Lattice vs continuum perturbative series}

\begin{figure}[]
	\includegraphics[angle=0,width=0.5\textwidth]{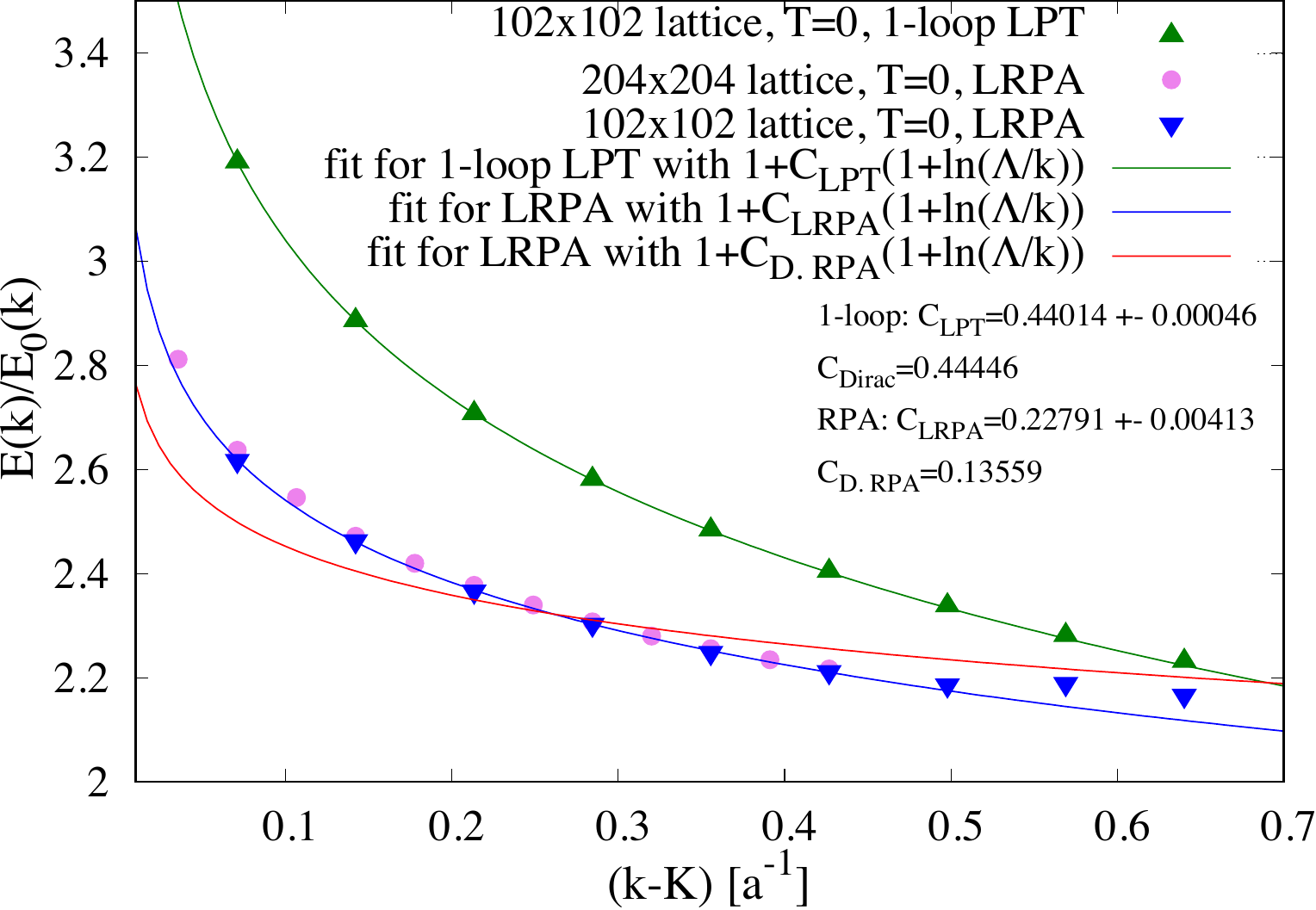}
	\caption{A comparison of lattice and continuum perturbation theory in the strongly-coupled regime. The on-site interaction is $U_0=9.2$ eV, all other potentials are defined by the Coulomb tail with  $\gamma=1.548$. For reference, we also include the values of the $C$ coefficients in EFT at one-loop and RPA level.  Lattice RPA data is shown on both $102 \times 102$ and $204 \times 204$ lattice to show the absence of any noticeable finite-lattice size effects.}
	\label{fig:energy_test_largeU}
\end{figure}

Next we turn to the discrepancy between EFT curves and LPT/QMC data. Since the Feynman diagrams included in LPT and EFT data are exactly the same, there are two possible sources for this discrepancy: first, the influence of lattice-scale physics, most notably, the presence of a physical cutoff accompanied by the curvature of the electronic dispersion away from the Dirac points; second, the differences between lattice potentials and continuous Coulomb interaction at small distances. 

Since LPT takes into account the latter effect, we can try to isolate the reasons for the discrepancies between the EFT and LPT data through a comparison with LPT at different orders of perturbation theory. One-loop LPT yields an almost identical $C$ as the one-loop EFT: compare $C=0.599$ in LPT for potential variant II and $C=0.626$ in EFT for Coulomb potential with the same long-range tail (see table \ref{tab:data_tableLPT}). This demonstrates that cRPA modifications to electron-electron interaction potentials at short distances do not influence the value of $C$ substantially. 

In order to present even more clear comparison of EFT and LPT, we eliminate the cRPA modifications to the Coulomb interaction completely and employ a two-body interaction, $V_{x,y}$, where all elements except the on-site one are defined by the Coulomb tail. Here we take $\gamma=1.548$, which corresponds to an effective fine structure constant of $\alpha=1.778$.  Corresponding comparison between LPT and continuum EFT at the one-loop and RPA level is shown in Fig. \ref{fig:energy_test_largeU}. We again observe that LPT and the continuum theory are basically identical at the one-loop level. However, they deviate substantially at the RPA level since it is impossible to describe the lattice data using $C$ taken from the continuum RPA expression and leaving only $\Lambda$ as a free parameter: $C$ coefficients (listed on the Fig. \ref{fig:energy_test_largeU}) are now too different from each other.  The same is also true for RPA-level diagrams for real potentials in suspended graphene:  a comparison of the $C$ coefficients extracted from LPT and the ones from EFT at the RPA level shows substantial differences: $C=0.343$ for LPT and $C=0.149$ for EFT for potentials variant II (see tables \ref{tab:data_tableDirac}  and \ref{tab:data_tableLPT}).

\begin{figure}[!t]
    \centering
	\includegraphics[angle=0,width=0.50\textwidth]{ 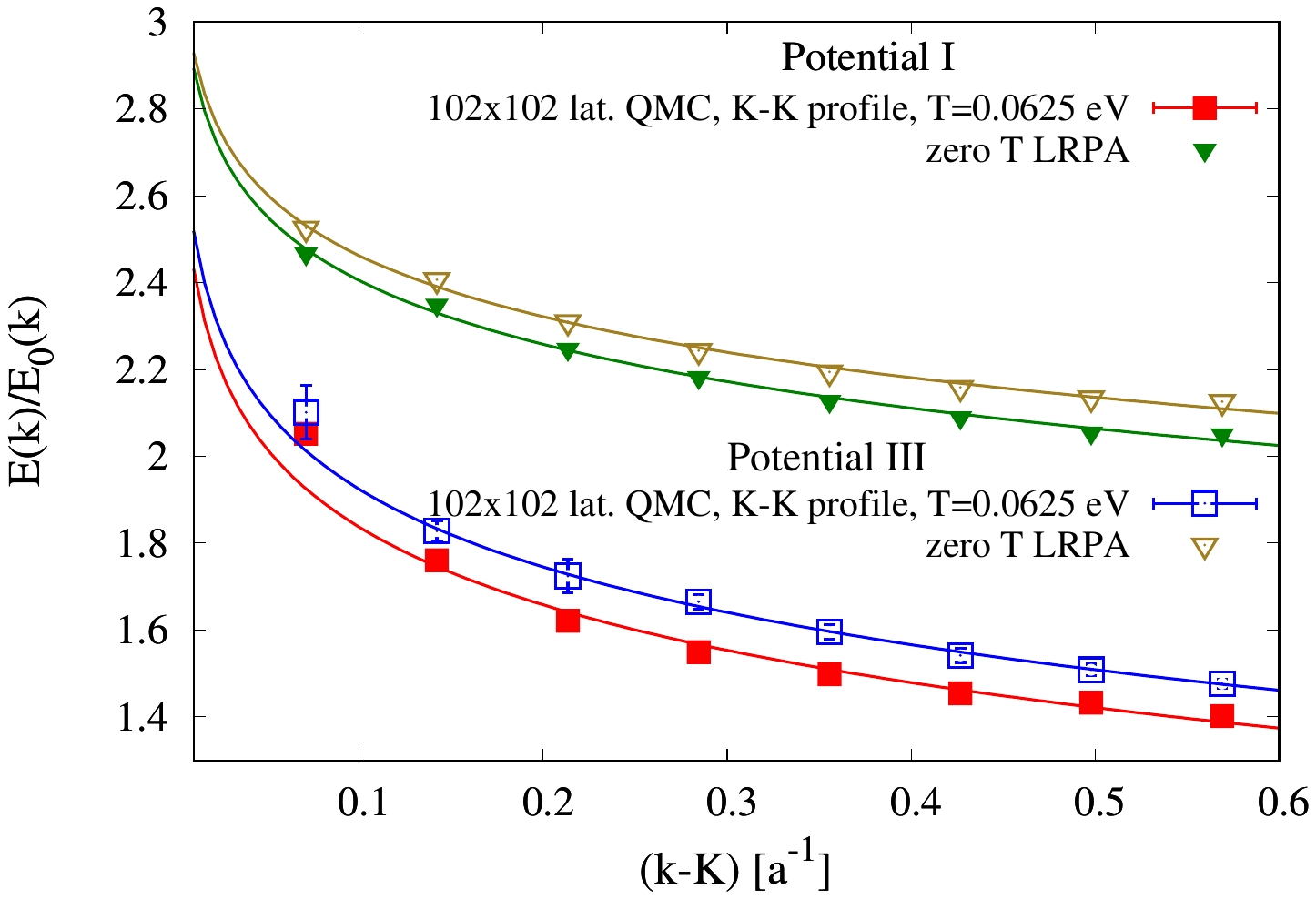}
	\caption{The role of the short-range couplings in determining the renormalized dispersion relation. We compare potentials PI and PIII, which differ only in the interaction between nearest-neighbors and next-to-nearest neighbors. The logarithmic fits are shown for all QMC and LPT data sets (the first points are omitted in the fits for finite-temperature profiles). The results of fitting the QMC data using function from Eq. \ref{eq:E_renorm}: $C$ coefficients for potentials PI and PIII respectively are equal to $C_{I}= 0.258 \pm 0.014 $ and $C_{III}= 0.258 \pm 0.006$; cutoff values for the same curves are equal to $\Lambda_{I}= 0.950 \pm 0.109$ and $\Lambda_{III}= 1.318 \pm 0.06 $.}
	\label{fig:compare_two_potentials}
\end{figure}

The influence of short-range couplings is further highlighted in Fig. \ref{fig:compare_two_potentials}, where we compare potential variant I and potential variant III, which differ only in the nearest-neighbor and next-to-nearest-neighbor couplings. If we look at the LPT data, the effect is clearly confined to the change of the cutoff, with the coefficient in front of the logarithm remaining constant. As a result, the dispersion curves are uniformly shifted vertically with respect to each other. The effect is exactly the same for QMC data, since the $C$ coefficients for two curves are almost identical (see the caption to the Fig. \ref{fig:compare_two_potentials}). This result is consistent with the general renormalization group (RG) point of view on the role of the couplings at different length scales. The short-range interactions in graphene are irrelevant at the infrared fixed point, and thus they can not change the qualitative behavior of $v_F$ at low momentum. However, they can still influence the results at a quantitative level by changing the parameters which are not fixed by the RG flow, such as the cutoff. Furthermore, these results show that fixing the cutoff is not a trivial procedure, as it is not solely defined by the bandwidth, but depends on interactions which are irrelevant in the RG sense. 

Summarizing, these two indications combined suggest that the main origin of discrepancies can not lie in cRPA corrections to interaction potentials at short distances, which leaves the lattice scale physics as the only source of discrepancies between LPT and EFT at RPA level.

\begin{figure*}[!t]
    \centering
	\includegraphics[angle=0,width=1.0\textwidth]{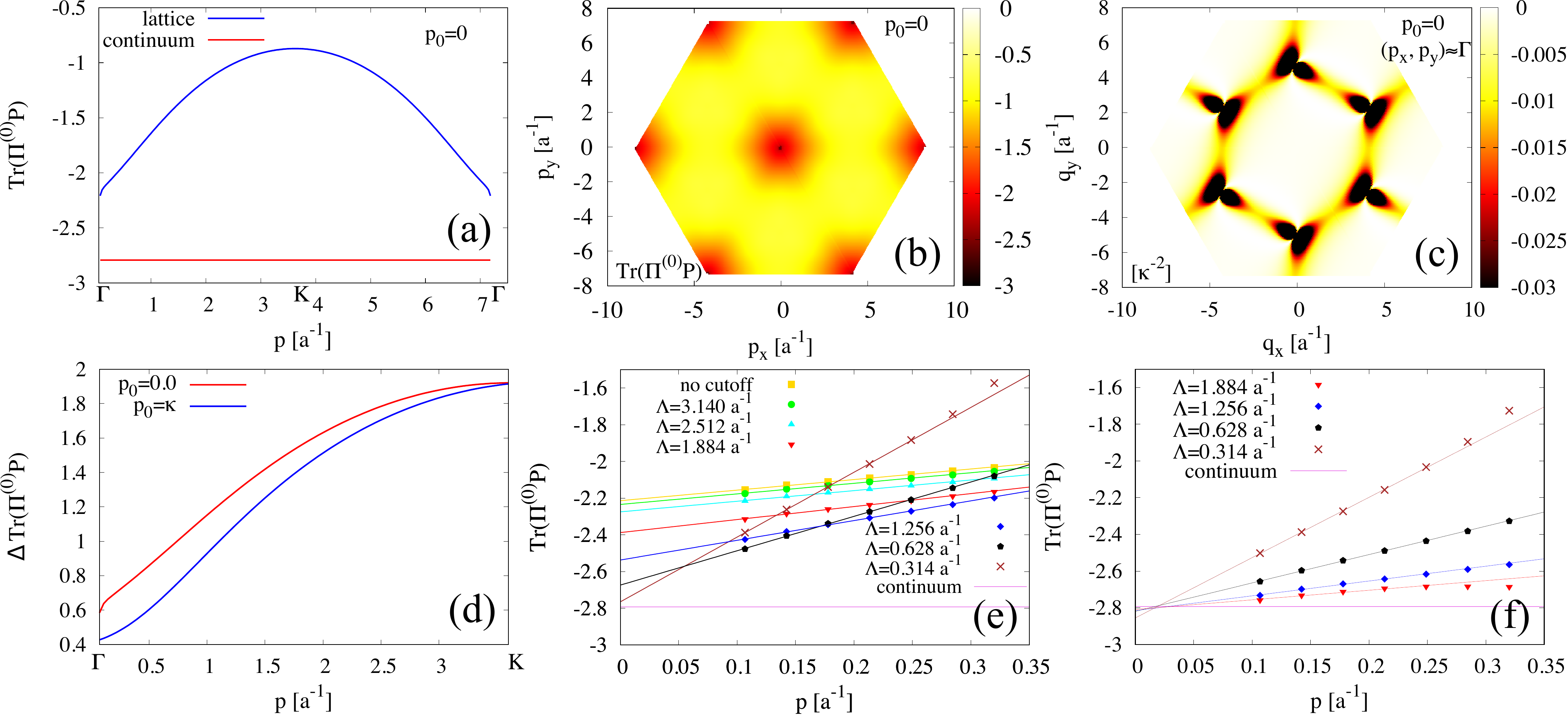}
	\caption{\textbf{(a)} Comparison of $\Tr(\Pi^{(0)} P)$ computed in LPT and in EFT. The external frequency is set to zero and the external momentum runs along the $\Gamma-K$ line. \textbf{(b)} Map of $\Tr(\Pi^{(0)} P)$ computed in LPT within the Brillouin zone on a $102\times102$ lattice at zero external frequency. \textbf{(c)}  Map of the integrand for the lattice polarization bubble within the Brillouin zone in LPT. The external momenta is slightly shifted away from the $\Gamma$ point and the external frequency is set to zero. The integrand is concentrated around the K points, as expected. However, there is a noticeable contribution from the regions between non-equivalent K points. \textbf{(d)} Difference between $\Tr(\Pi^{(0)} P)$ computed in LPT and in EFT at two values of the external frequency. External momentum runs along the $\Gamma-K$ line. \textbf{(e)} Behavior of $\Tr(\Pi^{(0)} P)$ on the lattice when a cutoff in momentum around the Dirac points is introduced in the polarization loop. External frequency set to zero, and external momentum is shown in the vicinity of $\Gamma$ point. For each cutoff value $\Lambda$ we perform a linear fit in the vicinity of zero external momentum. This shows that the continuum value of the trace, and thus the continuum dressed RPA interaction, is restored at zero momentum in the limit of small cutoff. \textbf{(f)} Same as in (e) but with $\Pi^{(0)}$ computed using a linearized dispersion relation for the electrons in the polarization bubble $P$ and using the continuum Coulomb propagator in $\Pi^{(0)}$. The continuum behavior is now restored at all momenta in the limit of large cutoff.}
	\label{fig:polarization_all}
\end{figure*}

The deviations of the continuum RPA results from their LPT counterparts can be attributed to differences between lattice and continuum polarization functions.
One recalls that the polarization bubble is only naively divergent in $(2+1)$-QED and can be computed using one of the conventional continuum regularization schemes to take care of the ultraviolet divergences, and thus there are no additional free parameters that can be added in the renormalization scheme in EFT to incorporate lattice corrections.

In order to show the differences between EFT and LPT explicitly, we consider the dressed bosonic propagator (see the diagram in Fig. \ref{fig:RPA}\textcolor{red}{(c)}):
\begin{equation} \label{eq:BosonicDysonEquation}
\hat \Pi(p_0,\vec{p}) = \hat \Pi^{(0)}(\vec{p}) \left( \mathbb{I} - \hat{P}(p_0,\vec{p}) \hat \Pi^{(0)}(\vec{p}) \right)^{-1},
\end{equation}
where $\hat \Pi^{(0)}(\vec{p})$ is the free bosonic propagator and $\hat{P}(p_0,\vec{p})$ is the polarization operator.

In EFT, both the free bosonic propagator, $\Pi^{(0)} \sim 1/|\vec p|$, and the polarization bubble,
\begin{equation} \label{eq:PolarizationContinuum}
P(p_0,\vec p) = \frac{e^2 N_f}{8} \frac{\vec p^2}{\sqrt{p_0^2 + v_{F,0}^2 \vec p^2}},
\end{equation}
are scalar quantities. Here, $N_f = 2$ denotes the number of Dirac fermion flavors in the EFT. In LPT, both polarization and free bosonic propagator are $2\times2$ matrices in sublattice space (see Supplementary material for the explicit expressions). According to the analysis above, the combined low-frequency and low-momentum limits of the free bosonic propagator coincide in EFT and LPT. Thus, the difference lies in the second term in the inverse operator in the eq. \ref{eq:BosonicDysonEquation}. Due to aforementioned sublattice matrix structure of LPT propagators, we should study $\mathcal{P}=\Tr(\Pi^{(0)}P)$ to compare LPT and EFT at RPA level. Results are shown in  Fig. \ref{fig:polarization_all}. 

First, one notices that according to  \ref{eq:PolarizationContinuum}, $\mathcal{P}$ is simply a constant in the limit of zero external frequency $p_0$ in EFT. This constant behavior at zero frequency is not seen on the lattice, where one observes a constant shift at the $\Gamma$ point with a subsequent linear dependence on external momentum $\vec{p}$ (figures \ref{fig:polarization_all}\textcolor{red}{a} and \textcolor{red}{b}). These two features are more or less independent of the external frequency, as seen in Fig. \ref{fig:polarization_all}\textcolor{red}{d}, where we plot difference between lattice and EFT versions of $\mathcal{P}$. To investigate this further, we plot the integrand in the LPT polarization bubble for external frequency $p_0=0$ and external momenta $\vec{p}\approx0$. We see from Fig. \ref{fig:polarization_all}\textcolor{red}{c} that the main contribution to the integral, which defines the lattice polarization, is still concentrated near the Dirac point, which is expected. However, there is a noticeable contribution coming from regions in between the non-equivalent Dirac points. One can further introduce an ultraviolet cutoff in the integral over loop momentum such that only momenta within a radius $\Lambda$ of the Dirac points remain. The behavior of lattice version of $\mathcal{P}$ in the vicinity of the $\Gamma$ point at $p_0=0$ as a function of the cutoff is shown in Fig. \ref{fig:polarization_all}\textcolor{red}{e}. The constant shift at the $\Gamma$ point disappears in the limit of small cutoff once $\Lambda$ excludes the region between the neighboring $K$ points (see \ref{fig:polarization_all}\textcolor{red}{c}). However, the linear asymptote grows with decreasing $\Lambda$. Finally, we compute $\mathcal{P}$ on the lattice, using the linear approximation for the dispersion relation of the electrons near the Dirac points. The results are shown in Fig. \ref{fig:polarization_all}\textcolor{red}{f}. Unlike \ref{fig:polarization_all}\textcolor{red}{e}, the constant shift at the $\Gamma$ point does not appear at larger $\Lambda$, and the linear term also decreases with increasing cutoff, so that we finally reproduce the continuum result. 

This analysis illustrates two distinct effects: first, a constant shift of $\mathcal{P}$ at the $\Gamma$ point due to inter-valley scattering, which manifests itself in the integrand in the polarization loop having non-zero values in the regions between non-equivalent Dirac points; second, a cutoff effect, which results in a linear correction to $\mathcal{P}$ around the $\Gamma$ point even if the loop integrals over the two valleys are completely independent. Both of these effects can be included in the continuum EFT, but at a large cost.
Inter-valley scattering can be described by the modified expression for the charged current in the EFT: 
\begin{eqnarray}
\tilde {j}_0={\bar\psi_a} \gamma_0 [I+ \\ 
\nonumber  \sum_j{\exp{(-i \vec x \vec {\Delta \mathcal{K}_j} \tau_3/2 )} \tau_1 \exp{(i \vec x \vec { \Delta \mathcal{K}_j} \tau_3/2 )} }]{\psi_a},
\label{eq:j}
\end{eqnarray}
where $\tau_j$ are Pauli matrices in valley space and $\vec  { \Delta \mathcal{K}_j}$, $j=1,2,3$ are lattice vectors connecting a single Dirac point with the three neighboring non-equivalent Dirac points (see Supplementary material for the derivation). This modified vertex should be accompanied by the addition of  higher-dimensional kinetic terms to the QED Lagrangian in order to reproduce the non-linear dispersion relation away from the Dirac points.  Moreover, we should employ a hard cutoff procedure in all loop integrals. An example of such a calculation for the polarization can be found in the supplementary material and we indeed get a term $\sim |\vec p|/\Lambda$ in $\mathcal{P}$. Interestingly, it is possible to show that both inter-valley scattering and the hard UV cutoff induce much smaller corrections in the one-loop electron self-energy, which is consistent with our observation that the discrepancies between LPT and the EFT appear only at the level of RPA. Further details regarding procedures for modifying the continuum EFT are discussed in details in the supplementary material.

Summarizing, we have demonstrated that finite BZ effects are responsible for the differences between LPT and EFT results starting at the RPA level. These differences arise from the loop integral in the polarization function and physically correspond to inter valley scattering processes. A rigorous incorporation of these processes into the EFT framework, with the aim of reproducing the LPT results beyond the one-loop self-energy correction, appears to be technically challenging. Therefore, whenever the interaction becomes sufficiently strong for loop corrections to play a significant role, the appropriate course of action is to switch to lattice perturbation theory.

\section{\label{sec:conductivity}Optical conductivity of suspended graphene}

  \begin{figure}[]
   \centering
   \subfigure[] {\label{fig:condutivityA}\includegraphics[width=0.45\textwidth,clip]{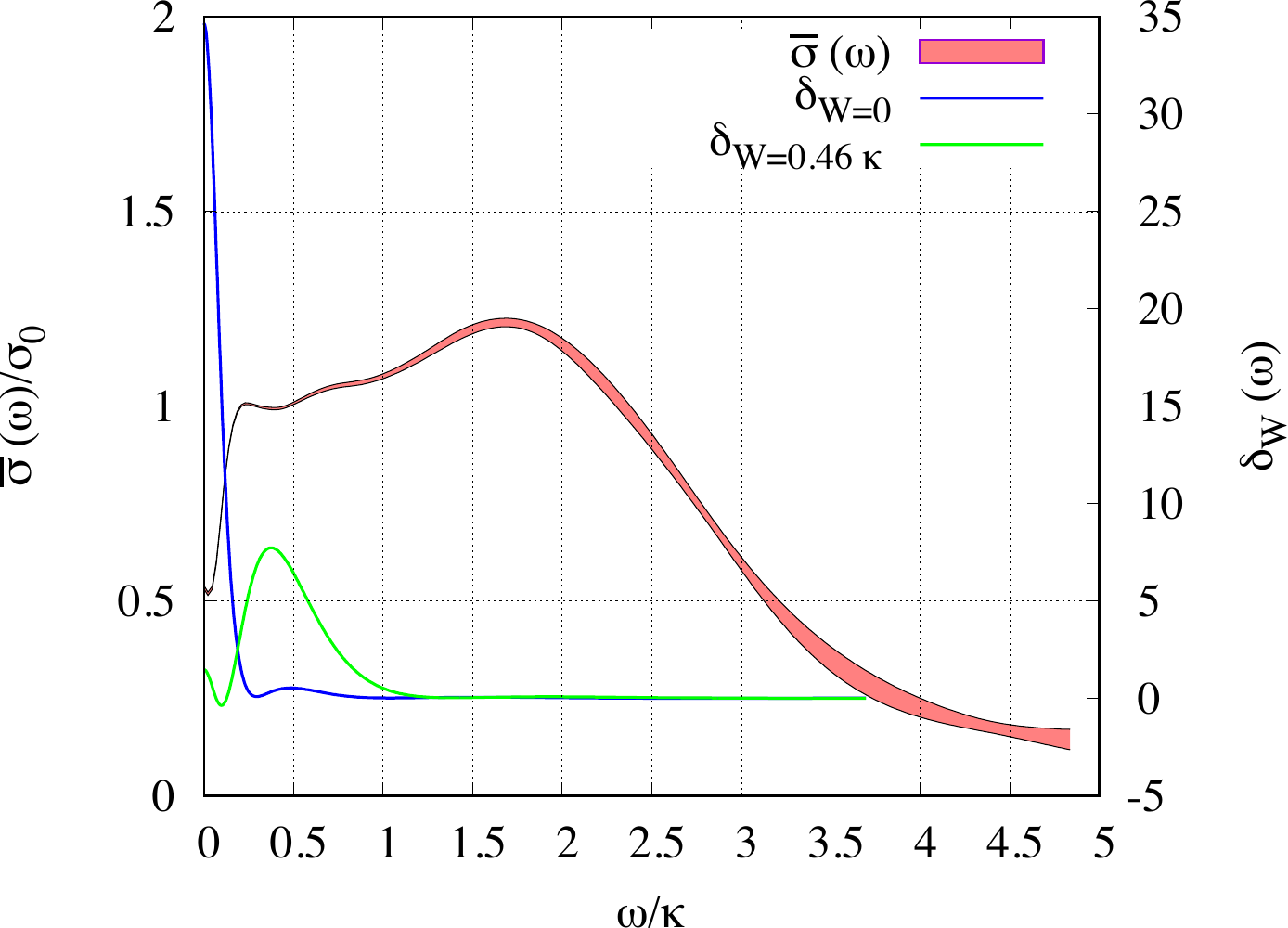}}
   \subfigure[]  {\label{fig:conductivityB}\includegraphics[width=0.45\textwidth,clip]{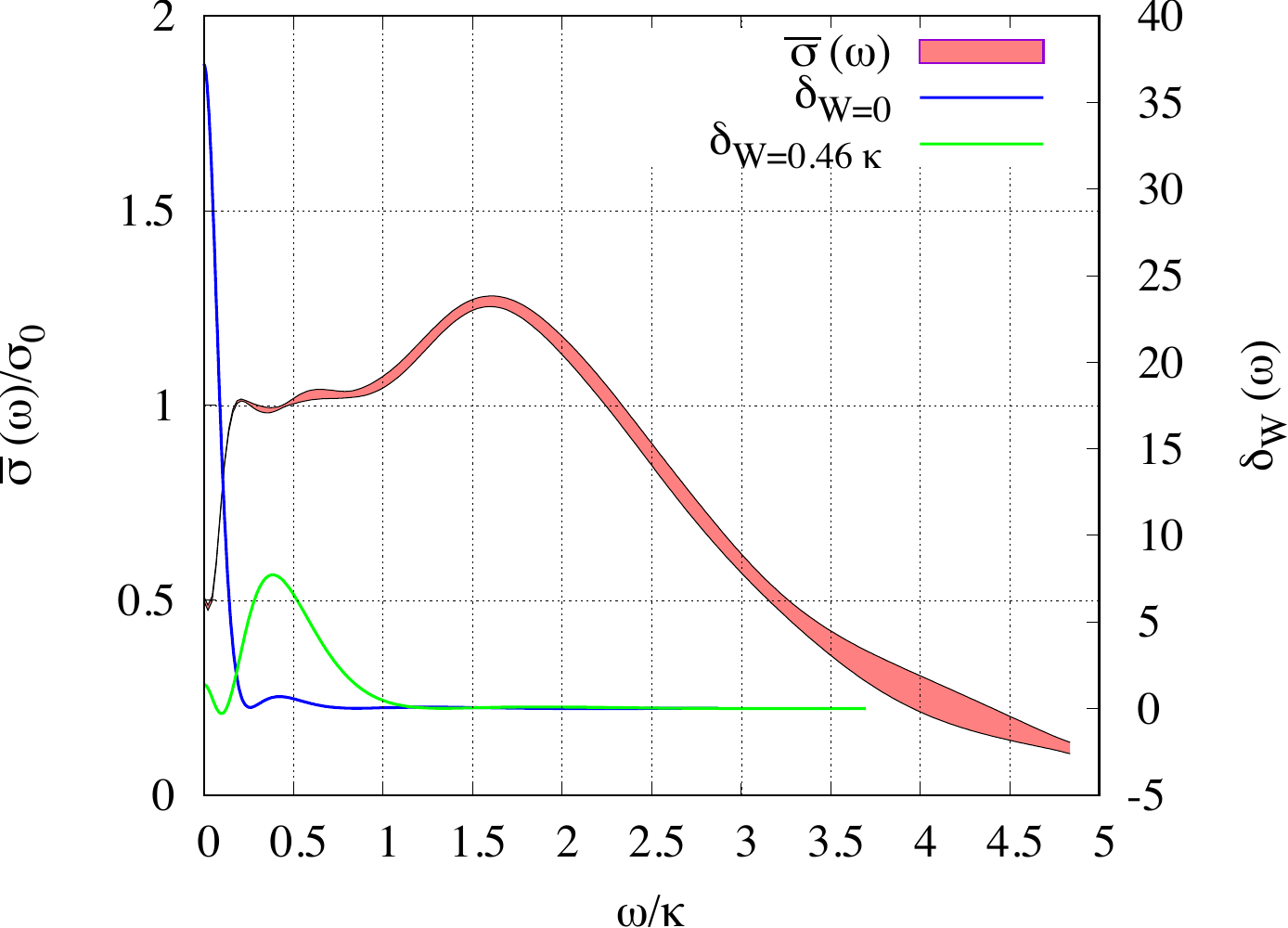}}
     \caption{Averaged conductivity $\bar \sigma (W)$ profiles from QMC data at $T=0.0625$ eV for \textbf{(a)} potential variant I and \textbf{(b)} potential variant II (left axis).  The shaded area represents statistical uncertainty of the averaged conductivity. Alongside with the averaged conductivity profiles, we have also plotted two examples of Backus-Gilbert resolution functions for centers at frequencies $W=0$ and  $W=0.46 \kappa$.  $\bar \sigma (W)$ is plotted in the units of $\sigma_0=e^2/ (4 \hbar)$. }
	\label{fig:conductivity}
\end{figure} 

In this section, we demonstrate one possible application of the techniques employed in this paper: a precision study of transport properties in quantum systems. As an example, we consider the optical conductivity of suspended graphene, which has been the subject of extensive discussion in the literature. Given that electron–electron interactions strongly modify the single-particle spectrum, it would be natural to expect that transport properties are also significantly affected by long-range Coulomb interactions. However, experimental data \cite{conductivity_experiment1, conductivity_experiment2, PhysRevLett.101.196405} show no such renormalization of the optical conductivity: it is equal to its non-interacting value $\sigma_0=e^2/ (4 \hbar)$ within error bars of few percent. Considerable theoretical effort has been devoted to explaining these results, with the literature broadly divided into two groups advocating either large \cite{PhysRevLett.100.046403, PhysRevB.82.235402, PhysRevLett.110.066602}  or small \cite{Mishchenko_2008, Teber_2014, PhysRevB.93.235447} renormalization of the optical conductivity, depending on the computational scheme employed. A more detailed review of these theoretical approaches can be found in \cite{PhysRevB.94.085421}. The issue has also been investigated in QMC simulations \cite{PhysRevB.94.085421, Stauber_2017fuj} performed on smaller lattices and at higher temperatures, where almost no renormalization was reported. However, those lattice sizes were insufficient to simultaneously resolve the renormalization of the dispersion relation induced by Coulomb interactions. Therefore, one could argue that the full set of experimental observations was not completely reproduced in those studies.

Here we revisit this issue by computing the optical conductivity within the same QMC simulations in which the strong renormalization of the Fermi velocity is clearly observed. According to linear response theory, the conductivity spectral function can be obtained from the Euclidean time current–current correlator $G(\tau)$ via the Green–Kubo relation,
\begin{equation}
G(\tau) = \int_0^\infty \sigma(\omega) \frac{\omega \cosh \bigl(\omega (\beta/2-\tau)\bigr)}{\pi \sinh (\omega \beta/2)} d\omega.
\label{eq:GK}
\end{equation}
We therefore compute the current–current correlator and solve the Green–Kubo relation to extract $\sigma(\omega)$. Since the integral equation \ref{eq:GK} is an ill-posed mathematical problem, an appropriate regularization procedure must be employed. In this work, we use the Backus–Gilbert (BG) method \cite{PhysRevD.92.094510}, which has been shown to be particularly well suited for this type of problem \cite{TRIPOLT2019129}. The reason for that is our specific goal being to recover the low-frequency limit of the Dirac plateau in the spectral function $\sigma(\omega)$, i.e., the frequency interval $T \ll \omega \ll \kappa$, where $\sigma$ is nearly frequency independent. The BG method yields the averaged conductivity
\begin{equation}
\bar \sigma (W) = \int_0^\infty \sigma(\omega) \delta_W (\omega) d\omega,
\label{eq:average_cond}
\end{equation}
where the resolution function $\delta_W (\omega)$ is centered at frequency $W$. If the width of the resolution function is smaller than the width of the plateau, the latter can be reliably identified from the profile of $\bar \sigma (W)$.

The results are shown in Fig. \ref{fig:conductivity} for potentials variants I and II. We display both the averaged conductivity and representative examples of the resolution functions in order to demonstrate that their widths are sufficiently small to resolve the Dirac plateau. Indeed, one can clearly distinguish the initial low-frequency region, followed by the Dirac plateau, then the peak associated with the van Hove singularity in the density of states of the hexagonal lattice, and finally the high-frequency decay due to the finite bandwidth. 
The initial suppression of the conductivity can be attributed to the Coulomb blockade of electron transport through the finite system: it appears since our simulations are done at temperatures, which are low comparably to even the lowest electronic excitations and Coulomb interactions are completely unscreened. Subsequently, we observe precisely the regime of interest: a frequency region around $\omega/\kappa = 0.2$, which corresponds to the low-frequency limit of the Dirac plateau. Since the temperature is $T = 0.0625$ eV $= 0.02315 \kappa$, this frequency range satisfies both inequalities $T \ll \omega$ and $\omega \ll \kappa$. Therefore, it corresponds exactly to the optical conductivity limit discussed in the literature.

As one can see, QMC data show effectively no renormalization of $\sigma$ in this limit. Thus our QMC simulations exactly reproduce the experimental situation: interactions dramatically modify the dispersion relation, yet they do not alter the optical conductivity. A minimal explanation is that the effective coupling constant entering the expressions for the conductivity should be taken as the renormalized one, which accounts for the logarithmic increase of the Fermi velocity:
\begin{equation}
\alpha^{(R)}=\frac{c}{v^{(R)}_F}\frac{1}{137}.
\label{eq:alpha_R}
\end{equation}
In this case, the renormalized optical conductivity can be written as
\begin{equation}
\frac{\sigma(\omega)}{\sigma_0}
= 1 + C_\sigma \alpha^{(R)} + O\left( {\alpha^{(R)}}^2 \right).
\label{eq:sigma_R}
\end{equation}
In addition to the reduction of $\alpha^{(R)}$ due to the renormalization of the Fermi velocity, the coefficient $C_\sigma$ can be also very small \cite{Mishchenko_2008, Teber_2014}, which indeed leads to the negligible observed renormalization of optical conductivity. However, an even stronger statement may be possible: the theorem on the non-renormalization of the optical conductivity for the Hubbard model on the hexagonal lattice \cite{mastropietro} might be extendable to the case of long-range interactions. We leave this question for future investigation.

\section*{Conclusions and Outlook}

For the first time, we have demonstrated the logarithmic divergence of the Fermi velocity in graphene in fully non-perturbative QMC simulations. In agreement with RG arguments, the long-range Coulomb tail defines the coefficient in front of the logarithm, while short-range interactions are responsible for non-universal features such as the value of cutoff $\Lambda$ in the logarithm. The logarithmic fit of the QMC data for potential variant II shows good agreement with experiment. This result, combined with the absence of any renormalization of the optical conductivity (also consistent with experimental observations), indicates that the interacting tight-binding model with interaction potentials described in \cite{Wehling_2011} and \cite{PhysRevB.89.195429} provides a reliable microscopic description of suspended graphene.

After verifying the microscopic model, we used the QMC data for direct comparisons with various implementations of perturbation theory. This approach offers significantly more flexibility than a direct comparison between experiment and perturbative results, since QMC simulations allow systematic variation of key parameters such as temperature and the strength of electron–electron interactions at different distances. Although the initial comparison with RPA-level corrections to the dispersion relation computed within EFT showed seemingly good agreement, a subsequent comparison with LPT (which explicitly incorporates lattice scale physics) revealed that the apparent validity of the RPA-level EFT corrections is largely a numerical coincidence. This discrepancy arises because EFT deviates from LPT once the interaction strength becomes sufficiently large for the polarization bubble to play an important role in the perturbative corrections to the fermionic self-energy.
Further analysis shows that the deviation between EFT and LPT in the polarization function is caused by inter valley scattering processes. These effects are difficult to rigorously incorporate within the EFT framework. Therefore, in the strongly correlated regime, LPT remains the only reliable perturbative counterpart for comparison with QMC data with EFT only being able to provide qualitative scaling laws. 

In general, comparison with perturbative data shows that the effective strength of the Coulomb interaction in suspended graphene is sufficiently large that even RPA-level diagrams fail to provide an adequate quantitative description of the renormalization of the Fermi velocity. This suggests that the system may already be approaching a regime in which the perturbative series begin to break down. In this context, it would be particularly interesting to investigate whether the agreement between QMC and LPT results improves upon including diagrams beyond the simple RPA level shown in Fig. \ref{fig:RPA}. If no systematic improvement is observed, this would constitute strong evidence for the breakdown of the perturbative expansion. We leave this study for future work.

From an experimental point of view, the clearest way to demonstrate the importance of higher-order corrections is to estimate the temperature corrections. If one measures $v_F$ in the ``high-temperature limit" at momenta corresponding to the energy in the free dispersion relation $E_0(k)\leq2...3T$, then the deviations of the Fermi velocity upwards from the logarithmic curve can be regarded as a qualitative sign of higher-order perturbative corrections beyond RPA. Potentially, such a measurement combined with aforementioned higher-order LPT calculations can demonstrate the onset of the divergence of the asymptotic series of strongly-correlated QED.  

Technically, sufficiently large lattice sizes to reliably fit the QMC data with logarithmic curves could only be achieved by employing the HMC algorithm with a stochastic evaluation of the fermionic determinant. As demonstrated by our calculation of the optical conductivity, this technique can serve as a powerful tool for investigating transport properties of strongly correlated quantum systems.
Also, the verified microscopic Hamiltonian can be used to provide reliable reference data for transport calculations within Boltzmann transport theory, a program that has already been initiated in QMC studies of hydrodynamic flow in half-filled graphene \cite{PhysRevLett.134.206303}. In this way, one could potentially verify the approximations used for the collision integral through direct comparison with QMC simulations. Unlike in experiments, QMC allows one to selectively switch on and off different scattering mechanisms, thereby generating a broad set of reference data for comparison across different scattering regimes. 

Another possible direction of research is a study of quasiparticle excitations such as plasmons with high momentum resolution, again approaching experimental scales. As was recently shown \cite{PhysRevResearch.4.043107}, plasmonic contribution can substantially alter characteristics of electronic flow in hydrodynamic regime, thus detailed characterization of these quasiparticles will help to more precisely estimate the viscosity of electronic fluid in graphene.

\section*{Acknowledgments}
This work mainly benefited from access to the Ir\`ene Joliot-Curie supercomputer of the Tr\`es grand centre de calcul (TGCC) of the Commissariat \`a l’\'energie atomique et aux \'energies alternatives (CEA) in France as part of a ``grand challenge" project (project id: gch413) awarded by GENCI (Grand Equipement National de Calcul Intensif) as well as through project gen2271. It also benefited from access to the Jean Zay supercomputer at the Institute for Development and Resources in Intensive Scientific Computing (IDRIS) in Orsay, France and from access to the Occigen supercomputer under projects A0080511504 and A0080502271, hosted by the Centre Informatique National de l’Enseignement Sup\'erieur (CINES) at Montpellier, France. It was also partially supported by the HPC Center of Champagne-Ardenne ROMEO and by the GPU cluster of the Institute of Theoretical Physics of the University of Heidelberg. Additional resources were provided by the Gauss Centre for Supercomputing e.V. (www.gauss-centre.eu) through the John von Neumann Institute for Computing (NIC)
on the GCS Supercomputer JUWELS~\cite{JUWELS} at J\"ulich Supercomputing Centre (JSC). 
FFA acknowledge financial support from the DFG through the W\"urzburg-Dresden Cluster of Excellence on Complexity and Topology in Quantum Matter - ct.qmat (EXC 2147, project-id 39085490),  through the SFB 1170 ToCoTronics. MU  thanks  the  DFG   for financial support  under the projects UL444/2-1.  CW acknowledges support by the Deutsche Forschungsgemeinschaft (DFG, German Research Foundation) through the CRC-TR 211 'Strong-interaction matter under extreme conditions'– project number 315477589 – TRR 211.
 The authors would like to thank Eugene Mishchenko for insightful discussions regarding the perturbative calculations.  FFA and MU  would  like to thank Shaffique Adam  for discussions during initial stages of this project.  
 FFA  would like to thank H.K.  Tang, J. N. Leaw, J. N. B. Rodrigues, I. F. Herbut,  P. Sengupta  and S. Adam   for previous collaborations on the subject.
  We also thank T. Stauber for providing the access to experimental data.


\bibliography{localbib}


\clearpage
\onecolumngrid
\setcounter{section}{0}
\renewcommand{\thesection}{S\Roman{section}}
\renewcommand{\thefigure}{S\arabic{figure}}

\topmargin 0.0cm
\oddsidemargin 0.2cm
\textwidth 16cm 
\textheight 21cm
\footskip 1.0cm

\begin{center}

{\Large\bfseries
Supplementary Material for:\\[0.4cm]
Bridging the gap between numerics and experiment in free standing graphene
}

\vspace{1cm}

{\normalsize
Maksim Ulybyshev$^{1\ast}$,
Savvas Zafeiropoulos$^{2}$,
Christopher Winterowd$^{3}$,\\
Fakher Assaad$^{1,4}$
}

\vspace{0.5cm}

{\small
$^{1}$Institute for Theoretical Physics, Julius-Maximilians-Universit\"at W\"urzburg,\\
Am Hubland, D-97074 W\"urzburg, Germany\\[0.2cm]

$^{2}$Aix Marseille Univ, Universit\'e de Toulon, CNRS, CPT,\\
Marseille, France\\[0.2cm]

$^{3}$Johann Wolfgang Goethe-Universit\"at Frankfurt am Main,\\
Frankfurt am Main, Germany\\[0.2cm]

$^{4}$W\"urzburg-Dresden Cluster of Excellence ct.qmat,\\
Am Hubland, 97074 Würzburg, Germany\\[0.3cm]

$^\ast$To whom correspondence should be addressed;\\
E-mail: maksim.ulybyshev@uni-wuerzburg.de
}

\end{center}

\vspace{1cm}

\section{Numerical setup}

In this section we briefly describe our lattice quantum Monte Carlo (QMC) setup and the calculation of the relevant observables. For a more detailed discussion we refer the reader to \cite{PhysRevLett.111.056801,PhysRevB.89.195429,PhysRevB.96.205115}. The goal of hybrid Monte Carlo is to efficiently and faithfully sample the functional integral representation of the partition function, $Z= \Tr e^{-\beta \hat{\mathcal{H}}} $. The trace is performed using fermionic coherent states by first breaking up the continuous Euclidean time interval, $[0, \beta)$, into $N_{\tau}$ slices. In order to compute matrix elements of the two-body interaction term, $\hat{\mathcal{H}}_{\text{int}}$, we must introduce a scalar bosonic field, $\phi$, on each time-slice and  lattice site. This is done through the Hubbard-Stratonovich transformation
\beq \label{eq:HubbardStrat}
e^{\delta_{\tau} \sum_{x,y} V_{x,y} \hat{q}_x \hat{q}_y} \cong \int \mathcal{D} \phi e^{-\frac{1}{2\delta_{\tau}} \sum_{x,y} \phi_x V^{-1}_{x,y} \phi_y } e^{i\sum_x \phi_x \hat{q}_x},
\eeq
\beq
\langle \mathcal{O} \rangle =  \frac{ \Tr \left( \hat O e^{-\beta \hat H} \right)} {\Tr \left(  e^{-\beta H} \right)} = \frac{\Tr  \left( \hat O \prod  e^{\delta_{\tau} \hat H_{tb}} e^{\delta_{\tau} \hat H_{int}} \right)} { \Tr \left(  \prod  e^{\delta_{\tau} \hat H_{tb}} e^{\delta_{\tau} \hat H_{int}} \right)}
\eeq

where the interaction is now bilinear in the fermion operators through the charge operator $\hat q_x \equiv \hat a^\dag_{x,\uparrow} \hat a_{x,\uparrow} + \hat a^\dag_{x,\downarrow} \hat a_{x,\downarrow}-1$. In the current paper, we have adopted a more complicated version of the Hubbard-Stratonovich decomposition, which introduces two separate scalar fields which couple to spin and charge density respectively. This is done by using the following identity for the on-site term
\beq
\frac{V_{x,x}}{2}\hat q_x^2 = \frac {\eta V_{x,x}}{2}\hat q_x^2 - \frac{(1-\eta) V_{x,x}}{2} \hat s_x^2 + (1-\eta) V_{x,x} \hat s_x,
\label{eq:split_int}
\eeq
where $\hat s_x \equiv \hat a^\dag_{x,\uparrow} \hat a_{x,\uparrow} - \hat a^\dag_{x,\downarrow} \hat a_{x,\downarrow}$ is the spin operator and $\eta \in [0,1]$ is a free parameter. 
Further details regarding the application of this technique to the case of the long-range Coulomb tail can be found in \cite{PhysRevB.99.205434}. This is done in order to avoid possible ergodicity issues in the sampling of the functional integral. This procedure is extensively discussed in \cite{PhysRevD.101.014508}, which we refer to for further details. 

In the functional integral formulation, the thermal expectation values of observables are computed in the usual way,
\beq \label{eq:EVObservable}
\vev{\mathcal{O}} = \frac{1}{Z} \int \mathcal{D} \phi    \det M[\phi] ^2\mathcal{O} e^{-S_B[\phi]},
\eeq
where $M$ is the bilinear fermion operator and $S_B$ is the quadratic bosonic action that results from applying (\ref{eq:HubbardStrat}) at each time slice. 
After  carrying out a   particle-hole transformation one can show that the action, for each Hubbard-Stratonovich  field,  is invariant under time-reversal symmetry. This  symmetry  leads to the absence of negative sign problem \cite{Wu04}, and one will show that the fermion determinant in the above equation is real.   To evaluate the expectation value  one can adopt the Blankenbecler, Scalapino, Sugar  (BSS) algorithm  \cite{Blankenbecler81} and  work with the  determinant directly.  Such an  approach  invariably leads to  a  computational  time that scales as   $N_\tau  N^{3} $   for a single sweep \cite{Hohenadler14,Tang570}.    Here, $N$  corresponds to the number of spatial sites in a single  Euclidean time-slice.  

There exists an alternative technique for this procedure, namely, hybrid Monte Carlo \cite{degrand2006lattice} (HMC). This algorithm is typically used in lattice quantum chromodynamics (LQCD) calculations, but can be adapted to condensed-matter systems. One can evaluate the determinant stochastically, 
\begin{equation} 
	 \det M[\phi]^2  \propto \int \mathcal{D}   \left\{ \eta, \eta^{\dagger} \right\}    e^{- \eta^{\dagger} \left(   M[\phi] M^{\dagger}[\phi]  \right)^{-1}  \eta}
\end{equation}
and  sample both over the  Hubbard-Stratonovich   and  pseudo-fermion  fields.  In principle, we add more statistical fluctuations to the system, thus increasing the prefactor in the scaling of the algorithm with the system size. However, when the fermion matrix is not too ill-conditioned, HMC  becomes very appealing  and outperforms the BSS approach. This  turns out to be the case  here, where the scaling of the computational time with volume and number of time slices  is $(N N_\tau)^{1.5}$, thus larger prefactor is compensated by better scaling for large systems. 

In the QMC calculations performed for this study, we have only slightly modified the algorithm used in the previous functional integral studies of graphene (e.g. in \cite{Stauber_2017fuj}) by employing a different symplectic integrator in the molecular dynamics \cite{PhysRevB.100.075141}.

\begin{figure}[t]
    \centering
	\includegraphics[angle=0,width=1.0\textwidth]{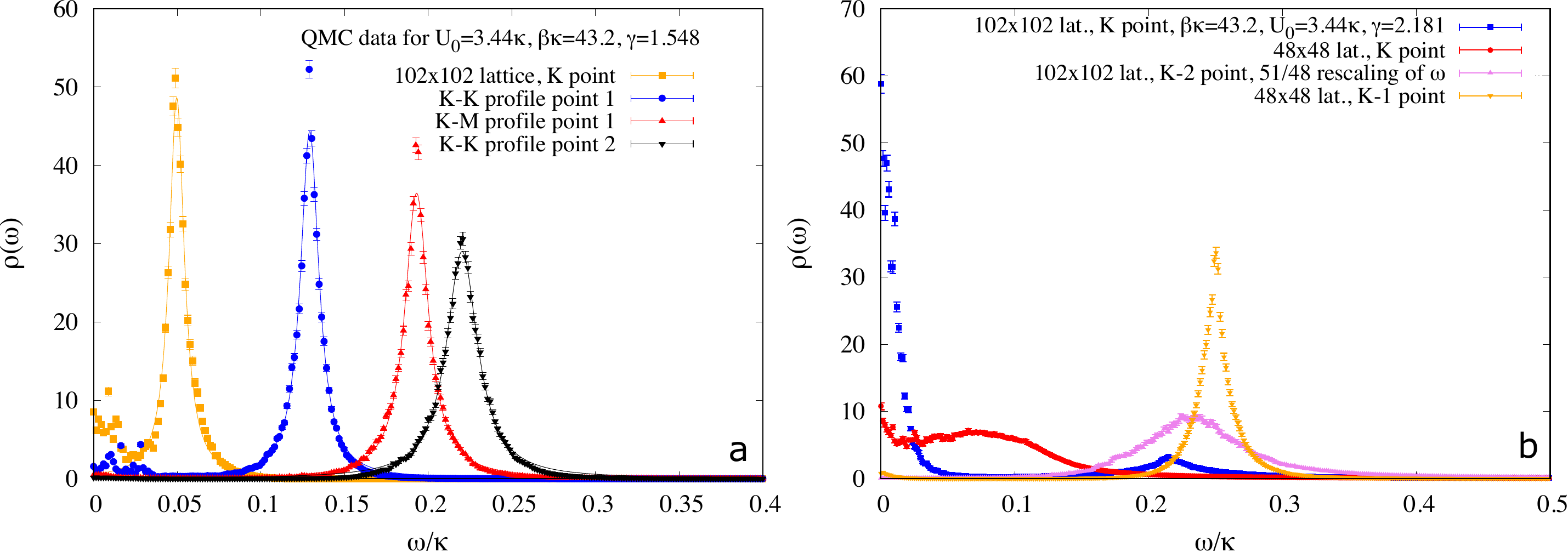}
	\caption{\textbf{(a)} Several spectral functions, reconstructed with the help of stochastic MEM at various points in the BZ. The data are shown for potential variant I with no additional rescaling in order to reside firmly within the strongly-coupled regime. We display the spectral functions for the K point and the three other points closest to it. The resonances are fitted with a Lorentzian ansatz, ${Z}/{((\omega-\omega_0)^2+\Gamma^2)}$. We note that a gap appears at the Dirac point. This is indicated by the presence of a large peak in the spectral function at the Dirac point which is shifted from the origin, with a smaller peak existing at zero frequency. \textbf{(b)} Spectral functions at the Dirac point and in its vicinity. We compare the results obtained on the $102\times102$ lattice (at the Dirac point and at the second point in momentum along the K-K line in the BZ) with those for the $48\times48$ lattice (again at the Dirac point and at the first point in K-K profile). The data shown in the plots were produced with potential variant II (see main text for definition). A comparison of the $102\times102$ lattice data with the spectral function in figure (a) at the K point shows that the gap disappears with an increased Coulomb tail, which drives us away from the antiferromagnetic phase transition (see \cite{Tang570}). Moreover, the comparison of the spectral functions at the K point, computed on the $48\times48$ and $102\times102$ lattices, shows that the gap we sometimes see is merely a finite-volume artifact. The double-peak structure seen at smaller lattice sizes disappears as the number of sites is increased.}
	\label{fig:examples_stMEM}
\end{figure}

The renormalized dispersion of the quasiparticles is obtained through the fermion propagator, which is defined as
\beq \label{eq:FermionPropagatorDefinition}
G(x,\tau) = \frac{\Tr \{ \hat{a}_x e^{-\tau \hat{\mathcal{H}}}\hat{a}^{\dagger}_0e^{-(\beta-\tau) \hat{\mathcal{H}}} \}}{Z},
\eeq
where the operators appearing in the numerator are in the Heisenberg representation and we have assumed translational invariance in both space and in Euclidean time. In our functional integral setup, (\ref{eq:FermionPropagatorDefinition}) can be recast in terms of elements of the inverse fermion operator
\beq \label{eq:FermionPropagatorPIMomentum}
G(\vec k,\tau) = \sum_x e^{-i \vec k \cdot \vec x} \vev{ M^{-1}_{x,\tau;0,0} },
\eeq
where the expectation value is taken as in (\ref{eq:EVObservable}), and we have projected to a momentum $k$ residing within the BZ. In practice, $G(\vec k,\tau)$ is $2\times2$ matrix in sublattice space and we are using its trace for further processing. In order to express this quantity in terms of the eigenvalues of the many-body Hamiltonian, we introduce the well-known spectral representation of the propagator 
\beq \label{eq:SpectralFermionPropagator}
G(\vec k, \omega) = \int^{+\infty}_{-\infty} \frac{\mathrm{d}\omega'}{2\pi}\frac{\rho(\vec k,\omega')}{-i\omega+\omega'},
\eeq 
where we have have also introduced the density of states, which is given as
\beq \label{eq:DensityOfStatesFermion}
\rho(\vec k, \omega) = \frac{2\pi}{Z} \sum_{n,m} \delta(E_m-E_n-\omega) |\braket{m|\hat{a}^{\dagger}_k|n}|^2 \left( e^{-\beta E_n} + e^{-\beta E_m} \right).
\eeq
Here $E_n$ and $E_m$ refer to eigenvalues of the Hamiltonian whose corresponding states have total momentum which differ by $k$. This analysis is completely generic, and for any Euclidean correlation function there will exist a corresponding spectral function. On the lattice, we compute observables in the Euclidean time representation, and thus one can transform (\ref{eq:SpectralFermionPropagator}) to obtain the Green-Kubo relation
\beq \label{eq:GreeenKubo}
G(\vec k,\tau) = \frac{1}{\pi} \int^{+\infty}_{0} \mathrm{d}\omega K(\omega, \tau) \rho(\vec k,\omega),
\eeq
where the kernel is given by $K(\omega, \tau) = \cosh \left[ \omega ( \tau - \beta/2 )\right]/\cosh \left[ \omega \beta/2 \right]$. Obtaining the density of states through the inversion of (\ref{eq:GreeenKubo}) is an ill-defined problem, as one is attempting to reconstruct an analytic function of frequency from $N_{\tau}$ data points and their associated statistical errors. The inverse problem can be dealt with by applying the stochastic maximum entropy method (MEM) \cite{2004cond.mat..3055B}. In this work, we have used the publicly available algorithms for lattice fermions (ALF) code, which contains an implementation of the stochastic MEM \cite{ALF}. In Fig. (\ref{fig:examples_stMEM}), one can see some of the results of this spectral reconstruction. Using the reconstructed spectral function, one can obtain estimate for the energy of the quasiparticle excitations $\omega_0(\vec k)$ by finding the peak of the spectral function. The cost of performing the inverse problem is that one must have data with very small statistical errors.

\begin{figure}[]
    \centering
	\includegraphics[angle=0,width=0.95\textwidth]{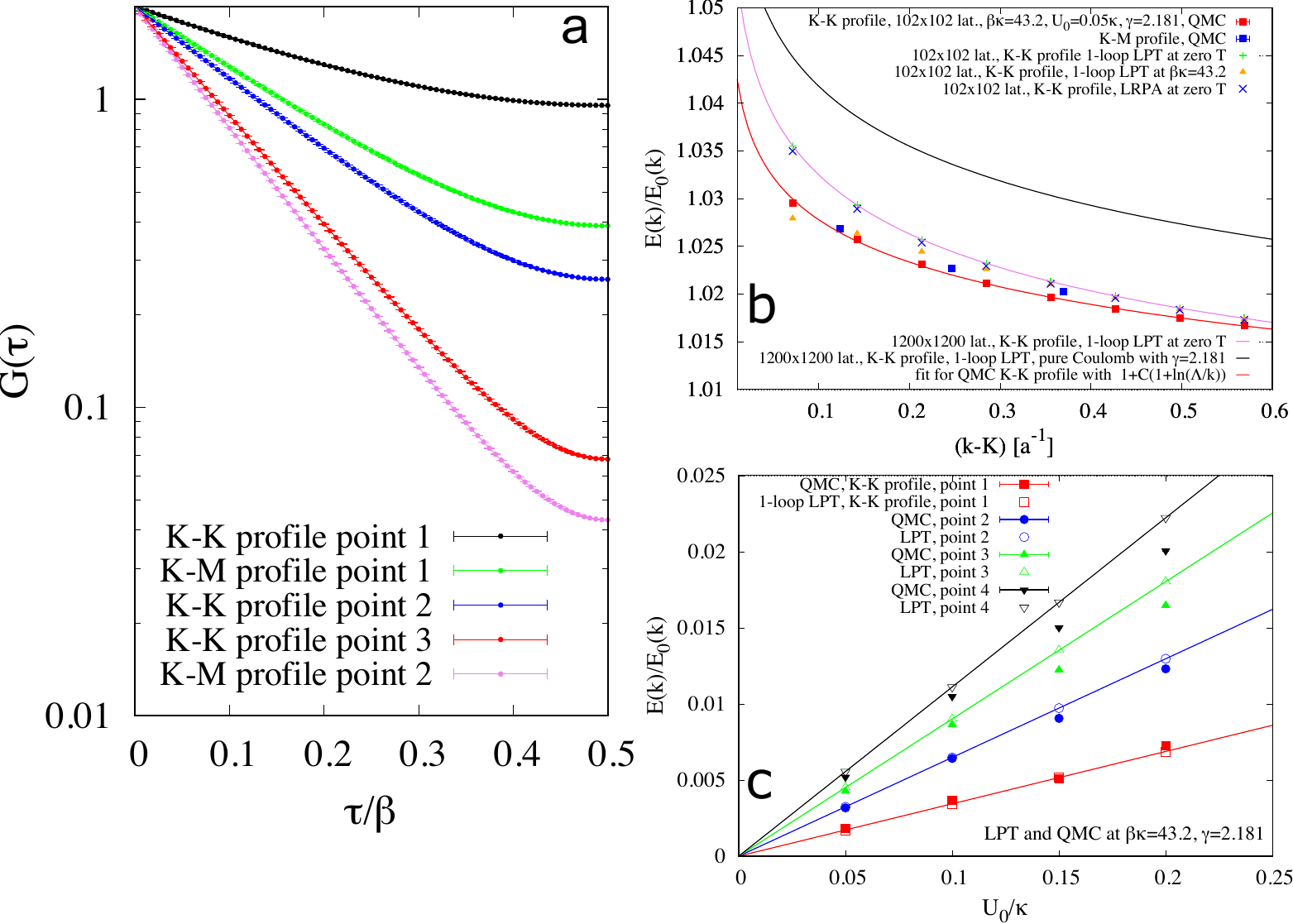}
	\caption{\textbf{(a)} QMC data for the fermion propagator in Euclidean time, $G(k,\tau)$. The computations have been performed with the interaction potential variant II, which has been rescaled in order for the on-site interaction to take the value $0.05\kappa$ (to be firmly within the weak-coupling regime). 
	Fitting is made with the lattice ansatz (\ref{eq:PropagatorFitAnsatzLattice}). \textbf{(b)} Renormalized dispersion at the same coupling. In order to cross-check the QMC data, we have added the LPT results. The plot exhibits several key features. Firstly, the absence of differences between the $102\times102$ and $1200\times1200$ LPT data clearly indicates that the thermodynamic limit has been achieved. Secondly, the RPA corrections can be observed but remain quite small at this coupling. Finally, the QMC data substantially deviates from the one-loop LPT result at zero temperature. This deviation can be accounted for by finite-temperature corrections in the calculation of the one-loop electronic self-energy on the lattice. In addition to these results, we also display the LPT results at  $1200\times1200$ for an alternate two-body potential with identical on-site interaction and $\gamma$, but all other couplings except the on-site one are defined by the Coulomb tail. This setup differs from potential variant II at intermediate distances. Nevertheless, this difference is significant enough to produce different results for the renormalized dispersion even at one-loop. This comparison serves as an additional piece of evidence that the details of the two-body interaction are important if one wishes to obtain quantitative precision. \textbf{(c)} Dependence of the renormalized dispersion on the interaction strength in the weakly-coupled regime (we rescale the potential and plot the data with respect to the rescaled on-site interaction, $U_0$). The QMC data clearly converge to the finite-temperature, one-loop LPT result in the limit of small interaction.
	}
	\label{fig:QMC_justification}
\end{figure}

\section{Justification of QMC}

\subsection{Small interaction limit}

In order to prove the correctness of the numerical procedures, we check that at small interaction the QMC results converge towards lattice perturbation theory. In order to do so, we take the potential variant II (see the main text for details) and uniformly rescale it so that the on-site interaction takes the value $U_0=0.05\kappa$. We then compute the fermion propagator $G(\vec k,\tau)$ (\ref{eq:FermionPropagatorPIMomentum}), using the QMC procedure described above. The energy renormalization is quite small in the weak-coupling regime, and thus quite high precision is needed in the measurements, typically at the level of 0.1\% or more. To achieve this precision, we need to take into account corrections introduced by the Euclidean time discretization in the partition function. This discretization has consequences on the extraction of quasiparticle energies, both in terms of the reconstruction of the spectral function, as well as in terms of fitting the exponential decay of the correlators. In order to take into account the finite lattice spacing in time, we start from the following expression for the ``discrete'' version of the fermion propagator:
\beq \label{eq:FermionPropagatorDescrete}
G_D(x,\tilde{\tau}) = \frac{\Tr \{ \hat{a}_x (1-\delta_{\tau} \hat{\mathcal{H}})^{\tilde{\tau}}\hat{a}^{\dagger}_0 (1-\delta_{\tau} \hat{\mathcal{H}})^{(N_\tau-{\tilde{\tau}})}\}}{ \Tr \{  (1-\delta_{\tau} \hat{\mathcal{H}})^{N_\tau} \} },
\eeq
where $\tilde{\tau}$ is the integer-valued discrete Euclidean time, $\tau=\delta_{\tau} \tilde{\tau}$, where $\delta_{\tau}\equiv \beta/N_\tau$ is the step in Euclidean time. Equation (\ref{eq:SpectralFermionPropagator}) can then be replaced by
\beq \label{eq:SpectralFermionPropagatorDiscrete}
G_D(\vec  k, \omega_n) = \int^{+\infty}_{-\infty} \frac{\mathrm{d}\omega'}{2\pi}\frac{\rho(\vec  k,\omega')}{1-e^{i \omega_n \delta_{\tau}}   (1-\delta_{\tau}\omega')},
\eeq
where the fermionic Matsubara frequencies are defined as
\beq \label{eq:Matsubara}
\omega_n = 2\pi \frac{n+1}{N_\tau \delta_{\tau}}, \, n \in \mathbb{Z}.
\eeq
This leads to the replacement of the kernel in the Green-Kubo relations by the function
\beq \label{eq:GreenKuboLatticeKernelDiscrete}
K^{(D)}(\tau,\omega) = \frac{(1-\delta_{\tau}\omega)^{\tilde{\tau}}}{1+(1-\delta_{\tau}\omega)^{N_{\tau}}}.
\eeq
The observed resonances are very narrow at small interaction strength and the QMC data quality is good enough to directly fit the correlators with the lattice versions of the exponents in the familiar K\"all\'en–Lehmann representation. Taking into account the particle-hole symmetry, and neglecting corrections of $O(\delta^2_{\tau})$, equation (\ref{eq:GreenKuboLatticeKernelDiscrete}) leads to the following ansatz for the fermion propagator
\beq \label{eq:PropagatorFitAnsatzLattice}
G_D(\vec  k,\tau) = \mathcal{C} \left[ (1-\delta_{\tau} E)^{\tilde{\tau}} + (1+\delta_{\tau} E)^{N_{\tau}-\tilde{\tau}} + (1+\delta_{\tau} E)^{\tilde{\tau}} + (1-\delta_{\tau} E)^{N_{\tau}-\tilde{\tau}} \right]
\eeq
where $E$ and $\mathcal{C}$ are fit parameters. Several examples of the fermion propagator at weak coupling and their fits using eq. (\ref{eq:PropagatorFitAnsatzLattice}) are shown in Fig. \ref{fig:QMC_justification}\textcolor{red}{a}. We have also checked that indeed, the lattice version of the fit performed better than its continuum counterpart.
In the strong-coupling regime, the fitting form ($\ref{eq:PropagatorFitAnsatzLattice}$) becomes less reliable due to the broadening of the resonances (i.e. decrease of the particle's lifetime) and one must resort to the spectral reconstruction described in the previous section. 

\begin{figure}[t]
    \centering
	\includegraphics[angle=0,width=1.0\textwidth]{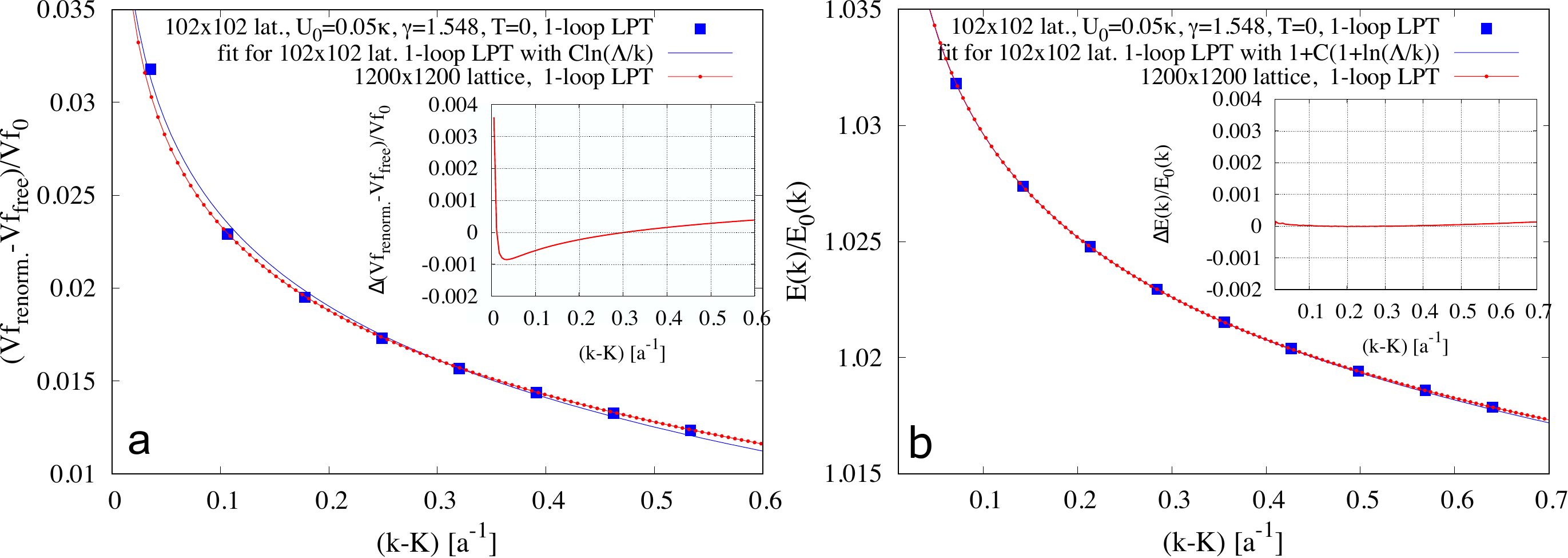}
	\caption{\textbf{(a)} Calculation of the renormalized Fermi velocity in lattice perturbation theory for small interaction: $U_0=0.05\kappa$ and $\gamma=1.548$. The renormalized Fermi velocity is determined through a finite difference of neighboring points in the lattice dispersion. The difference between the logarithmic fit of the $102\times102$ and $1200\times1200$ lattice data (shown in the inset) demonstrates that there are noticeable differences arising due to the volume-limited resolution in momentum space.  \textbf{(b)} The same data as in figure (a), but we now plot the dispersion relation itself. As there is no need for numerically estimating a derivative, the discrepancies between results obtained on lattices of different sizes disappear almost completely. The inset again shows the difference between the logarithmic fit of the $102\times102$ and $1200\times1200$ lattice data.}
	\label{fig:energy_Vf_fitting}
\end{figure}

After fitting the correlator to (\ref{eq:PropagatorFitAnsatzLattice}), one obtains the values of the renormalized energies, $E(\vec k)$. One way to obtain $v_F$ is through numerical differentiation
\beq \label{eq:numDiff}
v_F\left(\frac{k_1+k_2}{2}\right) = \frac{E(k_1)-E(k_2)}{k_1-k_2},
\eeq
where the momentum varies along a specific direction in the BZ. However, using this formula we introduce a systematic error of the order of $(k_1-k_2)^2$ and also increase the statistical uncertainties, as one is computing the difference of fluctuating quantities whose mean values only differ slightly . The analysis of the systematic errors connected with the numerical differentiation is presented in the figure \ref{fig:energy_Vf_fitting}.

An alternative is to look directly at the dispersion relation $E(\vec k)$ and modify the fitting functions. At weak coupling (and presumably at strong coupling too), the renormalization of the Fermi velocity can be decomposed as 
\beq \label{eq:vf_renorm}
v_F(\vec k) = v_{F,0}(\vec k) \left[1+ C \ln \frac{\Lambda}{|\vec k|}\right],
\eeq
where $v_{F,0}(\vec k)$ comes from the dispersion relation for free fermions on the hexagonal lattice. As we will work on quite large lattices, the correction to the linear dispersion will disappear once we approach the Dirac point:
\beq \label{eq:vf_free}
v_{F,0}(\vec k) = v^{C}_{F,0}+\Delta_v (\vec k),
\eeq
where $v^{C}_{F,0}$ is constant and $\Delta_v (\vec k)$ is a small correction with $\lim_{|\vec k|\to 0} \Delta_v (\vec k)=0$ (the Dirac point is placed at the origin). For most cases, we can neglect $\Delta_v (\vec k)$ and assume $v_{F0}(\vec k) = v^{C}_{F0}$, but at weak coupling it is actually important to take the corrections introduced by $\Delta_v (\vec k)$  into account due to the fact that the renormalization is quite small and is comparable with the curvature of the free band. 
One then integrates eq. (\ref{eq:vf_renorm}) along a direction in the BZ starting from the Dirac point, taking into account that $\Delta_v (\vec k)=O(k)$ in its vicinity to obtain the following expression for the renormalized energy 
\beq \label{eq:energy_free}
\frac{E(k)}{E_0(k)}=1+ C\left(1 +\ln{\frac{\Lambda}{k}}+ O(k^3 \ln k) \right).
\eeq
As the leading correction becomes small when approaching the Dirac point, it can be neglected and the first two terms can be used as a fitting function for the ratio of the renormalized to free dispersion. We note that if the logarithmic form (\ref{eq:vf_renorm}) is violated for some reason and we have a power law for the renormalized $v_F$, then the renormalized dispersion should also exhibit a power law. This would then become apparent when fitting the data.
Unfortunately, we only have access to the renormalized Fermi velocity when comparing with experiment, thus we employ the procedure in eq. (\ref{eq:numDiff}) in that case.

The results for the renormalized dispersion at small interaction are shown in Fig. \ref{fig:QMC_justification}\textcolor{red}{b}. As one can clearly see, perturbation theory works well in this regime as the RPA corrections to the one-loop result are quite small. The QMC data deviate substantially from the zero-temperature one-loop result, but coincide nicely with the finite-temperature one-loop lattice perturbation theory. 
In Fig. \ref{fig:QMC_justification}\textcolor{red}{c}, one can see the evolution of the renormalized dispersion with uniformly rescaled potentials. Clearly, QMC data converge to the finite-temperature one-loop approximation in the limit of small interaction.

\subsection{Finite-temperature effects}

\begin{figure}[!t]
    \centering
	\includegraphics[angle=0,width=0.95\textwidth]{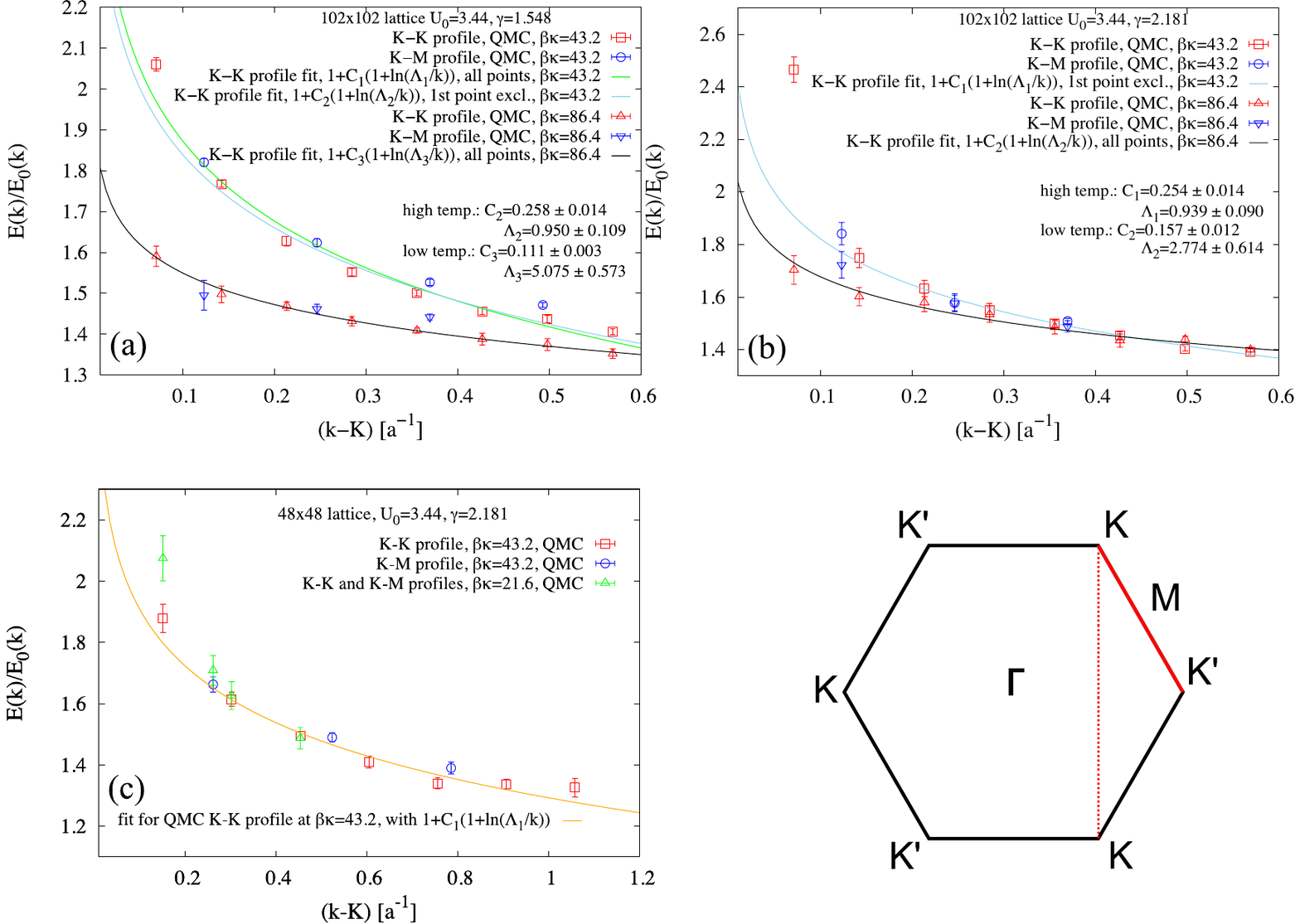}
	\caption{The renormalized dispersion of electrons at strong coupling using the electron-electron potential variant I (figure \textbf{(a)}) and the potential variant II (\textbf{(b)} and \textbf{(c)}). \textbf{(a)} and \textbf{(b)}: QMC results on $102\times102$ lattice. \textbf{(c)} QMC results on $48\times48$ lattice.  In all cases we fit the QMC data along the K-K line in the Brillouin zone (BZ) and the K-M profile is included for completeness. The horizontal axis is the distance from the Dirac point in units of inverse lattice step. The same units are used for the displayed values of the cutoff $\Lambda$. Description of the potentials of electron-electron interaction used in the simulation can be found in the main text. }
	\label{fig:energy_fitting}
\end{figure}

The renormalized dispersion relation for potential variants I and II (see the main text for the their definition) is shown in Fig. \ref{fig:energy_fitting}. Unlike previous QMC studies \cite{Tang570}, the lattice size appears to be large enough to clearly observe the non-linear dispersion relation. First we look at the results on a $102\times102$ lattice for both smaller (Fig. \ref{fig:energy_fitting}\textcolor{red}{a}) and larger (Fig. \ref{fig:energy_fitting}\textcolor{red}{b}) Coulomb tails. The results at intermediate temperature $T=0.02315 \kappa$, which corresponds to inverse temperature $\beta = 43.2/ \kappa$ show sizeable finite-temperature effects: 
while all points except the first one are aligned along the logarithmic curve, the points at the lowest momenta are shifted upwards for both a large and small Coulomb tail. This   feature  of  the  data cannot be attributed to  finite-size effects:  it is the most pronounced at lower energies  such  that  lower temperatures should lead to the enhancement of this effect. We however  observe  the opposite behaviour: the low-temperature data for the $102\times102$ lattice is perfectly aligned according to the logarithmic fit for both sets of potentials.  

A similar effect can be also observed on the  $48\times48$ lattice  if we compare the two large temperatures: $T=0.02315 \kappa$ and $T=0.04630 \kappa$ (Fig. \ref{fig:energy_fitting}\textcolor{red}{c}). Once we reduce both the inverse temperature and the lattice size by a factor of two, we observe a similar  enhancement  of  the  first  data  point with respect to the logarithmic curve, in comparison to  the same lattice at a lower temperature. Notably, further points do not experience significant finite-temperature effects in perfect agreement with the $102\times102$ lattice data at momentum $|k-K|>0.3a^{-1}$. Hence, we  conclude  that    these  finite-temperature effects, that drive the dispersion relation (hence also $v_F$) upwards with respect to the logarithmic curve, are not simulation artefacts and  should  be  observed  in  experiments.

\section{Perturbation theory on the lattice}
\subsection{Introduction and Feynman rules}
In this appendix, we discuss the details of our perturbative calculations carried out on the hexagonal lattice at finite volume, $L\equiv N_xN_y$. We start with the introduction of the lattice Hamiltonian and the Feynman rules in momentum space. We then discuss the one-loop self-energy and how we obtain the renormalized dispersion, $E_R(p)$. After this, we discuss the random phase approximation (RPA) \cite{PhysRev.92.609}, and how we have applied it to the present problem.

After performing a canonical particle-hole transformation on the fermion operators, the lattice Hamiltonian in position space can be written as
\beq
\label{eq:Hamiltonian_graphene_el_hol_position}
  \!\!\!\hat{{H}}\! =\! -\kappa \sum_{\langle x,y\rangle} (  \hat a^\dag_{x} \hat a_{y} \!+\! \hat b^\dag_{x} \hat b_{y} \!+\! \mbox{h.c} ) \!+\! \frac{1}{2}\sum_{x,y} V_{x,y}\hat q_x \hat q_y,
\eeq 
where we have introduced creation operators for electrons and holes, $\hat a^\dag_{x}$ and  $\hat b^\dag_{x}$. The charge operator appears in the interaction term and in terms of the electron and hole operators can be written as $\hat q_x = \hat a^\dag_{x} \hat a_{x} - \hat b^\dag_{x} \hat b_{x}$. For perturbative calculations, it is convenient to go to momentum space, where (\ref{eq:Hamiltonian_graphene_el_hol_position}) becomes 
\beq
\label{eq:Hamiltonian_graphene_el_hol_momentum}
\!\!\!\hat{{H}}\! = \sum_{\vec{k}} \hat \Psi^{\dagger}_{\vec{k}} \hat{H}_0(\vec{k}) \hat \Psi_{\vec{k}} + \frac{1}{2} \sum_{\vec{k},\mu,\nu} V_{\mu,\nu}({\vec{k}})~ \hat q_{\mu}(-\vec{k}) ~\hat q_{\nu}(\vec{k}),
\eeq
with the two-body potential matrix in momentum space given by
\beq \label{eq:MomentumSpacePotential}
V_{\mu,\nu}({\vec{k}}) \equiv  \sum_{\vec{r}} V_{0,\vec{r}} ~e^{-i \vec{k} \cdot \vec{r}}.
\eeq 
The sum in (\ref{eq:MomentumSpacePotential}) is over the Bravais lattice vectors and the indices specify the locations of the source and sink with respect to the two sublattices. For a finite spatial volume, periodic boundary conditions in space must be imposed and this modifies the infinite-volume prescription given in (\ref{eq:MomentumSpacePotential}).
The charge operator in momentum space takes the form
\beq
\hat q_{\mu}(\vec{k}) = \sum_{\vec{q}} \left( \hat a^\dag_{\mu, \vec{q}-\vec{k}} \hat a_{\mu, \vec{q}} - \hat b^\dag_{\mu,\vec{q}-\vec{k}} \hat b_{\mu, \vec{q}} \right),
\eeq
where we have introduced the sublattice index $\mu$. In (\ref{eq:Hamiltonian_graphene_el_hol_momentum}), we have written the hopping term using the following vector
\beq
\Psi^{\dagger}_{\vec{k}} =
\begin{pmatrix}
\hat a^{\dagger}_{1,k}, & a^{\dagger}_{2,k}, & b^{\dagger}_{1,k}, & b^{\dagger}_{2,k}  \end{pmatrix},
\eeq
with $a^{\dagger}_{\mu,k}$ and $b^{\dagger}_{\mu,k}$ representing particle and hole creation operators in momentum space on a given sublattice characterized by the index $\mu=1,2$. In this basis, the single-particle Hamiltonian takes the form 
\beq
&&\hat{H}_0(\vec{k})_{a,\mu;b,\nu} = \delta_{a,b} \hat{\tilde{H}}_0(\vec{k})_{\mu,\nu}, \\ \label{eq:FreeHamitlonianMatrixElements}
&&\hat{\tilde{H}}_0(\vec{k})_{1,2} = \left(\hat{\tilde{H}}_0(\vec{k})_{2,1}\right)^* = -\kappa f(\vec{k}),
\eeq 
where Latin indices represent the particle-hole (p-h) degree of freedom and Greek indices represent the sublattice degree of freedom. 
In (\ref{eq:FreeHamitlonianMatrixElements}), we have introduced the hexagonal lattice structure factor 
\beq \label{eq:StructureFactorHexagonal}
f(\vec{k}) \equiv \sum_{i=1,2,3} e^{i \vec{k} \cdot \vec{\delta}_i},
\eeq
where the sum runs over the three nearest-neighbor lattice vectors.

\begin{figure} 
\centering
        \includegraphics[ angle=0,width=0.5\textwidth]{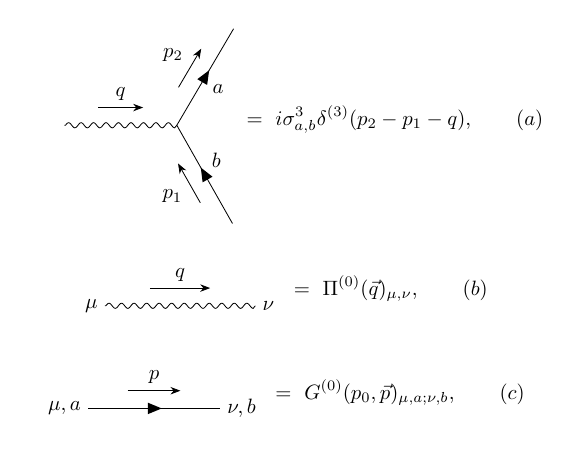}
        \caption{Feynman rules in momentum-frequency representation for the action in (\ref{eq:ActionRealSpace}). The three-component notation is used to denote the frequency and momentum dependence of each line, i.e. $p \equiv (p_0, \vec{p})$. The electron-boson vertex is shown in (a), where the Pauli matrix acts in p-h space and both momentum and frequency are conserved between the incoming and outgoing lines. The wavy line depicted in (b) corresponds to the free boson propagator and the solid line in (c) corresponds to the free fermion propagator.}
\label{fig:FeynmanRules}
\end{figure}

As is done on the lattice for QMC calculations, we formulate the Feynman rules using the functional integral formalism. We start with the partition function in Euclidean time after performing the Hubbard-Stratonovich transformation
\beq \label{eq:PartitionFunctionLinearized}
&&Z = \int \mathcal{D} \phi \mathcal{D} \bar\psi \mathcal{D} \psi e^{-S[\bar\psi, \psi, \phi]}, \\ \label{eq:ActionSumofTerms}
&&S = S_0 + S_{\rm {int.}} + S_{\rm{B}}, \\ \nn && S = \int^{\beta}_0 \mathrm{d}\tau \bigg[ -\sum_{x,a} \bar\psi_{x,a} \partial_{\tau} \psi_{x,a} + \sum_{x,a;y,a} \bar\psi_{x,a} (H_0)_{x,a;y,b} \psi_{y,b} \\ \label{eq:ActionRealSpace} && \qquad \qquad \qquad + i \sum_{x,a,b}\phi_x \bar\psi_{x,a} \sigma^z_{a,b}\psi_{x,b} - \frac{1}{2} \sum_{x,y}\phi_x V^{-1}_{x,y}\phi_y \bigg],
\eeq 
where $\bar \psi$ and $\psi$ are Grassmann fields and $\phi$ is a scalar field. We note that here, unlike in HMC simulations, we are working with only one, charge-coupled Hubbard field. The $U(1)$ gauge symmetry is violated by the potential matrix, $V_{x,y}$. In the limit where the $U(1)$ symmetry is restored, the Hubbard field $\phi$ becomes the electric field. Although we have suppressed the Euclidean time dependence of the fields in (\ref{eq:ActionRealSpace}), the usual (anti-)periodicity applies to ($\psi$)$\phi$. We now write down the expressions for the propagators. The fermion propagator is defined by the expression
\beq \label{eq:FermionPropagatorFullPathIntegral}
G_{x,a;y,b}(\tau) = \frac{1}{Z} \int \mathcal{D} \phi \mathcal{D} \bar\psi \mathcal{D} \psi ~ \psi_{x,a}(\tau_1) \bar\psi_{y,b}(\tau_2) e^{-S[\bar\psi, \psi, \phi]},
\eeq 
where $\tau\equiv\tau_1 - \tau_2$. For the free propagator, one can ignore both the interaction term and the bosonic action in (\ref{eq:ActionSumofTerms}). Going to the momentum-frequency representation, we obtain
\beq \label{eq:FreeFermionPropagator1}
G^{(0)}_{a,\mu;b,\nu}(p_0, \vec{p}) = \delta_{a,b} (\mathcal{M}^{-1}(p_0, \vec{p}))_{\mu,\nu},
\eeq 
where we see that the propagator is diagonal in particle/hole-space, and the matrix $\mathcal{M}$, which is in sublattice space, is given by
\beq
 \label{eq:FermionBilinearOperator}
\mathcal{M}(p_0, \vec{p}) = \begin{pmatrix} 
ip_0 & -\kappa f(\vec{p})  \\ 
-\kappa f^*(\vec{p}) & ip_0  
\end{pmatrix}.
\eeq
Plugging (\ref{eq:FermionBilinearOperator}) into (\ref{eq:FreeFermionPropagator1}) we obtain
\beq 
G^{(0)}_{a,\mu;b,\nu}(p_0, \vec{p}) = \frac{-\delta_{a,b}}{p^2_0 + \kappa |f(\vec{p})|^2}  \begin{pmatrix}  ip_0 & \kappa f(\vec{p})  
\\ \kappa f^*(\vec{p}) & ip_0  
\end{pmatrix}_{\mu,\nu}.
\eeq 
The free fermion propagator will be denoted by a solid line which carries spatial momentum $\vec{p}$ and frequency $p_0$ and is depicted in Fig.~\ref{fig:FeynmanRules}\textcolor{red}{c}.

The two-body interaction in our theory is mediated by the bosonic field, $\phi$, which satisfies periodic boundary conditions in Euclidean time. The propagator associated with this field is defined by
\beq \label{eq:BosonPropagatorFullPathIntegral}
\Pi_{x,y}(\tau) = \frac{1}{Z} \int \mathcal{D} \phi \mathcal{D} \bar\psi \mathcal{D} \psi ~ \phi_{x}(\tau_1) \phi_{y}(\tau_2) e^{-S[\bar\psi, \psi, \phi]},
\eeq 
where $\tau\equiv\tau_1 - \tau_2$. The free bosonic propagator is obtained by ignoring all terms in the action except $S_{\rm B}$, which is Gaussian. This leads to the following expression in the momentum-frequency representation
\beq \label{eq:FreeBosonic}
\Pi^{(0)}_{\mu, \nu}(\vec{p}) = V_{\mu,\nu}(\vec{p}),
\eeq 
where, in the infinite-volume limit, the two-body potential in momentum space which we consider satisfies $V_{\mu,\nu}(\vec{p}) = (V_{\nu,\mu}(\vec{p}))^*$ and $V_{\mu,\nu}(\vec{p}) = (V_{\mu,\nu}(-\vec{p}))^*$. We note that (\ref{eq:FreeBosonic}) is independent of frequency. Here the indices refer to the sublattice degree of freedom. The free bosonic propagator will be denoted by a wavy line which carries spatial momentum $\vec{p}$ and frequency $p_0$, see Fig.~\ref{fig:FeynmanRules}\textcolor{red}{b}.

We finally come to the electron-boson interaction vertex which is determined by $S_{\rm {int}}$. It consists of an incoming fermion line with p-h (particle-hole) index $a$, an outgoing fermion line with p-h index $b$, and a boson line. Frequency and momentum are conserved among the three lines. The vertex, depicted in Fig.~\ref{fig:FeynmanRules}\textcolor{red}{a}, is trivial in sublattice space and includes a factor of $i \sigma^z_{a,b}$ which acts in p-h space (sum over contracted p-h indices).

We comment here that so far this derivation has applied to the Euclidean version of the theory. In this case, the time variable $\tau$, runs over a compact interval, $[0,\beta)$, and the frequencies become discrete multiples of $\pi/\beta$ (even multiples for bosons and odd multiples for fermions). For the fermions, at $T=0$, the only thing that would change is the replacement $p_0 \to p_0 \pm i \eta$, where $\eta = 0^+$. The infinitesimal quantity $\eta$ displaces the poles from the real axis in a direction which depends on whether we are dealing with the positive or negative energy states of $\hat H_0$, when working in the eigenbasis. In what follows, we work in the $T=0$ but also give the results for expressions in the Matsubara formalism.

\begin{figure} 
\centering
        \includegraphics[ angle=0,width=0.5\textwidth]{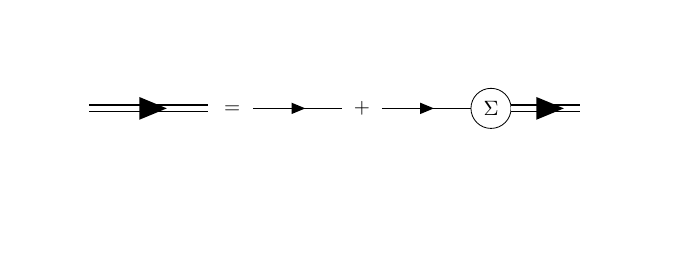}
        \caption{Diagrammatic expression of Dyson's equation for the fermion propagator. The thick line denotes the full propagator, $G$, the thin line denotes the free propagator, $G^{(0)}$, and the blob represents the self-energy $\Sigma$.}
        \label{fig:DysonEquationFermion}
\end{figure}

\subsection{Renormalized dispersion at one-loop}
To obtain the renormalized energy, and thus the effects of interactions on the Fermi velocity within perturbation theory, we must find the poles of the full fermion propagator in momentum-frequency representation. The full propagator is represented by all connected Feynman diagrams with one incoming and one outgoing fermion line. Typically, one considers irreducible diagramatic quantities. For the fermion propagator, one is interested in one-particle irreducible (1PI) diagrams. These are connected diagrams with a single incoming and outgoing fermion lines which have been amputated, and which do not become disconnected after cutting a single internal fermion line. These diagrams constitute what is known as the self-energy, $\hat \Sigma(p_0,\vec{p})$, which is a matrix that is diagonal in p-h space and is generally not diagonal in sublattice space. Using Dyson's equation, one can express the full propagator as
\beq \label{eq:DysonEquationFermion}
\hat{G}^{-1}(p_0,\vec{p}) = \hat{G}^{-1}_0(p_0,\vec{p}) - \hat \Sigma(p_0,\vec{p}).
\eeq 
In terms of Feynman diagrams, (\ref{eq:DysonEquationFermion}) is depicted in Fig. \ref{fig:DysonEquationFermion}.
The poles of $\hat{G}$ determine the quasiparticle's dispersion and this leads to the following equation
\beq \label{eq:PoleFermionPropagator}
\begin{pmatrix} ip_0 & (\hat{G}^{-1}_0 - \hat \Sigma)_{1,2} \\ (\hat{G}^{-1}_0 - \hat \Sigma)_{2,1} & ip_0 \end{pmatrix} = 0,\quad p_0 = iE_R(\vec{p}).
\eeq
Typically, this equation must be solved numerically for a fixed momentum $\vec{p}$ in the BZ.

\begin{figure} 
\centering
        \includegraphics[ angle=0,width=0.5\textwidth]{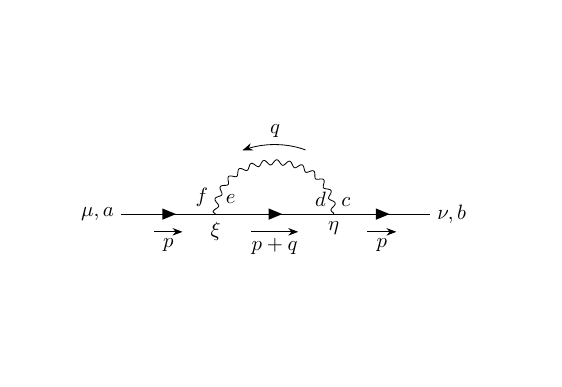}
        \caption{Diagrammatic expression for the quantity $G^{(0)}\Sigma^{(1)}G^{(0)}$, where $\Sigma^{(1)}$ is the bare first-order self-energy.}
\label{fig:BareFirstOrder}
\end{figure}

\begin{figure} 
\centering
        \includegraphics[ angle=0,width=0.5\textwidth]{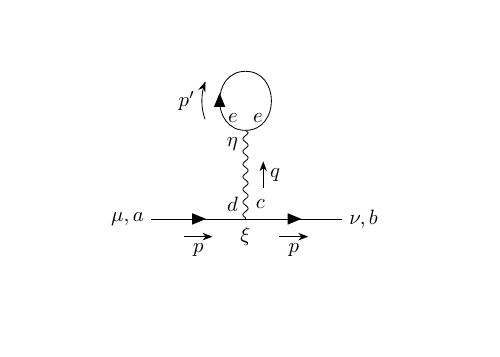}
        \caption{Diagrammatic expression for the quantity $G^{(0)}\Sigma^{(H)}G^{(0)}$, where $\Sigma^{(H)}$ is the Hartree self-energy.}
\label{fig:Hartree}
\end{figure}

The self-energy, to lowest-order, is given by the Fock diagram displayed in Fig.~\ref{fig:BareFirstOrder}. The expression for the first-order self-energy is given by
\beq \nn
&& \Sigma^{(1)}(p_0, \vec{p})_{f,\xi;c,\eta} =  -\frac{\delta_{c,f}}{4 \pi L^2} \int \mathrm{d}q_0 \sum_{\vec{q}} \bigg[ \Pi^{(0)}_{\eta, \xi}(\vec{q}) G^{(0)}_{\xi,\eta}(q_0, \vec{p}+\vec{q}) \\ \label{eq:SigmaFirstOrderBare} &&\qquad \qquad \qquad \qquad \qquad + \Pi^{(0)}_{\xi,\eta}(\vec{q}) G^{(0)}_{\xi,\eta}(q_0, \vec{p}-\vec{q}) \bigg].
\eeq
The sum over spatial momentum ranges over the entire Brillouin zone and the integration over all frequencies can be performed analytically. Noting that (\ref{eq:SigmaFirstOrderBare}) does not depend on $p_0$, one can easily obtain the expression for the renormalized dispersion 
\beq \label{eq:RenomarlizedDispersion} 
&&E_R(\vec{p}) = | G^{(0)}_{1,2} - \Sigma^{(1)}_{1,2} |, \\ &&= \left| - \kappa f(\vec{p}) + \frac{1}{4 L^2}\bigg[\sum_{\vec{q}} \Pi^{(0)}_{2, 1}(\vec{q})  e^{i \theta(\vec k+\vec q)} + \sum_{\vec{q}} \Pi^{(0)}_{1, 2}(\vec{q})  e^{i \theta(\vec k-\vec q)}\bigg] \right|,
\eeq 
where $e^{i \theta(\vec q)} \equiv \frac{f(\vec{q})}{|f(\vec{q})|}$. This is what we will refer to as the ``bare" one-loop renormalized dispersion. The finite-temperature generalization can be obtained by inserting factors of \\ $\tanh(|\kappa f(k\pm q)|/(2T))$ into the corresponding sums over momentum in (\ref{eq:RenomarlizedDispersion}). We note that the Hartree diagram (depicted in Fig.~\ref{fig:Hartree}), which is formally the contribution of the same order as the Fock diagram, vanishes due to p-h symmetry. This symmetry would be violated with the introduction of a chemical potential or taking into account next-to-nearest neighbor hopping in the single-particle Hamiltonian, thus modifying the form of (\ref{eq:PoleFermionPropagator}). The results of the first-order calculation in the weak-coupling regime are shown in Fig. \ref{fig:energy_Vf_fitting} for various volumes. One can clearly see the logarithmic behavior of both the Fermi velocity as well as the dispersion.

\begin{figure} 
\centering
        \includegraphics[ angle=0,width=0.5\textwidth]{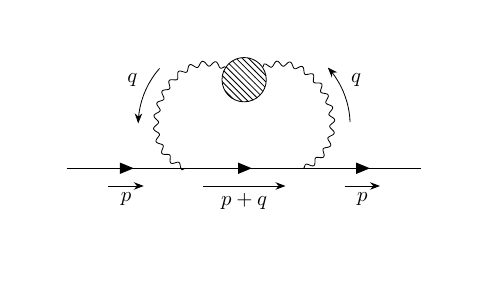}
        \caption{Diagram corresponding to the RPA self-energy. Here we have omitted the sublattice and p-h indices.}
        \label{fig:RPASelfEnergy}
\end{figure}

\begin{figure} 
\centering
        \includegraphics[angle=0,width=0.7\textwidth]{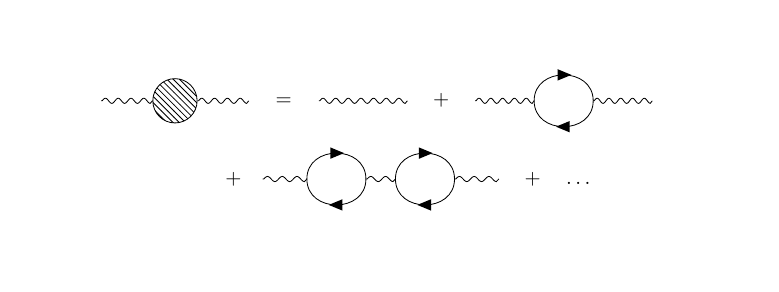}
        \caption{Diagram expression for the bosonic propagator in the RPA approximation. The polarization is calculated to first-order, which corresponds to the particle-hole bubble.}
        \label{fig:DysonEquationBoson}
\end{figure}

\begin{figure} 
\centering
        \includegraphics[angle=0,width=0.5\textwidth]{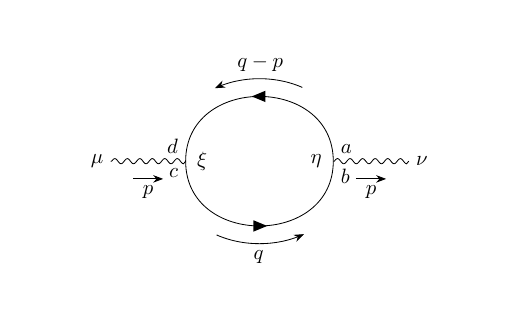}
        \caption{Diagram expression for the particle-hole bubble which is the lowest order approximation to the polarization, $\hat P$.}
        \label{fig:Polarization}
\end{figure}

\subsection{Lattice RPA approximation}
Although it is unreasonable to assume that one can obtain the propagator, and thus the dispersion, to all orders in perturbation theory, there are several resummations which provide good approximations. One such approximation is RPA, whereby one ``dresses" the bare bosonic propagator and computes the diagram with the same topology as the bare self-energy (see Fig.~\ref{fig:RPASelfEnergy}). The bosonic propagator is dressed by computing the 1PI diagrams and then applying Dyson's equation which reads
\beq \label{eq:BosonicDysonEquation}
\hat \Pi(p_0,\vec{p}) = \hat \Pi^{(0)}(\vec{p}) \left( \mathbb{1} - \hat{P}(p_0,\vec{p})  \hat \Pi^{(0)} (\vec{p}) \right)^{-1},
\eeq
where the matrix-valued quantity $\hat P$ is known as the polarization and contains the bosonic 1PI diagrams with two external vertices. This relation between the polarization and full bosonic propagator is depicted in Fig.~\ref{fig:DysonEquationBoson}.
To lowest-order, the polarization can be represented by the particle-hole bubble depicted in Fig.~\ref{fig:Polarization}. The expression for this quantity reads
\beq \nn
&& P^{(1)}(p_0,\vec{p})_{\eta,\xi} = \frac{1}{2\pi L^2} \int \mathrm{d}q_0 \sum_{\vec{q}} \bigg[ G^{(0)}_{\xi,\eta}(q_0, \vec{q})  G^{(0)}_{\eta,\xi}(q_0+p_0, \vec{p}+\vec{q}) \\ \label{eq:ParticleHoleBubble} && \qquad \qquad \qquad \qquad +  G^{(0)}_{\eta,\xi}(q_0, \vec{q})  G^{(0)}_{\xi,\eta}(q_0-p_0, \vec{q}-\vec{p}) \bigg].
\eeq 
The diagonal elements of $\hat G^{(0)}$ are identical, and thus we can write
\beq \label{eq:ParticleHoleBubbleDiagonal}
P^{(1)}(p_0,\vec{p})_{1,1} = P^{(1)}(p_0,\vec{p})_{2,2} = -\frac{\kappa}{L^2} \sum_{\vec{q}}{}^{'} \frac{(|f(\vec{q})| + |f(\vec{p}+\vec{q})|)}{p^2_0 + \kappa^2 (|f(\vec{q})| + |f(\vec{p}+\vec{q})|)^2},
\eeq 
where the frequency integral in (\ref{eq:ParticleHoleBubble}) has been performed analytically
\beq \nn 
I_{1,1} = I_{2,2} &\equiv& \int^{+\infty}_{-\infty} \mathrm{d}q_0 \frac{q_0(q_0+p_0)}{\left(q^2_0 + \kappa^2 |f(\vec{q})|^2 \right)\left((q_0 + p_0)^2+ \kappa^2 |f(\vec{p}+\vec{q})|^2 \right)}, \\ \label{eq:FrequencyIntegralDiagonal} &=& \frac{\pi \kappa \left(|f(\vec{q})| +  |f(\vec{p}+\vec{q})| \right)}{p^2_0 + \kappa^2\left(|f(\vec{q})|+|f(\vec{p}+\vec{q})| \right)^2}.
\eeq 
The off-diagonal elements are given by 
\beq \label{eq:ParticleHoleBubbleOffDiagonal}
P^{(1)}(p_0,\vec{p})_{1,2}  = \frac{\kappa}{L^2} \sum_{\vec{q}}{}^{'} \frac{e^{-i\theta(q)}e^{i\theta(q+p)}(|f(\vec{q})| + |f(\vec{p}+\vec{q})|)}{p^2_0 + \kappa^2 (|f(\vec{q})| + |f(\vec{p}+\vec{q})|)^2},
\eeq 
where $P^{(1)}_{1,2} = (P^{(1)}_{2,1})^*$ and the frequency integral has also been performed analytically.
\beq \nn 
I_{1,2} &\equiv& \int^{+\infty}_{-\infty} \frac{\mathrm{d}q_0}{\left(q^2_0 + \kappa^2 |f(\vec{q})|^2 \right)\left((q_0 + p_0)^2+ \kappa^2 |f(\vec{p}+\vec{q})|^2 \right)}, \\ \label{eq:FrequencyIntegralOffDiagonal} &=& \frac{\pi \left(|f(\vec{q})| +  |f(\vec{p}+\vec{q})| \right)}{\kappa|f(\vec{q})||f(\vec{p}+\vec{q})|\left(p^2_0 + \kappa^2\left(|f(\vec{q})|+|f(\vec{p}+\vec{q})| \right)^2 \right)}.
\eeq 
The phases $\theta(q)$ and $\theta(p+q)$ are not well-defined at the Dirac point, where the structure factor vanishes, and thus in (\ref{eq:ParticleHoleBubbleDiagonal}) and (\ref{eq:ParticleHoleBubbleOffDiagonal}), the sum over loop momenta excludes these points.

When generalizing the zero-temperature result for the polarization to $T\neq 0$, we replace the integrals over the frequency $q_0$ with sums over the fermionic Matsubara frequencies, $\omega_n = (2n+1)\pi/\beta$, which circulate in the particle-hole bubble. To evaluate these sums, one first expresses the sum as a contour integral in the complex plane which encloses the Matusbara frequencies on the imaginary axis. Deforming the contour in the usual such that the poles of the fermion propagators are enclosed, gives the final expression in terms of the residues at these poles. The result for the diagonal elements of the polarization tensor is given by
\beq \nn  
&& P^{(1)}(i\Omega_n,\vec{p})_{T,\text{diag}} = -\frac{1}{L^2} \sum_{\vec{q}}{}^{'} \bigg[ -\frac{\left( i\Omega_n + E_0(\vec{q}) \right) n_F(E_0(\vec{q}))}{\left( i\Omega_n + E_0(\vec{q}) \right)^2 - E^2_0(\vec{q}+\vec{p})} \\ \nn && \qquad - \frac{\left( i\Omega_n - E_0(\vec{q}) \right) n_F(-E_0(\vec{q}))}{\left( i\Omega_n - E_0(\vec{q}) \right)^2 - E^2_0(\vec{q}+\vec{p})} + \frac{\left( i\Omega_n + E_0(\vec{p}+\vec{q}) \right) n_F(-E_0(\vec{p}+\vec{q}))}{\left( i\Omega_n + E_0(\vec{p}+\vec{q}) \right)^2 - E^2_0(\vec{q})} \\ \label{eq:ParticleHoleBubbleDiagonalNonzeroTDiag} && \quad + \frac{\left( i\Omega_n - E_0(\vec{p}+\vec{q}) \right) n_F(E_0(\vec{p}+\vec{q}))}{\left( i\Omega_n - E_0(\vec{p}+\vec{q}) \right)^2 - E^2_0(\vec{q})} \bigg],
\eeq 
where $P_{1,1} = P_{2,2} \equiv P_{\text{diag}}$, $E_0(\vec{q}) = \kappa | f(\vec{q}) |$, $n_F$ is the Fermi function, and $\Omega_n = 2n \pi /\beta$ is a bosonic Matsubara frequency. For the off-diagonal element we obtain
\beq \nn
&& P^{(1)}(i\Omega_n,\vec{p})_{T,\text{off}} = -\frac{\kappa^2}{L^2} \sum_{\vec{q}}{}^{'} f(\vec{p}+\vec{q})f^*(\vec{q}) \bigg[ \frac{n_F(E_0(\vec{q}))}{E_0(\vec{q})\left[ \left( i\Omega_n + E_0(\vec{q}) \right)^2 - E^2_0(\vec{q}+\vec{p})\right]} \\ \nn && \quad 
-\frac{n_F(-E_0(\vec{q}))}{E_0(\vec{q})\left[ \left( i\Omega_n - E_0(\vec{q}) \right)^2 - E^2_0(\vec{q}+\vec{p})\right]} + \frac{n_F(E_0(\vec{p}+\vec{q}))}{E_0(\vec{p}+\vec{q})\left[ \left( i\Omega_n - E_0(\vec{p}+\vec{q}) \right)^2 - E^2_0(\vec{q})\right]} \\ \label{eq:ParticleHoleBubbleDiagonalNonzeroTNonDiag} &&  -\frac{n_F(-E_0(\vec{p}+\vec{q}))}{E_0(\vec{p}+\vec{q})\left[ \left( i\Omega_n + E_0(\vec{p}+\vec{q}) \right)^2 - E^2_0(\vec{q})\right]},
\eeq 
where $P_{1,2}=P^*_{2,1} = P_{\text{off}}$. One can verify that these expressions reproduce (\ref{eq:ParticleHoleBubbleDiagonal}) and (\ref{eq:ParticleHoleBubbleOffDiagonal}) in the limit $T\to 0$.

Putting all of this together, we can write the fermion self-energy in the zero-temperature RPA approximation as 
\beq \nn 
&& \Sigma^{({RPA})}(p_0,\vec{p})_{f,\xi;c,\eta} = -\frac{\delta_{f,c}}{4\pi L^2} \int \mathrm{d}q_0 \sum_{\vec{q}} \bigg[ \Pi(q_0,\vec{q})_{\eta,\xi} G^{(0)}(p_0+q_0,\vec{p}+\vec{q})_{\xi,\eta} \\ \label{eq:SigmaRPADef}&& \qquad \qquad \qquad \qquad \qquad + \Pi(q_0,\vec{q})_{\xi,\eta} G^{(0)}(p_0-q_0,\vec{p}-\vec{q})_{\xi,\eta} \bigg], 
\eeq 
where as in the bare case, the RPA self-energy is also diagonal in p-h space. From (\ref{eq:SigmaRPADef}), one sees that the frequency integral needs to be evaluated numerically. It will be convenient to rewrite  (\ref{eq:SigmaRPADef}) in the form
\beq
&&\Sigma^{({RPA})}(p_0,\vec{p})_{f,\xi;c,\eta} = \Sigma^{(1)}(\vec{p})_{f,\xi;c,\eta} + \delta \Sigma^{({RPA})}(p_0,\vec{p})_{f,\xi;c,\eta},\\ \nn
&& \delta \Sigma^{({RPA})}(p_0,\vec{p})_{f,\xi;c,\eta} \equiv-\frac{\delta_{f,c}}{4\pi L^2} \int \mathrm{d}q_0 \sum_{\vec{q}} \bigg[ F(q_0,\vec{q})_{\eta,\xi} G^{(0)}(p_0+q_0,\vec{p}+\vec{q})_{\xi,\eta} \\ \label{eq:DeltaSigmaRPADef}&& \qquad \qquad  \qquad \qquad + F(q_0,\vec{q})_{\xi,\eta} G^{(0)}(p_0-q_0,\vec{p}-\vec{q})_{\xi,\eta} \bigg], 
\eeq 
where we have introduced the matrix, $\hat F \equiv \hat \Pi - \hat \Pi^{(0)} $. The entire frequency dependence of $\hat \Sigma^{({RPA})}$ is now contained in (\ref{eq:DeltaSigmaRPADef}). Unlike the case of the bare first-order self-energy, finding the pole of the propagator will require multiple numerical evaluations of (\ref{eq:DeltaSigmaRPADef}) for fixed $\vec{p}$.

From (\ref{eq:ParticleHoleBubbleDiagonal}) and (\ref{eq:ParticleHoleBubbleOffDiagonal}), one sees that the dressed bosonic propagator depends on frequency. Before we use this in the evaluation of the RPA self-energy, we determine the frequency dependence for asymptotically large values of $p_0$. We start by writing
\beq \nn
\mathcal{F}(p_0,\vec{p}) &\equiv& F(q_0,\vec{q})_{1,2} =  \Pi(p_0,\vec{p})_{1,2} - \Pi^{(0)}(\vec{p})_{1,2}, \\ \label{eq:DefinitionF}
&=& \frac{P^{(1)}_{1,2} \left[ (\Pi^{(0)}_{1,1})^2 - (\Pi^{(0)}_{1,2})^2 \right] -\Pi^{(0)}_{1,2} \left( \det \hat B - 1 \right)}{\det \hat B },   
\eeq 
where we have defined the matrix, $\hat B \equiv \mathbb{1} - \hat{P}\hat \Pi^{(0)}$, and have used (\ref{eq:BosonicDysonEquation}) in expressing the matrix elements of $\hat F$ in terms of the various components of $\hat P$ and $\hat \Pi$. Using the frequency dependence of $\hat P$, one can determine that 
\beq
\mathcal{F}(p_0,\vec{p}) \propto \frac{1}{p^2_0 + \mathcal{C}}, \qquad p_0 \to \pm \infty
\eeq 
 where $\mathcal{C}$ is a frequency-independent constant. Thus the frequency integral is convergent and can be evaluated numerically using standard techniques.
 
 At nonzero-$T$, it turns out that performing the Matsubara sum in (\ref{eq:DeltaSigmaRPADef}) is rather straightforward. This results from the fact that the determinant of the matrix $\hat{B}$, which appears in the denominator of (\ref{eq:DefinitionF}), has no zeros in the complex frequency plane for any value of the momentum $\vec{p}$. This can be understood physically in terms of collective charge excitations in monolayer graphene \cite{PhysRevB.75.205418}. 

As we did in the bare case, we make the replacement $p_0 = iE_R(\vec{p})$. Here we note that this is, in fact, an assumption that the pole is real and thus the lifetime of the particle is infinite. This is a feature of the RPA approximation that was seen in previous studies \cite{DasSarma:PhysRevB2007}. The expression in (\ref{eq:DeltaSigmaRPADef}), after combining the two terms, can be written as 
\beq \label{eq:DeltaSigmaRPAFinal}
\delta \Sigma^{({RPA})}_{f,1;c,2}(E_R,\vec{p}) = -\frac{\delta_{f,c}}{4\pi L^2} \int \mathrm{d}q_0 \sum_{\vec{q}} \frac{ \kappa f(\vec{p}+\vec{q}) \mathcal{F}^*(q_0,\vec{q})(q^2_0+\kappa^2|f(\vec{p}+\vec{q})|^2-E^2_R)}{(q^2_0 + \kappa^2|f(\vec{p}+\vec{q})|^2-E^2_R)^2 + 4q^2_0E^2_R},
\eeq 
where one can show that $\delta \Sigma^{({RPA})}_{2,1} = \big(\delta \Sigma^{({RPA})}_{1,2}\big)^*$. Similarly, the expression for nonzero-temperature can be written as 
 \beq  \label{eq:DeltaSigmaRPAFinalNonzeroT}
&&\delta \Sigma^{({RPA,T})}_{f,1;c,2}(E_R,\vec{p}) = -\frac{\delta_{f,c}}{L^2} \sum_{\vec{q}} e^{i \theta(\vec{p}+\vec{q})} \sum_{\sigma=\pm 1,\tau = \pm 1} (-)^{\sigma} \mathcal{F}^*(\epsilon_{\sigma,\tau},\vec{q}) n_F(\epsilon_{\sigma,\tau}), \\ 
 &&\epsilon_{\sigma,\tau} = \epsilon_{\sigma,\tau}(\vec{q}+\vec{p}) \equiv (-)^{\sigma}E_R + (-)^{\tau} E_0(\vec{p}+\vec{q}).
 \eeq 
Using these expressions for the zero- and nonzero-temperature self-energy, the equation which determines the pole of the fermion propagator becomes
\beq \label{eq:RenormalizedDispersionRPA}
E_R(\vec{p}) = | G^{(0)}_{1,2} - \Sigma^{(1)}_{1,2} - \delta \Sigma^{({RPA})}_{1,2} |,
\eeq 
where as is clear from (\ref{eq:DeltaSigmaRPAFinal}), the RPA correction also depends on $E_R$. The solution to (\ref{eq:RenormalizedDispersionRPA}) can be obtained without much difficulty using fixed-point iterations for root-finding.

\section{Continuum RPA approximation}

 The well-known continuum RPA results \cite{GonzalezPhysRevB1999,SonPhysRevB2007} describe $N_f=2$ flavors of four-component Dirac fermions in two spatial dimensions coupled to a scalar potential, $A_0$, which lives in all three spatial dimensions. The action is as follows
\beq 
 S_{\rm E} = - \int \mathrm{d}t \mathrm{d}^2x (\bpsi_\alpha \gamma_0 \partial_0 \psi_\alpha + v_{0,F}\bpsi_\alpha \gamma_i \partial_i \psi_\alpha + iA_0 \bpsi_\alpha \gamma_0 \psi_\alpha ) + 
 \frac{1}{2e^2} \int \mathrm{d}t \mathrm{d}^3x ( \partial_i A_0 )^2,
 \label{eq:ContinuumDiracAction}
\eeq
where $v_{0,F}$ is the bare Fermi velocity and the four-dimensional gamma matrices satisfy the Clifford algebra $\{ \gamma_{\mu}, \gamma_{\nu} \} = 2\delta_{\mu \nu}$.
The resulting Fermi velocity renormalization in RPA approximation is also a logarithmic similar to eq. (\ref{eq:vf_renorm}) with the coefficient in front of the logarithm being equal to
\beq \label{eq:SonRG}
C = \frac{4}{\pi^2 N_f} \( F_1(\lambda) - F_0(\lambda) \),
\eeq
where the rescaled coupling is $ \lambda \equiv e^2 N_f/(16 v_F)$  (we are using the notation of \cite{SonPhysRevB2007}) and
\beq \label{eq:F1SonRG}
F_1(\lambda) = \left\{ \begin{array}{ll}
-(\sqrt{1-\lambda^2}/\lambda)\arccos \lambda -1 +\pi/(2\lambda) & \lambda < 1 \\
-(\sqrt{\lambda^2-1}/\lambda) \log \left( \lambda + \sqrt{\lambda^2-1} \right) -1 +\pi/(2\lambda)  & \lambda > 1
\end{array} \right.,
\eeq 

\beq \label{eq:F0SonRG}
F_0(\lambda) = \left\{ \begin{array}{ll}
-((2-\lambda^2)/(\lambda\sqrt{1-\lambda^2}))\arccos \lambda -2 +\pi/\lambda & \lambda < 1 \\
-((\lambda^2-2)/(\lambda\sqrt{\lambda^2-1})) \log \( \lambda + \sqrt{\lambda^2-1}\) -2 +\pi/\lambda  & \lambda > 1
\end{array} \right.
\eeq

\section{Inter-valley scattering in EFT}
In passing to the continuum theory, each individual fermion creation and annihilation operator in the lattice Hamiltonian can be expressed in terms of the appropriate low-energy modes. We thus introduce the following representation for the annihilation operator
\beq \label{eq:ModeExpansionDirac1}
\hat{a}_{\vec x,\alpha,\sigma}  \approx \frac{1}{L} e^{i\vec{\mathcal{K}_1} \cdot \vec x} \sum_{|\vec k| < \Lambda} e^{i\vec k \cdot \vec x} \hat{a}_{\vec k+ \vec{\mathcal{K}_1},\alpha,\sigma} +  \frac{1}{L} e^{i\vec{\mathcal{K}_2} \cdot \vec x} \sum_{|\vec k| < \Lambda} e^{i\vec k \cdot \vec x} \hat{a}_{\vec k-\vec{\mathcal{K}_2},\alpha,\sigma},
\eeq
where $\alpha$ represents the sublattice and $\sigma$ represents the electron's spin. Here we have expanded the Fourier sum around the two non-equivalent Dirac points  $\vec{\mathcal{K}_1}$ and $\vec{\mathcal{K}_2}$ and have cut each sum off at some large value, $\Lambda$. The expression in (\ref{eq:ModeExpansionDirac1}) allows us to introduce new operators which vary slowly over the unit cell
\beq \label{eq:DiracSpinorBasis}
\hat a_{\sigma,\vec{r},\nu}=e^{i \vec{\mathcal{K}_1}\cdot \vec r}\hat\psi^1_{\sigma,\vec r,\nu}+e^{i \vec{\mathcal{K}_2} \cdot \vec r}\hat\psi^2_{\sigma,\vec r,\nu},
\eeq 
where $\hat\psi^\mu_{\sigma,\vec r,\nu}$ is a component the four-component Dirac spinor
\beq \label{eq:ModeExpansionDirac2}
\hat\psi_{\sigma,\vec r}=(\hat\psi^1_{\sigma,\vec r,1} ,\hat\psi^1_{\sigma,\vec r,2} ,\hat\psi^2_{\sigma,\vec r,1} ,\hat\psi^2_{\sigma,\vec r,2} ),
\eeq
with $\nu$ representing the sublattice and the superscript denoting the valley. In order to describe the ``tunneling'' between a given Dirac point and all three of its neighbouring, non-equivalent Dirac points (see Fig. \textcolor{red}{9c} in  the main text), we rewrite (\ref{eq:ModeExpansionDirac2}) in the form
\beq \label{eq:ModeExpansionDiracFull}
\hat a_{\sigma,\vec{r},\nu}=\frac{1}{3} (e^{i \vec{\mathcal{K}_1}\cdot \vec r}+e^{i \vec{\mathcal{K'}_1}\cdot\vec r}+e^{i \vec{\mathcal{K''}_1}\cdot\vec r})\hat\psi^1_{\sigma,\vec r,\nu}+\frac{1}{3} (e^{i \vec{\mathcal{K}_2}\cdot\vec r}+e^{i \vec{\mathcal{K'}_2}\cdot\vec r}+e^{i \vec{\mathcal{K''}_2}\cdot\vec r})\hat\psi^2_{\sigma,\vec r,\nu},
\eeq 
where $\vec{\mathcal{K'}_i}=\vec{\mathcal{K}_i} + \vec b_1$ and $\vec{\mathcal{K''}_i}=\vec{\mathcal{K}_i} - \vec b_2$ are equivalent Dirac points.

\begin{figure} 
\centering
        \includegraphics[angle=0,width=0.5\textwidth]{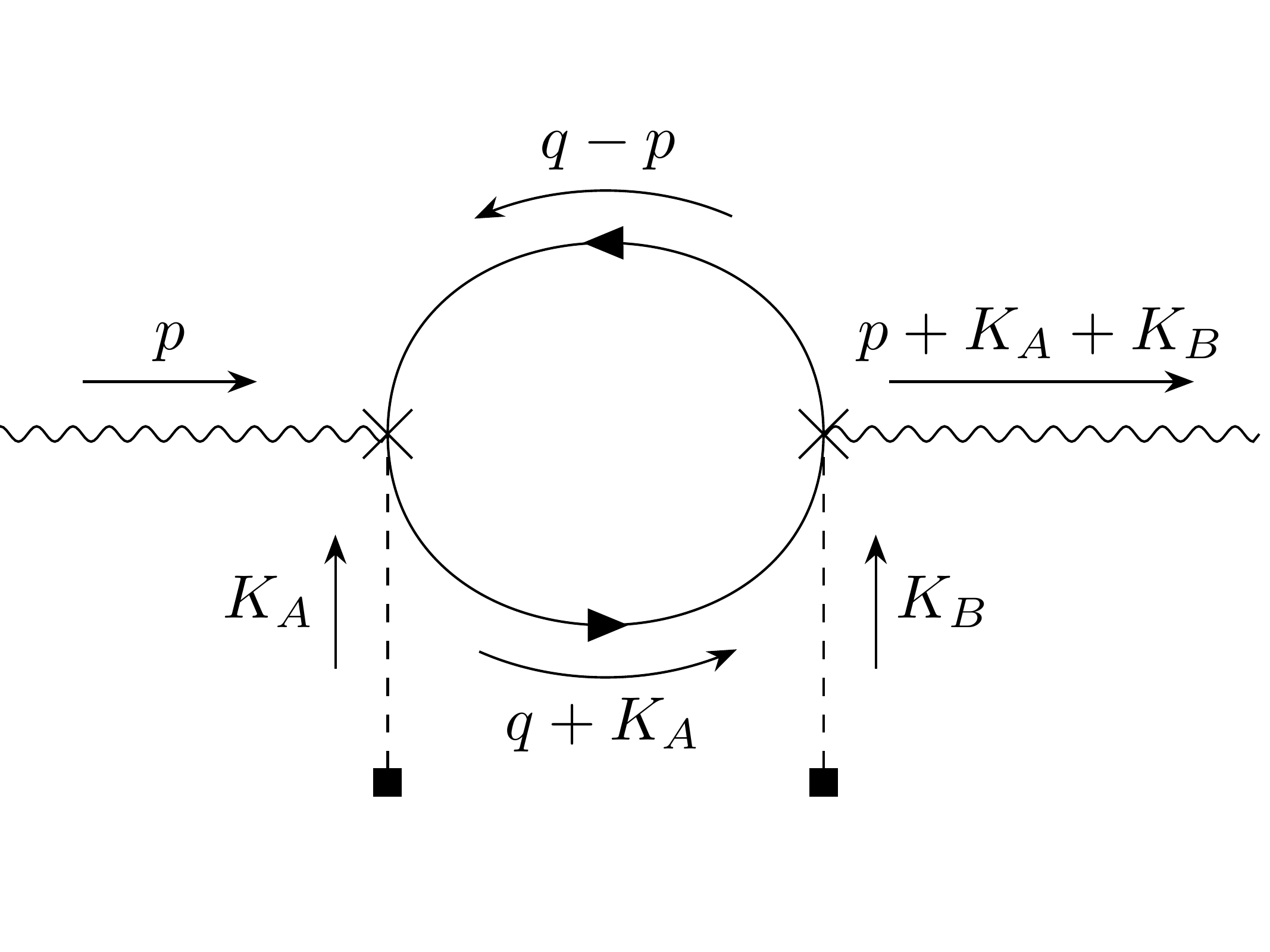}
        \caption{Correction to the polarization $P$, arising from inter-valley scattering.}
        \label{fig:PolarizationShifted}
\end{figure}

\begin{figure} 
\centering
        \includegraphics[angle=0,width=0.5\textwidth]{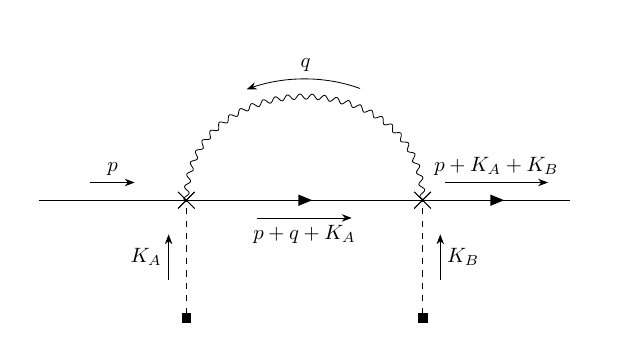}
        \caption{Correction to the self energy, arising from inter-valley scattering.}
        \label{fig:SelfEnergyShifted}
\end{figure}

\begin{figure} 
\centering
        \includegraphics[angle=0,width=0.7\textwidth]{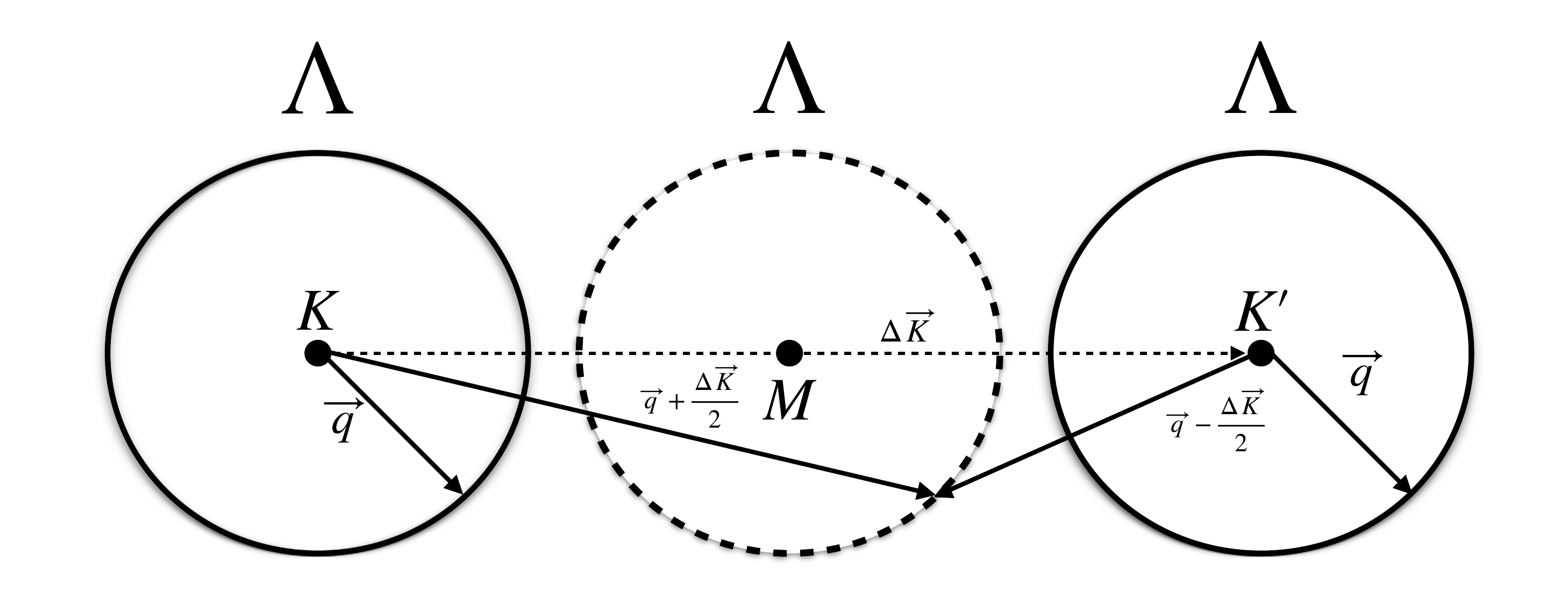}
        \caption{Integration region for the contribution to the polarization stemming from the modified continuum charged current. Each circle represents a region in momentum space of radius $\Lambda$, centered around the two Dirac points and the$ M$ point, respectively.}
        \label{fig:momentumShift}
\end{figure}

We apply this transformation to the electron-electron interaction between unit cells separated by a distance $\vec \Delta$
\beq \label{eq:cellInteraction}
\sum_{\nu,\nu'} V_{\vec x,\vec x+\vec \Delta} ~q_{\vec x, \nu} q_{\vec x+\vec \Delta, \nu'},
\eeq 
where we assume that $|\vec \Delta|$ is large compared with the lattice spacing. In this case, the diagonal and off-diagonal elements of the interaction in sublattice space are nearly identical and thus we can sum over the sublattice indices, $\nu$ and $\nu'$.  This procedure leads to the modification of the expression for the charged current in the EFT 
\beq \label{eq:additionalCurrent}
\tilde {j}_0={\bar\psi_a} \gamma_0(I+\sum_j{e^{(-i \vec x \cdot \vec {\Delta \mathcal{K}_j} \tau_3/2 )} \tau_1 e^{(i \vec x \cdot \vec { \Delta \mathcal{K}_j} \tau_3/2 )} }){\psi_a},
\eeq
where $\tau_j$ are Pauli matrices in valley space and $\vec  { \Delta \mathcal{K}_j}$, $j=1,2,3$ are the lattice vectors which connect a single Dirac point with its three neighbouring, non-equivalent Dirac points.

We finally insert this expression into the polarization in order to see how the continuum expression is modified. We note that the additional term in the current can be written in a more compact way
\beq \label{eq:currentConversion}
e^{(-i \vec x \cdot \vec {\Delta \mathcal{K}_j} \tau_3/2 )} \tau_1 e^{(i \vec x \cdot \vec { \Delta \mathcal{K}_j} \tau_3/2 )} = \tau_+ e^{i \vec x \cdot \vec {\Delta \mathcal{K}_j}} + \tau_- e^{-i \vec x \cdot \vec {\Delta \mathcal{K}_j}},
\eeq
where $\tau_+=\frac{1}{2}(\tau_1+\tau_2)$ and $\tau_-=\frac{1}{2}(\tau_1-\tau_2)$ are matrices acting in valley space.  
The additional term generates a shift of the momentum at the vertex due to the oscillating exponents, which is accompanied by a jump between the valleys. The corresponding diagram is depicted in Fig. \ref{fig:PolarizationShifted} with $\vec K_A, \, \vec K_B=\pm \vec { \Delta \mathcal{K}_j}$. We are, however, only interested in the diagram where the total momentum is conserved, and thus $\vec K_A=-\vec K_B$. The additional terms in the polarization which are generated by the modified vertex are given by
\beq \label{eq:PolarizationShifted1}
\delta P^{(I)}_j(p_0, \vec p)=\int \frac{\mathrm{d}q_0 \mathrm{d}\vec q}{(2 \pi)^3} \Tr \left[{ \gamma_0 \tau_- G(p_0+q_0, \vec p + \vec q + \vec { \Delta \mathcal{K}_j}/2) \gamma_0 \tau_+ G(q_0, \vec q - \vec { \Delta \mathcal{K}_j}/2) }\right],  
\eeq
 where 
 \beq \label{eq:EFTProp}
 G(p_0,\vec p)=\frac{p_0\gamma_0 + \vec p \cdot \vec \gamma v_{F,0} }{p_0^2 + \vec p^2 v^2_{F,0} }
\eeq
is the continuum Euclidean fermion propagator. We choose the following basis for the $\gamma$ matrices 
\beq \label{eq:GammaMatrices}
\gamma_0=\begin{pmatrix}
\sigma_3 & 0 \\
0 & \sigma_3 
\end{pmatrix}, \, \gamma_1=\begin{pmatrix}
\sigma_2 & 0 \\
0 & -\sigma_2 
\end{pmatrix}, \, 
\gamma_2=\begin{pmatrix}
\sigma_1 & 0 \\
0 & \sigma_1 
\end{pmatrix}, \, 
\gamma_3=\begin{pmatrix}
0 & \sigma_2 \\
 \sigma_2 & 0 
\end{pmatrix},
\eeq
where the Pauli matrices $\sigma_i$ act in sublattice space.
The presence of the $\tau_\pm$ matrices implies that only inter-valley terms contribute when performing the trace. If take the origin of the momenta in the fermion propagators to reside at the neighbouring non-equivalent Dirac points (different valleys), the resulting integral will be centered exactly around the $M$ point, as shown in figure \ref{fig:momentumShift}. Evaluating (\ref{eq:PolarizationShifted1}) gives the expression
\beq \label{eq:PolarizationShifted2}
 \delta P^{(I)}_j(p_0, \vec p) \sim  \frac{p_0^2 + (p_1 + {(\Delta \mathcal{K}_j)}_1)^2}{\sqrt{p^2_0 + v^2_{F,0}(\vec p + \vec { \Delta \mathcal{K}_j})^2 }} + \frac{p_0^2 + (p_1 - {(\Delta \mathcal{K}_j)}_1)^2}{\sqrt{p^2_0 + v^2_{F,0}(\vec p - \vec { \Delta \mathcal{K}_j})^2 }}  
\eeq
which clearly violates the rotational symmetry of the EFT. The problem arises due to the fact that the linear, low-energy approximation to the lattice propagator $G(p_0, \vec p)$ (\ref{eq:EFTProp}) is not appropriate for (\ref{eq:PolarizationShifted1}), since shifts in momentum space $\vec { \Delta \mathcal{K}_j}$ are quite large and on the order of the cutoff. This means that we need to add higher-dimensional kinetic terms to the QED Lagrangian in order to reproduce the dispersion relation around the $M$ point and thus obtain a better description for inter-valley scattering. 

As, with the polarization, the modified current also gives a contribution to the one-loop self energy (Fig. \ref{fig:SelfEnergyShifted}) 
\beq \label{eq:SelfEnergyShifted}
\delta \Sigma^{(I)}_j(p_0, \vec p)=\int \frac{\mathrm{d}q_0 \mathrm{d}\vec q}{(2 \pi)^3}  \gamma_0 \tau_- G(q_0, \vec q) \gamma_0 \tau_+ \frac{1}{\vec p - \vec q - \vec { \Delta \mathcal{K}_j}} \sim p \ln \frac{\Lambda}{|\vec p -\vec { \Delta \mathcal{K}_j}  |}.  
\eeq
Again, we only consider diagrams where the total momentum is conserved, $\vec K_A=-\vec K_B$. Due to the momentum shift, the correction to $v_F$ is proportional to $ \ln \frac{\Lambda}{|\vec p -\vec { \Delta \mathcal{K}_j}  |}$, which goes like $\ln \frac{\Lambda}{ \Delta \mathcal{K}_j}+ O(|\vec{p}|/\Delta \mathcal{K}_j)$ in the limit $|\vec{p}| \rightarrow 0$. The constant can be absorbed into the cutoff, and the resulting corrections to the Fermi velocity are suppressed, as they are $O(|\vec{p}|/\ln |\vec{p}|)$ with respect to the leading logarithmic divergence of $v_F$. One can thus conclude that these corrections from the modified current are considerably smaller for the one-loop self-energy than for the polarization (\ref{eq:PolarizationShifted2}).

\section{Comment on regularization of UV divergences}
We comment here on the effects due to the finite cutoff and on the choice of regularization scheme used in continuum perturbative calculations. We also discuss the implications for lattice perturbation theory as well as QMC.

Previous studies in the continuum have typically relied on dimensional regularization for the calculation of divergent integrals. The most notable example is the polarization bubble, which is naively linearly divergent. On the lattice, however, one introduces an intrinsic ultraviolet cutoff of the order of the inverse lattice spacing, $\Lambda \sim a^{-1}$. It is well-known that in QED, such a hard cutoff introduces inconsistencies. This is due to the violation of the Ward identity which results from the photon receiving a mass proportional to the cutoff, $m_{\gamma} \propto e \Lambda$ in $(3+1)$ dimensions \cite{Peskin:257493}. As a result, dimensional regularization, or even Pauli-Villars, are preferable regularization schemes in this case. These same issues were debated previously in many-body perturbative calculations where the linear approximation to the dispersion around the Dirac points has been used \cite{Mishchenko_2008,PhysRevB.82.235402}, especially in the calculation of the conductivity of graphene. 

Here, motivated by the comparison with LPT and QMC data, we still argue that a hard cutoff scheme should be used. The violation of the Ward identities is not really a problem if we are interested in reproducing LPT results. In fact, the lattice polarization does not have the general form $P\sim \vec k^2 F(k)$, as it contains inter-valley scattering and finite-cutoff effects. To keep the theory self-consistent, we should modify the current according to   (\ref{eq:additionalCurrent}). It is thus natural to ignore the violation of the naive Ward identities as they hold in QED due to the fact that the relation $\partial_\mu j^\mu=0$ no longer holds. 

Below, we show that the introduction of a finite cutoff leads to exactly the same effects in the polarization which we observed during the comparison of the continuum and LPT results. 
We start from the ordinary polarization bubble (Fig. \ref{fig:Polarization}) in the EFT and first perform the integration over the loop frequency $q_0$ to get
\beq \label{eq:PolarizationContinuumInt}
P^{(\Lambda)}(p_0,\vec p) = 2 N_f v_{F,0} \int_0^\Lambda \frac{\mathrm{d}\vec q}{(2 \pi)^2}  \left( {\frac{|\vec q|+ v_{F,0} |\vec q + \vec p| }{p_0^2  + v^2_{F,0} (|\vec q| + |\vec q + \vec p|)^2} } \right)   \left( {1 - \frac{\vec q (\vec p + \vec q)}{|\vec q| |\vec q + \vec p|}} \right).
\eeq 
Note that the cutoff is imposed only on the spatial components of the loop momenta and the integral over frequencies is taken over the entire real axis.
Instead of exactly computing the integral in (\ref{eq:PolarizationContinuumInt}), we use the fact that it is convergent in the limit $\Lambda \rightarrow \infty$ which is evident from the continuum expression
\begin{equation} \label{eq:PolarizationContinuum}
P(p_0,\vec p) = \frac{e^2 N_f}{8} \frac{\vec p^2}{\sqrt{p_0^2 + v_{F,0}^2 \vec p^2}}
\end{equation}
Thus, we can write 
\beq \label{eq:PolarizationContinuumHardCutoff1}
P^{(\Lambda)}(p_0,\vec p) = P(p_0,\vec p) + \delta P^{(\Lambda)}(p_0,\vec p), 
\eeq 
where the first term on the right-hand side of (\ref{eq:PolarizationContinuumHardCutoff1}) is identical to (\ref{eq:PolarizationContinuum}), which was computed in dimensional regularization. The correction $\delta P^{(0)}_{\Lambda}(p_0,\vec p)$ can be expanded in powers of $\Lambda^{-1}$:
\beq \label{eq:PolarizationContinuumHardCutoff2}
\delta P^{(\Lambda)}(p_0,\vec p) = -\frac{N_{\rm{f}}}{v_{F,0}} \frac{7}{16\pi} \frac{\vec p^2}{\Lambda} + O\left(\Lambda^{-2} \right),
\eeq
where the first correction due to the hard cutoff is frequency independent, while the higher-order terms will, in general, depend on frequency. This correction multiplied by the bare Coulomb propagator, which is proportional to $\sim 1/|\vec p|$ gives the term proportional to $\sim |\vec p|$ near the $\Gamma$ point, exactly as we saw in the figures \textcolor{red}{9a} and \textcolor{red}{9d-f} in the main text. The coefficient in front of this term disappears in the limit $\Lambda \rightarrow 0$ exactly as predicted in the Fig. \textcolor{red}{9f} in the main text.

The same technique can be also applied to the bare first-order fermion self-energy. Note that we can not shift the loop momentum after the introduction of the Feynman parameters as is usually done. Performing this calculation we obtain the usual one-loop leading term 
\beq \label{eq:ContinuumBareSelfEnergy}
\Sigma(p_0,\vec p) = v_{F,0} \gamma_i p_i \frac{\alpha}{4} \log\left( \frac{\Lambda}{|\vec{p}|} \right),
\eeq
and corrections
\beq \label{eq:ContinuumBareSelfEnergyCorrections}
\delta \Sigma^{(\Lambda)}(p_0,\vec p) = v_{F,0} \gamma_i p_i \left( C_0 + C_1 \frac{|\vec{p}|^2}{\Lambda^2} + O\left(\frac{|\vec{p}|^4}{\Lambda^4}\right) \right).
\eeq
We note that the hard cutoff does not generate a term in the self-energy proportional to $k_0 \gamma_0$ at this order. The constant $C_0$ can be absorbed in the cutoff and thus the leading correction is $O(|\vec p|^2/\Lambda^2)$. It is again smaller then the correction to the polarization, which is consistent with the  appearance of the lattice corrections starting at the RPA level when comparing the continuum calculations to LPT and QMC data.

\begin{figure} 
\centering
        \includegraphics[angle=0,width=0.5\textwidth]{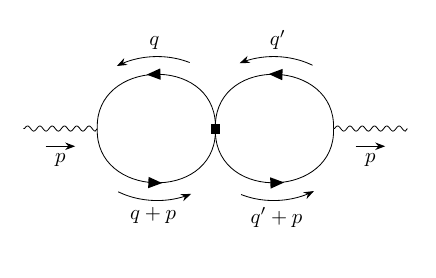}
        \caption{Lowest-order correction to the polarization $P$, arising due to the addition of four-Fermi terms in the low-energy continuum action.}
        \label{fig:PolarizationCorrectionFourFermi}
\end{figure}

\section{Four-Fermi terms in the continuum EFT}
Another way to modify the EFT is to add higher-dimensional operators \cite{DrutSonPhysRevB2008}. One can write down twelve four-Fermi operators, each having a mass-dimension of four and are thus ``irrelevant" in the RG sense. For completeness, we write down the general form of these terms in the Euclidean action
\beq \label{eq:FourFermiGeneral}
S_{\rm{E},\rm{int}} = \frac{u_i}{2N_f} \int \mathrm{d}t \mathrm{d}^2x (\bpsi_\alpha \Gamma_i \psi_\alpha )^2,
\eeq
where the coupling $u_i$ is proportional to the inverse UV cutoff, $\alpha$ is a flavor index, and $\Gamma_i$ determines the vertex structure in spinor space. Taking into account the structure of the four-component Dirac spinor (\ref{eq:DiracSpinorBasis}) and the fact that only the $\gamma_3$ matrix is non-diagonal in valley space (\ref{eq:GammaMatrices}), the obvious candidates for the the description of inter-valley scattering are the terms of the form $\Gamma_i = \gamma_3 \tilde{\Gamma}_i$, where $\tilde{\Gamma}_i \in \{ 1, \gamma_0, \gamma_i, \gamma_5, \gamma_0 \gamma_i, \gamma_0 \gamma_5 \}$ and $\gamma_5 \equiv \gamma_0 \gamma_1 \gamma_2$.

The polarization in the continuum Dirac field theory receives a contribution at linear order in each $u_i$, which is depicted in Fig. \ref{fig:PolarizationCorrectionFourFermi}. These diagrams represent two terms, a single-trace term and a double-trace term
\beq \label{eq:GeneralPolarizationCorrection}
P^{(u)}(p_0,\vec p) = P^{(u)}_a(p_0,\vec p) + P^{(u)}_b(p_0,\vec p), 
\eeq
where 
\beq \label{eq:DoubleTracePolarizationCorrection}
P^{(u)}_a(p) = u_i \frac{e^2}{{v_F}^2} N_f \int \frac{\mathrm{d}^3q}{(2\pi)^3} \Tr \left( G(q+p) \gamma_0 G(q) \Gamma_i \right) \int\frac{\mathrm{d}^3q'}{(2\pi)^3} \Tr \left( G(q'+p) \Gamma_i G(q') \gamma_0 \right), \\
\label{eq:SingleTracePolarizationCorrection}
P^{(u)}_b(p) = -u_i \frac{e^2}{{v_F}^2}  \int \int \frac{\mathrm{d}^3q}{(2\pi)^3} \frac{\mathrm{d}^3q'}{(2\pi)^3} \Tr \left( G(q) \Gamma_i G(q') \gamma_0 G(q'+p) \Gamma_i G(q+p) \gamma_0  \right).
\eeq 
Here we use the ``relativistic", four-vector notation $q=(q_0, v_{F} \vec q)$ for all momenta, while the trace is performed over spinor indices with the appropriate powers of $N_{\rm{f}}$ resulting from tracing over flavor indices.
We show in detail how the integrals are computed for the four-Fermi term associated with $\Gamma_i = \gamma_3$ and simply state the final results for the other terms.

One can easily show that (\ref{eq:DoubleTracePolarizationCorrection}) vanishes due to the fact that $\gamma_3$ is block off-diagonal while the remaining gamma matrices are all block-diagonal. This also holds true for all other couplings involving $\gamma_3$. To evaluate (\ref{eq:SingleTracePolarizationCorrection}), we note that the two loop integrals factor and thus the term takes the form
\beq 
P^{(u)}_b(p) = -u_i \frac{e^2}{{v_F}^2} \tilde{\epsilon}_{\mu \nu \kappa \lambda} I(p)_{\mu \nu} I(p)_{\kappa \lambda},
\eeq
where we have introduced
\beq \label{eq:EpsilonTrace}
\tilde{\epsilon}_{\mu \nu \kappa \lambda} \equiv \Tr \left( \gamma_{\mu} \gamma_0 \gamma_{\nu} \gamma_3 \gamma_{\lambda} \gamma_0 \gamma_{\kappa} \gamma_3\right), 
\eeq 
and
\beq \label{eq:IntegralTensor}
I(p)_{\mu \nu} \equiv \int \frac{\mathrm{d}^3q}{(2\pi)^3} \frac{(q+p)_{\mu}q_{\nu}}{(q+p)^2q^2}.
\eeq
Using the fact that the matrix $\gamma_3$ anticommuntes with all other gamma matrices together with the identity $\Tr (\gamma_{\mu} \gamma_{\nu} \gamma_{\kappa} \gamma_{\lambda}) = 4 ( \delta_{\mu \nu} \delta_{\kappa \lambda} - \delta_{\mu \kappa} \delta_{\nu \lambda} + \delta_{\mu \lambda} \delta_{\nu \kappa} )$, one can evaluate (\ref{eq:EpsilonTrace}). The result is given by
\beq \nn 
\tilde{\epsilon}_{\mu \nu \kappa \lambda} = -4\left( \delta_{\mu \nu} \delta_{\lambda \kappa} - \delta_{\mu \lambda} \delta_{\nu \kappa} + \delta_{\mu \kappa} \delta_{\nu \lambda} \right) + 8 \delta_{0 \lambda}  \left( \delta_{\mu \nu} \delta_{0 \kappa} - \delta_{\mu 0} \delta_{\nu \kappa} + \delta_{\mu \kappa} \delta_{\nu 0} \right) \\ \label{eq:EpsilonTraceResult} - 8 \delta_{0 \nu}  \left( \delta_{\mu \lambda} \delta_{0 \kappa} - \delta_{\mu 0} \delta_{\lambda \kappa} + \delta_{\mu \kappa} \delta_{\lambda 0} \right), 
\eeq 
To compute the integral over the loop momentum in (\ref{eq:IntegralTensor}) using dimensional regularization, one uses the standard procedure of combining the factors in the denominator using Feynman parameters, shifting the loop momentum to make the denominator an even function of $q$, and then dropping terms in the numerator which are linear in $q$. The result of this procedure can be written as
\beq \label{eq:IntegralTensorResult}
I(p)_{\mu \nu} = - \frac{\sqrt{p^2}}{64} \left( \delta_{\mu \nu} + \frac{p_{\mu} p_{\nu}}{p^2} \right).
\eeq 
Now, contracting this expression with the tensor (\ref{eq:EpsilonTraceResult}) we obtain the result
\beq \label{eq:PolarizationCorrectionResult}
P^{(u)}_{(\gamma_3)}(p_0,\vec p) = \frac{u_1 {e^2}}{256 {{v_F}^2}} \vec p^2.
\eeq
The correction in (\ref{eq:PolarizationCorrectionResult}) is quadratic in spatial momentum, thus it contributes to the linear term in $\mathcal{P}$, but does not affect the constant shift near the $\Gamma$ point (see Fig. \textcolor{red}{9a} in the main text). As we discussed previously, the linear term in the lattice polarization is a lattice cutoff effect, while inter-valley scattering is responsible for the constant shift at the $\Gamma$ point. All of this implies that the $\Gamma_i = \gamma_3$ four-Fermi term cannot describe the inter-valley scattering which we see in the lattice polarization.

We proceed with the contributions from the remaining five terms which couple the two Dirac points. The calculations are similar to those performed above and the results are 
\beq
P^{(u)}_{(\gamma_0 \gamma_3)}(p_0,\vec p) =  -\frac{u_2e^2}{256 {{v_F}^2}} (\vec p^2-p^2_0)\left( 1 - \frac{p^2_0}{p^2} \right), \\
P^{(u)}_{(i\gamma_3 \gamma_5)}(p_0,\vec p) = \frac{u_3 e^2}{256 {{v_F}^2}} \vec p^2, \\
P^{(u)}_{(i\gamma_3 \gamma_i)}(p_0,\vec p) = - \frac{2 u_4 e^2}{256 {{v_F}^2}} \vec p^2 \left( 1 - \frac{p^2_0}{p^2} \right), \\
P^{(u)}_{(i\gamma_3 \gamma_0 \gamma_5)}(p_0,\vec p) = \frac{ u_5 e^2}{256 {{v_F}^2}}  (\vec p^2-p^2_0) \left( 1 - \frac{p^2_0}{p^2} \right), \\
P^{(u)}_{(i\gamma_3 \gamma_0 \gamma_i)}(p_0,\vec p) = \frac{2 u_6 e^2}{256 {{v_F}^2}} \vec p^2 \left( 1 - \frac{p^2_0}{p^2} \right).
\eeq
Thus, one can see that these four-Fermi operators also contribute to the linear term in $\mathcal{P}$, but do not affect the constant shift at the $\Gamma$ point.

An additional problem is that the other six four-Fermi terms which do not involve $\gamma_3$ also contribute to the polarization at this order. For the operators corresponding to $\gamma_0$ and $\gamma_i$, both the single-trace term and double-trace term are nonzero. The final contributions to the polarization are as follows 
\beq
P^{(u)}_{(\mathbb{1})}(p_0,\vec p) = -\frac{u_7 e^2}{256 {{v_F}^2}} \vec p^2, \\
P^{(u)}_{(\gamma_0)}(p_0,\vec p) = \frac{u_8 e^2}{256 {{v_F}^2}} \left\{ -(\vec p^2-p^2_0) \left( 1 - \frac{p^2_0}{p^2} \right) + 4 N_f\vec p^2\right\}, \\
P^{(u)}_{(i\gamma_5)}(p_0,\vec p) = \frac{u_9 e^2}{256 {{v_F}^2}} \vec p^2, \\
P^{(u)}_{(i\gamma_0\gamma_5)}(p_0,\vec p) = -\frac{u_{10} e^2}{256 {{v_F}^2}} (\vec p^2-p^2_0) \left( 1 - \frac{p^2_0}{p^2} \right), \\
P^{(u)}_{(\gamma_i)}(p_0,\vec p) = \frac{u_{11} e^2}{256 {{v_F}^2}} \left( 2\frac{\vec p^4}{p^2} + 4N_f\frac{p^2_0 \vec p^2}{p2} \right), \\
P^{(u)}_{(\gamma_0 \gamma_i)}(p_0,\vec p) = \frac{2u_{12} e^2}{256 {{v_F}^2}} p^2 \left( 3 + \frac{p^4_0}{p^4} \right).
\eeq
Disentangling each of these contributions and their relation to the lattice polarization is a difficult task. For instance, it is not possible to distinguish the contributions $P^{(u)}_{(\mathbb{1})}$ and $P^{(u)}_{(\gamma_3)}$. One thus requires additional information in order to fix the inclusion of the four-Fermi terms in the continuum EFT.

It is known, however, that four-Fermi terms can actually play a different role in the EFT. Previous studies have shown that short-range interactions on the hexagonal lattice can generate some of the four-Fermi terms enumerated above once one passes to the low-energy continuum EFT \cite{PhysRevB.79.085116}. We now list the terms generated by the on-site Hubbard interaction as well as the nearest-neighbor interaction.

The Hubbard interaction on the hexagonal lattice can be written as 
\beq \label{eq:HubbardStandardFormHexagonal}
\hat{H}_U = \frac{U}{4} \sum_x \left\{ \left( \hat{n}_{x,A} + \hat{n}_{x,B}\right)^2 + \left( \hat{n}_{x,A} - \hat{n}_{x,B}\right)^2 \right\},
\eeq
where the sum runs over the unit cells of the lattice and we have introduced the number operator, $\hat{n}_{x,A} \equiv \hat{a}^{\dagger}_{x,A,\uparrow} \hat{a}_{x,A,\uparrow} + \hat{a}^{\dagger}_{x,A,\downarrow} \hat{a}_{x,A,\downarrow}$, restricted to sublattice $A$ with the same operator $\hat{n}_{x,B}$  on sublattice $B$.  As shown previously, the fermion operators can be expressed in terms of the degrees of freedom of the low-energy EFT (\ref{eq:DiracSpinorBasis}). We apply this transformation to (\ref{eq:HubbardStandardFormHexagonal}), and drop oscillating terms which contain $\cos(\vec  { \Delta \mathcal{K}_j} \vec x)$ or $\sin(\vec  { \Delta \mathcal{K}_j} \vec x)$, as they are sub-leading at low energies. The first term in (\ref{eq:HubbardStandardFormHexagonal}), can be expressed as 
\beq \label{eq:HubbardLowEnergy1}
\sum_{\alpha} \int \mathrm{d}^2x \left\{  (\bpsi_\alpha \gamma_0 \psi_\alpha )^2  - \frac{1}{2}(\bpsi_\alpha \gamma_2 \gamma_3 \psi_\alpha )^2 - \frac{1}{2}(\bpsi_\alpha \gamma_0 \gamma_1 \gamma_3 \psi_\alpha )^2 \right\},
\eeq
where the sum runs over spin and we have passed to the continuum limit by replacing the sum with an integral, $a^2\sum_x \sim \int \mathrm{d}^2x $, where $a$ is the lattice spacing of the hexagonal lattice. In a similar way, the second term in (\ref{eq:HubbardStandardFormHexagonal}) can be written as 
\beq \label{eq:HubbardLowEnergy2}
\sum_{\alpha} \int \mathrm{d}^2x \left\{  (\bpsi_\alpha \psi_\alpha )^2  - \frac{1}{2}(\bpsi_\alpha \gamma_0 \gamma_2 \gamma_3 \psi_\alpha )^2 - \frac{1}{2}(\bpsi_\alpha \gamma_1 \gamma_3 \psi_\alpha )^2 \right\}.
\eeq
Putting together (\ref{eq:HubbardLowEnergy1}) and (\ref{eq:HubbardLowEnergy2}), one can see that the following four-Fermi terms are added to the low-energy action by the Hubbard interaction 
\beq \label{eq:HubbardActionTerms}
S_U = G_U \sum_{\alpha} \int \mathrm{d}^2x \left\{  (\bpsi_\alpha \psi_\alpha )^2 +  (\bpsi_\alpha \gamma_0 \psi_\alpha )^2 - \frac{1}{2}(\bpsi_\alpha \gamma_i \gamma_3 \psi_\alpha )^2 - \frac{1}{2}(\bpsi_\alpha \gamma_0 \gamma_i \gamma_3 \psi_\alpha )^2 \right\},
\eeq 
where we have introduced the effective coupling $G_U$, which has a negative mass dimension and is proportional to the Hubbard interaction $U$ in (\ref{eq:HubbardStandardFormHexagonal}).

The nearest-neighbor density-density interaction also plays a role in the low-energy continuum action. The lattice interaction Hamiltonian can be written as follows
\beq
\label{eq:NearestNeighborStandardFormHexagonal}
\hat{H}_V = \frac{V}{4} \sum_{\vec \delta_i} \sum_x \left\{ \left( \hat{n}_{x,A} + \hat{n}_{x+\delta_i,B}\right)^2 - \left( \hat{n}_{x,A} - \hat{n}_{x+\delta_i,B}\right)^2 \right\},
\eeq
where the first sum runs over the three nearest-neighbor vectors $\delta_i$. Repeating the same steps as for the on-site interaction and then summing over the nearest-neighbor vectors, one obtains 
\beq \label{eq:NearestNeighborActionTerms}
S_V = G_V \sum_{\alpha} \int \mathrm{d}^2x \left\{  (\bpsi_\alpha \gamma_0 \psi_\alpha )^2  -  (\bpsi_\alpha \psi_\alpha )^2 \right\},
\eeq 
where we have introduced the effective coupling $G_V$ arising from the nearest-neighbor interaction.

We have thus seen how a subset of the allowed four-Fermi operators in the continuum effective field theory naturally arise from the short-range interactions inherent in our many-body Hamiltonian.
However, it has been previously shown that the fermion self-energy receives only small corrections due to local interactions and the higher-dimensional operators that they generate in the continuum \cite{PhysRevB.79.201403}. This is confirmed by QMC and higher-order LPT data, where we have found that these couplings modify the fitted value of the cutoff and do not affect the coefficient in front of the logarithm in the expression for the renormalized Fermi velocity.

Another, simpler approach that modifies the EFT is to simply absorb lattice-scale physics as well as possible higher-order perturbative corrections into a ``renormalized" coupling constant. This procedure excludes a mapping of the microscopic details of the lattice theory to the continuum theory. In general, with the modification of a single constant $\alpha$, one can not fully absorb all corrections coming from  lattice-scale physics. However, in the particular case of the renormalization of $v_F$, which boils down to a single coefficient in front of the logarithm, this procedure  seems reasonable. If we use the one-loop expression for $v_F$ in the EFT \ref{eq:ContinuumBareSelfEnergy} and match it to lattice RPA (potential variant II), we get $\alpha^{(1)}_{renorm.}=1.372$. Doing the same with QMC yields $\alpha^{(2)}_{renorm.}=0.626$, and thus we see a substantial reduction in the ``renormalized" coupling in comparison with its bare value $\alpha=2.504$ in suspended graphene. We can in principle perform this matching also at the level of the continuum RPA, providing that the QMC coefficient $C$ does not exceed $2/\pi^2$, which is the saturation value of the expression \ref{eq:SonRG} for continuum RPA. This is true for at least low-temperature data, though the high-temperature data falls outside of this limit (as seen in Fig.~2 of the main text). Thus it is unlikely that one can obtain a satisfactory fit of finite-temperature effects with this approach while also signaling that the continuum theory is not useful beyond the one-loop level. 


\end{document}